\documentclass[twocolumn]{aastex631}

\usepackage{textcomp}
\usepackage{amsmath}
\usepackage{xspace}
\usepackage{ulem}
\def\vplanet{\texttt{\footnotesize{VPLanet}}\xspace}
\def\volcgases{\texttt{\footnotesize{VolcGases}}\xspace}

\def\stellar{\texttt{\footnotesize{STELLAR}}\xspace}

\def\vsp{\texttt{\footnotesize{VSPACE}}\xspace}

\begin{document}

\title{Investigation of Venus’ thermal history, crustal evolution, and core dynamics with a coupled interior-lithosphere-atmosphere model}

\author{Rodolfo Garcia}
\affiliation{Department of Astronomy, University of Washington, 3910 15th Avenue NE, Seattle, WA 98195, USA} 

\author[0000-0001-6487-5445]{Rory Barnes}
\affiliation{Department of Astronomy, University of Washington, 3910 15th Avenue NE, Seattle, WA 98195, USA} 

\author{Peter E. Driscoll}
     \affiliation{Earth and Planets Laboratory, Carnegie Institution for Science, 5241 Broad Branch Rd, Washington, DC, 20015, USA}

\author{Victoria S. Meadows}
\affiliation{Astronomy Department, University of Washington, Box 315180, Seattle, WA 98195}

\author{Megan Gialluca}
\affiliation{Astronomy Department, University of Washington, Box 315180, Seattle, WA 98195}

\begin{abstract}
    We simulate Venus' evolution with a coupled one-dimensional solar-atmosphere-lithosphere-mantle-core model to predict currently unobservable features and its eruptive mass flux.
    We identified four distinct evolutionary pathways that simultaneously match the atmospheric abundances of water and carbon dioxide as well as the lack of a core dynamo. These scenarios are characterized by I) generally monotonic cooling, II) a low mantle melt fraction in which Venus' volcanically active phase is ending, III) a small inner core, and IV) oscillations of internal properties.
    Through random forest classification we determined that the key parameters that distinguish these types are the initial mantle water abundance, the mantle viscosity, the dehydration stiffening strength, the eruption efficiency, and the melting point of the core.
    In each of the plausible histories, Venus retains at least one Earth ocean's worth of water in its mantle and remains volcanically active today. Venus' lack of a current geodynamo allows thermal histories with an initially large inner core in our parameter sweep.
    In 88\% of plausible histories we found that Venus possessed a past magnetic field. 
    The results strongly disfavor recent high eruption rate estimates, but are consistent with lower estimates. Current resurfacing estimates also strongly disfavor the low melt scenario, implying that Venus is not nearly volcanically ``dead.'' These predictions are testable with anticipated data and the model can be applied to exoplanets to predict their properties.
\medskip
\end{abstract}

\section{Introduction}
Venus is broadly similar to Earth in terms of bulk density, mass, and orbit, but the two planets' present states are vastly different.
Most notably, present-day Earth maintains surface liquid water, has a 1 bar nitrogen/oxygen atmosphere, and a core-generated magnetic field, while Venus' surface is dry, with 92 bars of carbon dioxide in its atmosphere, and no detectable core dynamo.
These properties manifest another stark difference between the two planets: Earth incontrovertibly exhibits active biology, while Venus does not.
Despite decades of research, the origin of these  disparities remains a mystery, in part due to the sparse observational data pertaining to Venus' interior, which has resulted in a relative scarcity of planetary evolution models that explain the apparent divergence of Earth and Venus’ evolutions. 

Given the two planet's similarities in bulk properties and position in the Solar System, the different atmospheric states may just be due to divergent evolution \citep{kaula_1990, driscoll_divergent_2013,WellerKiefer2025}. In other words, the planets may have formed with similar bulk compositions and orbits, but small initial differences between them were amplified over the intervening 4.5 Gyrs. While reconstructing the history of any planet is difficult, Venus poses unique challenges due to a surface-obscuring cloud deck as well as surface temperatures and pressures that challenge detailed, long-term {\it in situ} analysis. In spite of this inaccessibility, modeling efforts have been able to reproduce many aspects of Venus' current state by relying on insights provided by laboratory experiments \citep[e.g.,][]{armann_tackley_2012,driscoll_divergent_2013,orourke_prospects_2018,Way2023}. 

The recent detections of surface morphological changes imply Venus remains tectonically active today \citep{herrick_surface_2023,sulcanese_evidence_2024}. Crucially, their estimated eruptive mass fluxes may also provide constraints on Venus' history. Thus, models of the eruptive mass over time may also provide insight into the origin, magnitude, and frequency of eruptions. A major goal of the research presented here is to generate a model that reproduces the estimated eruptive mass fluxes with a full geophysical and geochemical model.

To that end, we present a one-dimensional ``whole planet model'' \citep[e.g.][]{foley_whole_2016} of Venus that tracks its geodynamo, thermal evolution,
geochemistry, surface properties, and atmospheric abundances over the age of the Solar System and assumes Venus has always operated in a stagnant lid tectonic mode \citep{solomatov_moresi_1996}. Our model is able to reproduce several fundamental observations of Venus as an end state of four identified qualitatively different paths that Venus may have followed. These validated solutions also make falsifiable predictions that may be testable with future space missions such as {\it DAVINCI} \citep{Garvin_2022}, {\it VERITAS} \citep{smrekar_2022}, and {\it EnVision} \citep{ESA2021}.

For the rest of this introduction, we review the properties of modern Venus, as well as previous models of Venus' atmosphere, surface, and interior. Our model is presented in Sec.~\ref{sec:modeldesc} and our numerical procedures in Sec.~\ref{sec:methods}.
Section \ref{sec:results} describes our results and defines the four types of planetary evolution that reproduce some aspects of modern Venus. We discuss the application of these results to Venus and other planets in the context of previous work and future missions in Sec.~\ref{sec:discussion}. Finally, we summarize our results and conclusions in Sec.~\ref{sec:conclusion}.

\subsection{Modern Venus}

We begin by taking stock of modern Venus' properties. Its atmosphere is dominated by about 92 bars of carbon dioxide and about 3 bars of nitrogen. Additionally, the atmosphere contains trace amounts of water ($\sim$3 mbar), as well as other species such as sulfur dioxide, argon, carbon monoxide, and many others \citep{johnson_deoliveira_2019}. The relative abundances of these species place important constraints on geochemical cycling, photochemical processes, and atmospheric escape.

The surface temperature and pressure of Venus are estimated to be 760 K and 96.5 bars, respectively \citep{taylor_venus:_2018, 2017aeil.book.....C}. Significant research has explored the details of the atmosphere's structure, but a complete review of the Venusian atmosphere is not appropriate here. Instead we refer the reader to \citet{2017aeil.book.....C} and \citet{taylor_venus:_2018}.

The Venusian atmosphere is slowly being lost to space. {\it Pioneer Venus Orbiter} data reveal that Venus is losing about $7\times10^{26}$ ions/s \citep{brace_1980,brace_1982}. Later, the {\it Venus Express} mission detected the escape of H$^+$, He$^+$, and O$^+$ ions, with a total hydrogen ion loss estimated to be about $10^{25}$ per second \citep{barabash_2007}, a value similar to Earth \citep{taylor_venus:_2018}. These loss rates are negligible in relation to the mass of the Venusian atmosphere, but the loss of hydrogen ions does slowly remove a constituent of water molecules, which can be an important factor in the geodynamical evolution of Venus.

Measurements from the {\it Pioneer Venus Orbiter} could not detect a magnetic moment, $\mathcal{M}$, of Venus \citep{russell_1980,phillips_russell_1987}, providing only an upper limit of $\mathcal{M} < 8.4\times 10^{10}$ T m$^3$, or $<10^{-5}$ times the value of modern Earth's magnetic moment, $\mathcal{M_\oplus}$ \citep{tikoo_evans_2022}. This upper limit is typically interpreted to mean that there is no active dynamo in Venus' core today, which is an influential constraint on Venus' history, as shown below. Note that Venus' upper atmosphere does generate an ionospheric magnetic field induced by the solar wind \citep{bridge_1967}, but as this field is clearly not generated in the core, we neglect its presence in our models.

In addition to the atmospheric and magnetic properties, Venus' surface also displays numerous noteworthy features. For example, global maps of Venus generated by the {\it Magellan} orbiter \citep{ford_pettengill_1992} made it clear that the surface lacks many of the spatial and topographical features of a planet with a mobile lid: namely spreading ridges and subduction zones. Instead Venus' surface is mostly covered in basaltic plains \citep{ivanov_surface_2018} that are not broken into mobile plates, though there appear to be some regions that may have localized mobility \citep{byrne_globally_2021}.  The hypsometry of Venus is unimodal, consistent with a lack of continental plates, unlike the bimodal hypsometry of Earth \citep{smrekar_venus_2018}.

The {\it Magellan} surface maps also confirmed that Venus' surface possesses relatively few craters. Follow-up studies have found that the distribution of craters is consistent with a surface age of less than 1 Gyr  \citep{1997veii.conf..969M}. This relatively young age represents a strong constraint on any geodynamic model of Venus, requiring active resurfacing over the past Gyr. While the surface appears to mostly be smooth plains, about 10\% of the surface consists of the tesserae highlands \citep{basilevsky_head_1995}, which may be felsic in composition \citep{smrekar_venus_2018,Widemann2023}. This dichotomy also presents evidence of past processes and can constrain modeling efforts.

Venus' surface shows some evidence of ongoing tectonic activity, suggesting that the convecting mantle beneath has some surface expression. For example, \citet{davaille_experimental_2017} argue that plume-driven localized subduction is occurring at coronae, while \citet{Adams2022} and \citet{Adams2023} demonstrate that rift zones can also be important sites of delamination. The coronae on Venus in general present sites of complex thermal-lithosphere interactions \citep{gulcher_corona_2020} and, with Venusian rift zones, manifest high, Earth-like heat fluxes in contrast to regions with high lithospheric thickness \citep{smrekar_earth-like_2023}.

More recent observations have presented compelling evidence of active volcanism on the surface in the form of variations of sulfur dioxide \citep{marcq2013} and morphological changes to surface features over the past 10 years \citep{herrick_surface_2023,sulcanese_evidence_2024}, (see also \citet{smrekar_recent_2010} and \citet{filiberto_present-day_2020}). These observations support models that predict a volcanically active Venus today.

Rather than attempt to reproduce its detailed surface and atmospheric features, we focus on reproducing three global constraints: 1) the partial pressure of carbon dioxide, $p$CO$_2$, 2) the partial pressure of water, $p$H$_2$O, and 3) the magnetic moment of Venus. 
These constraints are summarized in Table 1.
While this goal may appear modest, reproducing all three constraints, within their uncertainties, after 4.5 Gyr of evolution still requires a complex and interdisciplinary model. 

\begin{deluxetable*}{llll} 
    \tablewidth{\linewidth}
    \tablecaption{Observed Properties of Venus Reproduced in this Study\label{constraint_table}}
    \tablehead{\colhead{Notation} & \colhead{Description} & \colhead{Value} & \colhead{Citation}}
\startdata
$p$CO$_2$ & Partial pressure of CO$_2$  & 92.254 $\pm$ 0.765 bar & \citet{2017aeil.book.....C} \\
$p$H$_2$O & Partial pressure of H$_2$O  & 2.868 $\pm$ 0.2868 mbar & \citet{arney_2014} \\
$\mathcal{M}$ & Magnetic moment & $<10^{-5}$ $\mathcal{M_\oplus}$ & \citet{phillips_russell_1987}\\ 
\enddata
\end{deluxetable*}

\subsection{Atmospheric Modeling}

Atmospheric models of Venus have tended to focus on its structure, photochemical processes, and atmospheric escape. As a first approximation, our model does not take into account the atmospheric structure, nor do we use the atmosphere to calculate a self-consistent surface temperature so a review of literature on the Venusian atmospheric structure and clouds is outside of our scope.
We assume that Venus has always been in the runaway greenhouse state, and thus water is able to reach the upper atmosphere and escape.
The photolysis of water and the escape of its constituent elements is central to the bulk evolution of Venus in our model, so we concentrate on this mechanism.

Many atmospheric modeling studies have focused on the enhanced deuterium abundance in the Venusian atmosphere as an indicator that Venus has lost approximately one terrestrial ocean-mass (1 TO) \citep{donahue_1982,kasting_1988,debergh_1991,donahue_1992,grinspoon_1993,bullock_grinspoon_2001,arney_2014}. This research provides circumstantial evidence that Earth and Venus formed with a similar volatile abundance, and that H escape has led to the vastly different climates we observe today.
However, as the initial D/H ratio of Venus is unknown; it remains possible that the high D/H ratio on Venus is a result of late cometary accretion of high D/H ratio volatiles \citep{Avice2022}.

One mechanism that could drive desiccation is the photoloysis of water by FUV photons, followed by ``energy-limited escape'' wherein the released hydrogen atoms achieve escape velocity after absorbing X-ray and UV (XUV) photons (1-1000\AA) \citep{watson_dynamics_1981,chassefiere_1996}. Additionally, non-thermal mechanisms could also remove hydrogen, such as electromagnetic interactions between ions in Venus' upper atmosphere and solar wind particles \citep{lammer_2006,Gunell2018}. Hydrogen loss to space acts as a permanent sink for planetary water, so,  barring a stochastic event like a cometary impact, a planet experiencing this process continuously loses water. The liberated oxygen atoms presumably reacted with the crust and ultimately contributed to oxidizing the interior \citep[see][]{krissansen-totton_was_2021}. 

In this context, the enhanced D/H ratio observed today likely results from the preferential escape of hydrogen over deuterium. As the latter isotope is twice as heavy as the former, gravity will tend to slow its escape. The details of this escape process, however, are also important. The evolution of the sun's XUV luminosity is not well-constrained \citep{ribas_2005,claire_2012}, nor is the frequency of flaring and coronal mass ejections that could strip the atmosphere non-thermally. 

Venus' thick carbon dioxide-rich atmosphere precludes stable liquid water on the surface today. As is now well-known, the rovibrational bands of CO$_2$ in the infrared slow the escape of thermal photons generated from the surface \citep{ingersoll_1969,nakajima_1992}. This greenhouse effect maintains the 760 K surface temperature that not only forces water to enter the vapor phase, but also produces an atmospheric temperature profile that allows water vapor to leak into the stratosphere where it can be photolyzed, leading to H escape. 

\subsection{Surface and Mantle Modeling}

Despite the limited knowledge of the interior of Venus, much effort has been devoted to understanding its so-called ``stagnant lid'' mode of tectonics \citep{solomatov_moresi_1996,solomatov_moresi_2000}.
In this mode, the cold and thick lithosphere does not participate in mantle convection and acts as a conductive barrier to convective heat transfer.
In this traditional view of the stagnant lid, the level of mobility of the lithosphere is negligible, with topographic changes usually a result solely of
mantle plumes generating uplift.

While the current geodynamic state can be modeled by a stagnant lid \citep[e.g.,][]{driscoll_divergent_2013,krissansen-totton_was_2021}, it is unknown as to whether this model applies for all of Venus' history \citep{rolf_dynamics_2022,smrekar_venus_2018} or even if other more exotic processes have operated, such as sagduction tectonism \citep{ParmentierHess1992}, large-scale resurfacing \citep{Reese1999}, or convective mobilization of the lithosphere \citep{noack_coupling_2012}.
Using a one-dimensional parameterized convection model, \citet{driscoll_thermal_2014} produced evolutionary histories consistent with Venus that assumed a stagnant lid for the entire evolution. \citet{krissansen-totton_was_2021}, however were able to produce evolutionary histories by instead including an explicit magma ocean model as well as imposing a transition from mobile to stagnant lid dynamics.
Such a transition between a mobile and stagnant lid model is usually accomplished by imposing an abrupt change in the mantle cooling efficiency, which results in a discontinuity in the solutions \citep{nimmo_why_2002}.
This approach illustrates a shortcoming of parameterized convection models. 

While the results of \citet{krissansen-totton_was_2021} represented a significant step forward in our understanding of Venus' history, their model did not self-consistently model the core, instead assuming a simple exponential decay model for the core's thermal evolution. Thus, it is currently unclear how the core's evolution actually affects the mantle, crust, and atmosphere. Furthermore, they did not include the lack of magnetic field as a constraint on the planet's evolution, so their results may exhibit a systematic error. A central goal of the work described here is to include the constraint on the core to refine our understanding of the mantle, surface, and atmosphere of Venus.

Two- and three-dimensional mantle models that explicitly model convective stress on the lithosphere predict that Venus' stagnant lid may be a snapshot of long-term episodic lid behavior, i.e., a transitional mode between mobile and stagnant lid \citep{armann_simulating_2012,gillmann_atmosphere/mantle_2014,orourke_prospects_2018,weller_physics_2020}. On the other hand, \citet{lourenco_plutonic-squishy_2020} integrates the role of highly intrusive magma in their lithosphere model to define a ``plutonic-squishy lid'' regime, which may be applicable to Venus.
While our work is not able to self-consistently model such regimes, our whole-planet approach is able to probe how the core could affect the mantle's evolution, and how the evolution of the mantle could change outgassing rates on Venus.

The diversity of plausible geodynamic regimes in Venus' history yields different explanations for the evidence of a relatively young surface. One idea is that a transient (and possibly episodic) period of plate tectonics  occurred 500--1000 Myr ago that destroyed the previous craters over the course of hundreds of millions of years \citep{turcotte_1993}. Another is that, a (near-)total overturn of the lithosphere occurred over a very short timescale, the so-called ``catastrophic resurfacing" hypothesis \citep{strom_1994,armann_tackley_2012,smrekar_venus_2018,rolf_dynamics_2022,WellerKiefer2025}. Finally, it may be that localized but widespread volcanism erodes away craters in the Venusian plains faster than craters can be generated by impacts \citep{guest_stofan_1999,orourke_venus_2014}.
These models could be tested by investigating the alteration histories of craters on the Venusian surface \citep{Basilevsky2003,orourke_venus_2014} as the resolution of the \textit{Magellan} data is not high enough to disambiguate between these different hypotheses \citep{Widemann2023}.
Our work presents an end member case exploration, assuming that Venus has been in a stagnant lid regime throughout its post-differentiation history.

One critically important element of parameterized stagnant lid models is the fraction of the melt that erupts to the surface, which determines the eruptive mass flux. Previous spatially resolved episodic lid and stagnant lid models, i.e., in 2D and 3D, assumed either a fully extrusive (``heat pipe'') volcanic mode \citep{armann_simulating_2012,gillmann_atmosphere/mantle_2014}, or a mixed mode in which some melt erupts intrusively into the lithosphere (the plutonic-squishy lid geodynamic regime) \citep{lourenco_plutonic-squishy_2020}.
Previous parameterized 1D convection models, such as \citet{driscoll_thermal_2014} and \citet{patocka_2020}, as well as ours, allow for the fraction of extrusive versus intrusive melt to be modeled as a free parameter.

Varying the fraction of volcanism that is extrusive, the ``eruption efficiency", can help elucidate the role that melting plays in Venus' thermal history.
\citet{driscoll_thermal_2014}, for example, showed that, in a stagnant lid regime characterized by inefficient convective cooling, the eruption efficiency plays a crucial role in the thermal evolution of the mantle, with high (low) eruption efficiencies resulting in a cooler (hotter) mantle. 
Eruptive cooling is thought to also influence the cooling rate of the core, and therefore the dynamo.
The nominal Venus model in \citet{driscoll_thermal_2014} adopts a an eruption efficiency of 50\%, which is low enough to cause the mantle to heat up due to radiogenic heating, remain relatively hot due to the inefficient cooling of the stagnant lid, and lose its dynamo by 4.3 Gyrs.

Another major unknown for Venus' mantle is
the role of volatile cycling. Currently it is thought that very little chemical weathering is occurring that could draw down atmospheric carbon dioxide, implying that, once outgassed, carbon dioxide remains in the atmosphere indefinitely. This situation is of course markedly different from Earth, where the carbonate-silicate cycle, enabled by liquid water and plate tectonics, inhibits carbon dioxide from building up in the atmosphere, and thus permits liquid water to remain stable on its surface for billions of years \citep{walker_negative_1981, graham_pierrehumbert_2020, Honing_2021}. Recently, it has been proposed that a carbon cycle is possible for stagnant lid planets \citep{foley_carbon_2018}, but it is unclear if this process has ever operated on Venus.

The recent discoveries of potential changes in surface features \citep{herrick_surface_2023,sulcanese_evidence_2024} present an exciting new opportunity to learn about Venus' interior and history. With just a few detections currently known and a large fraction of the planetary surface still unexamined, the true resurfacing rate and eruptive mass flux remain poorly constrained. \citet{sulcanese_evidence_2024} factor these uncertainties into a model that predicts a low eruption rate of $\sim 5$ km$^3$/yr and a high rate of $\sim 15$ km$^3$/yr. So far, relatively little theoretical work has attempted to determine if these observations are consistent with interior models. As we show below, these observations may be able to falsify some models that are otherwise consistent with core and atmospheric properties. We also show that our model strongly disfavor the high eruption rate, a prediction which could be verified as more of the surface is analyzed or the eruption rates can be constrained by spectroscopic measurements of its atmosphere \citep{Dias2025}.

\subsection{Venus Core Modeling}

Venus' current lack of a measurable magnetic field \citep{russell_1980,phillips_russell_1987} is in stark contrast to Earth's long-lived magnetic field. Terrestrial planet magnetic fields are generated by convection in a liquid iron-rich core \citep{busse_1976,olson_christenen_2006,driscoll_handbook_2018}. Owing to its bulk density, Venus is assumed to have a large iron-rich core, but its state (solid or liquid), composition, and structure are relatively unknown \citep{Shah2022}. Indeed, one hypothesis for Venus' lack of a geodynamo is that the core is stratified as a result of a slow formation process in which iron from impactors accrete onto the core's surface and inefficiently mix \citep{jacobson_formation_2017}. This putative stratification then suppresses outer core convection and hence magnetic field generation.

In rocky planets, the two most commonly invoked mechanisms to drive core convection are thermal and compositional buoyancy fluxes. To drive thermal convection, the core heat flow must exceed the amount of heat that can be conducted without motion, or, the conductive heat flow limit. 
This heat flow limit is set by the thermal conductivity of the core, an uncertain quantity even for Earth, with a plausible range of 40 to 200 W/m/K \citep{ohta_experimental_2016,konopkova_direct_2016,pozzo2022}. Another possibility is that a geodynamo can be generated by convection of liquid silicates under high pressure \citep{Stixrude2020,Blanc2020}, but we will not consider that possibility in this study.

Uncertainties in expected radioactive heating of the core also contribute to the core state.
For example, a $\sim$200 ppm mass fraction of potassium in the core has been invoked to accurately model the size of Earth's inner core \citep{nimmo_influence_2004,driscoll_thermal_2014}, but experiments show that the potassium concentration in the primordial core should be less than 35 ppm \citep{hirose_composition_2013,orourke_prospects_2018}. We consider up to 200 ppm of potassium in the core as an upper bound for our model, but most trials contain less than 100 ppm in the core. Uranium may also occur in the core as add an additional radiogenic heating source \citep{WohlersWood2015}, but we do not consider that possibility here.

The lack of an active Venusian dynamo today has been explained in three possible ways.
First, the core may be completely solid, thereby preventing convection \citep{dumoulin_tidal_2017}.
Second, the core may be wholly or partially liquid, but the core heat flow, proportional to the temperature difference between the core and mantle, is too small to induce convection \citep{nimmo_why_2002,olson_christenen_2006,driscoll_thermal_2014}.

Finally, the core-mantle temperature difference may be sufficient to induce a magnetic field, but a shrinking inner core results in the sum of the thermal and compositional buoyancy being negative \citep[see][]{hemingway_2021}.
Compositional core convection is generated by the release of a compositionally buoyant phase, which is produced in the Earth by the rejection of light elements during the solidification of the solid inner core. 
For our purposes, acceptable Venus thermal and magnetic histories must result in one of these states today.

While Venus is thought to lack a dynamo today, it may have maintained one in its past when the thermal and compositional state of the core was more favorable to dynamo generation \citep{driscoll_thermal_2014,orourke_prospects_2018}.
The existence of a past dynamo on Venus could be preserved in minerals on the surface \citep{orourke_2019}, suggesting that future orbiters, landers, and/or aerial platforms may be able to recover evidence for a past dynamo.
Whether remanent magnetization is detected on Venus or not, the difference between Earth and Venus' dynamo at present-day highlights the importance of understanding how planetary magnetic fields evolve \citep{orourke_prospects_2018}, and motivates whole planet modeling of Venus and other rocky planets. 

A common geodynamical explanation for the lack of a dynamo on present-day Venus \citep{nimmo_why_2002} evokes a stagnant lid that limits the core heat flow. 
The possible transition from mobile to stagnant lid is investigated in \citet{armann_simulating_2012} where the authors define an episodic lid regime that allows Venus to remove internal heat early in its evolution and suppress core cooling later to shut off a dynamo.
Conveniently, this mechanism could also drive a catastrophic resurfacing process that erased existing craters.

While the model described in \citet{nimmo_why_2002} provides a simple explanation for a young surface and no magnetic field, a huge range of plausible thermal histories remain. For example, \citet{driscoll_thermal_2014} and \citet{orourke_thermal_2015} show several stagnant lid models that can suppress a core dynamo due to inefficient cooling, resulting in low core heat flows and early magnetic field generation. However, \citet{armann_simulating_2012} find that, under certain conditions, a stagnant lid does not shut down the magnetic dynamo. In their model, they quenched the dynamo by invoking an episodic lid mode where radioisotope-bearing basalts subduct to the core-mantle boundary (CMB) and insulate the core.
Conversely, \citet{Marchi2023} find that a thermal evolution model with traditional stagnant lid parameterized convection can explain both the lack of a current dynamo on Venus as well as its young surface (without the need for lid overturn) if early impacts are taken into account.

The model of Venus in \citet{orourke_prospects_2018} considered both mantle evolution and the uncertainty of core parameters to investigate the potential magnetic history of Venus.
Their model connected the one-dimensional \citet{orourke_2017} core model to the two-dimensional \texttt{StagYY} mantle convection code \citep{armann_simulating_2012} to a one-dimensional grey radiative transfer model \citep{gillmann_atmosphere/mantle_2014} and included the brightening of the Sun.
By modeling the evolution of Venus across such large parameter space ranges, \citet{orourke_prospects_2018} outlined regions of parameter space that match Venus and argue that such remnant magnetization could potentially still be measured \citep[see][]{orourke_2019}.
They found that if Venus' core is still partially liquid, it should have a high core thermal conductivity ($>$ 100 W/m/K) and low amounts of potassium ($<200$ ppm) to prevent a dynamo today.
Our work builds on \citet{orourke_prospects_2018} by including volatile cycling, making different mantle dynamic assumptions (stagnant lid), and expanding the mantle viscosity parameter space. Note that none of these prior studies attempted to simultaneously reproduce core and atmospheric properties.

In summary, it remains unclear why Venus has no dynamo today while Earth does.
Numerous explanations are consistent with the (dearth of) observations. Most previous work on Venus has focused on either mantle-atmosphere evolution or core evolution. Crucially, the mantle and core are in direct contact and heat can flow between the two.
Thus, we will rely on a whole planet model to explore the history of the core by directly connecting the internal processes to the atmosphere, which introduces more observational constraints on the core.

\section{Model Description} \label{sec:modeldesc}
\begin{figure*}
    \includegraphics[width=\linewidth]{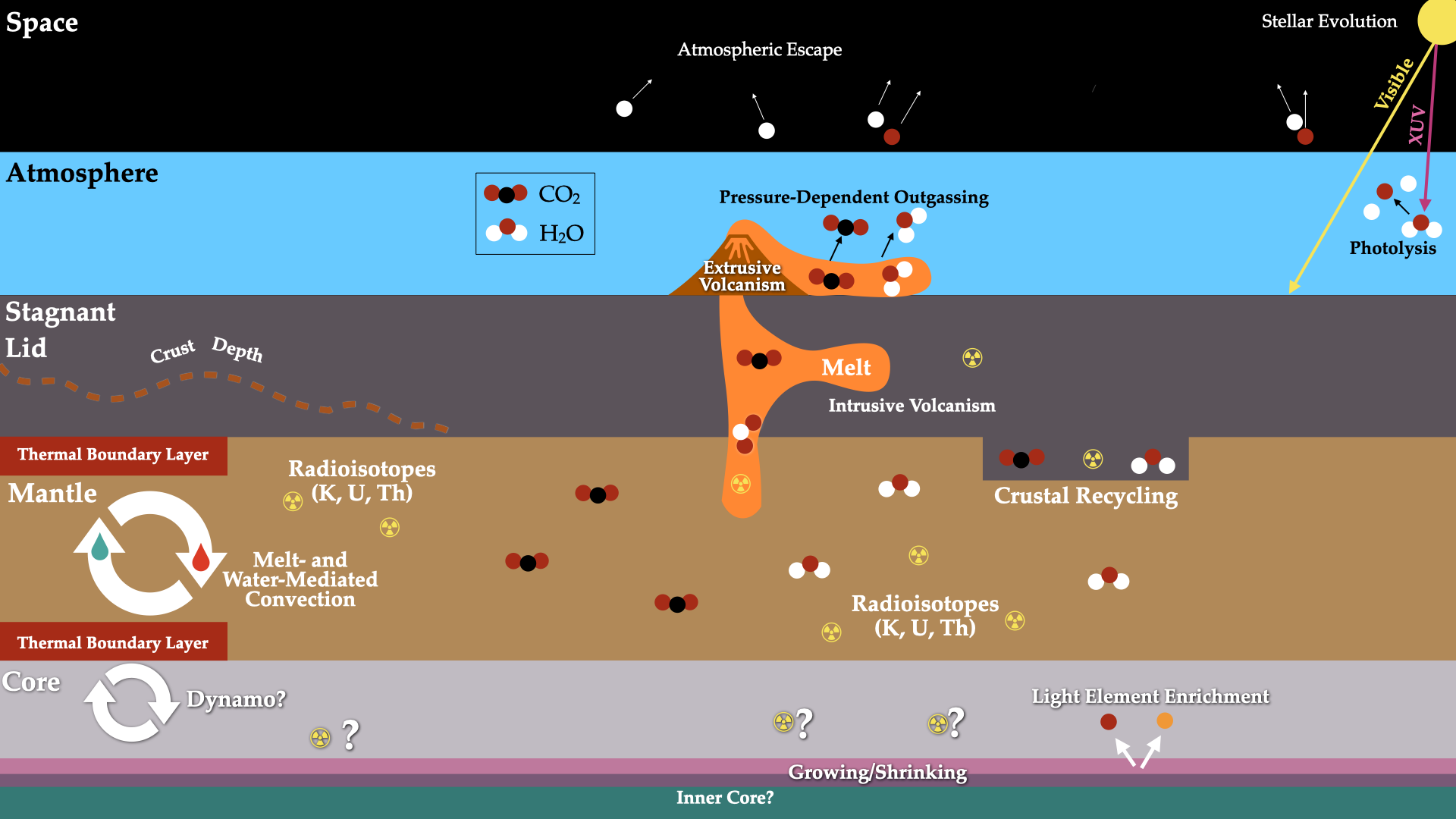}
    \caption{Illustration of the processes implemented in our model. Mantle convection is lubricated via the presence of water and melt; the vigor of the viscosity thus sets the boundary of the non-adiabatic thermal boundary layer and above that the conductive stagnant lid lithosphere, which consists of both a mantle and crust layer. Melting in the mantle transports radioisotopes and volatiles to the crust and atmosphere, where water is photolyzed and escapes. Radioisotopes and volatiles trapped in the crust due to intrusive volcanism are recycled back into the mantle through crustal delamination. The core is potentially heated by radioisotopes and cooled by convection with a vigor controlled by the temperature differential between the core-mantle boundary and the lower mantle. A cool enough core could potentially result in inner-core solidification, which, together with high enough convective vigor, can result in the generation of a core dynamo and planetary magnetic field. As the core solidifies, the outer core is enriched in light elements, resulting in a decrease in the outer core liquidus temperature. \label{fig:modelgraphic}}
\end{figure*}

Our coupled whole-planet model is based heavily on the thermal interior, atmospheric escape, and stellar evolution models described in \citet{driscoll_divergent_2013,driscoll_thermal_2014} and \citet{barnes_vplanet_2020}.
The differential equations describing each of these processes are coupled and simultaneously solved using \vplanet \citep{barnes_vplanet_2020}, a software package that can incorporate a variety of physical processes to model planetary evolution. This section focuses on describing our additions to \vplanet including a new model of convection under a stagnant lid modulated by the presence of water in the mantle, the changing thickness of that stagnant lid, a volatile cycling model that includes outgassing and crustal recycling, and a model of inner core growth that incorporates light element concentration in the outer core.
We also provide here a short review of the base mantle and core processes, details of which can be found in \citet{barnes_vplanet_2020}.
A schematic of the model is shown in Fig. \ref{fig:modelgraphic}.

To the thermal interior evolutionary model published in \citet{barnes_vplanet_2020}, we have added a parameterization of convection under a purely conductive stagnant lid lithosphere.
By doing so, we are also able to take into account the growth and shrinkage of both the mantle and crustal components of the lithosphere.
We also incorporate a geochemical transport model that simulates the melt-mediated transport of radioisotopes and volatiles to the crust and atmosphere.
The partitioning of radioisotopes and volatiles between the crust and the atmosphere is calculated based on the fraction of extrusive volcanism (eruption efficiency) and a surface pressure-dependent outgassing model.
The role of volatiles in geophysical processes is new as well, and includes water influencing the mantle viscosity with carbon dioxide acting as a medium for diffusion-limited atmospheric escape.

\subsection{Thermal Evolution of the Mantle and Core}
We trace the evolution of the mantle temperature by tracking the heating and cooling rates of the mantle over time.
In our model we consider heating of the mantle due to the decay of radioactive species $i$ ($Q_{i,\mathrm{man}}$), latent heat from mantle solidification ($Q_{L,\mathrm{man}}$), and heat flow from the core ($Q_\mathrm{CMB}$).
We consider cooling of the mantle due to convection ($Q_\mathrm{conv}$) and loss of magma from the mantle (through intrusive and extrusive volcanism) ($Q_\mathrm{melt}$). 
By balancing these heating and cooling rates with the secular cooling rate of the mantle, we derive an equation for the change in mantle temperature over time,
\begin{equation}
    \dot{T}_m = \frac{Q_{i,\mathrm{man}} + Q_{L,\mathrm{man}} + Q_\mathrm{CMB} - Q_\mathrm{conv} - Q_\mathrm{melt}}{c_m M_\mathrm{man}},
\end{equation}
where $c_m$ and $M_\mathrm{man}$ are the heat capacity and mass of the mantle, respectively.
Note that all constant values in our model, such as $c_\mathrm{man}$, can be found in Table \ref{consttable}.
Because the amount of crust changes over time, the mass of the mantle is time-dependent.
The $Q_\mathrm{CMB}$ term in the equation above represents the importance of modeling the core in addition to the mantle to derive a fully consistent thermal history for the planet as a whole \citep{foley_whole_2016}.
The rate at which the core loses heat to the lower mantle is based on the viscosity at the lower mantle, $\nu_\mathrm{LM}$, which we assume to be greater than the viscosity at the upper mantle, $\nu_\mathrm{UM}$ by a constant amount, $\nu_\mathrm{jump}$ due to higher pressures such that:
\begin{equation} \label{eq:viscjump}
    \nu_\mathrm{LM} = \nu_\mathrm{jump} \nu_\mathrm{UM}.
\end{equation}

\startlongtable
\begin{deluxetable*}{llll}
\centerwidetable
\tablecaption{Constant parameters used in this study\label{consttable}}
\tablehead{\colhead{Notation} & \colhead{Description} & \colhead{Value} & \colhead{Citation}}
\startdata
$A_{sol}$ & Solidus fit coefficient & -1.16$\times 10^{-16}$ K/m$^3$ & \citet{elkins-tanton_linked_2008} \\
$a_{CO_2}$ & Constant for solubility of CO$_2$ in mafic melts & 1 & \citet{wogan_abundant_2020} \\
$a_{H2_O}$ & Constant for solubility of H$_2$O in mafic melts & 0.54 & \citet{wogan_abundant_2020} \\
$\alpha$ & Thermal expansivity of the mantle & $3\times10^{-5}$ K$^{-1}$ & \citet{elkins-tanton_linked_2008} \\
$\alpha_c$ & Thermal expansivity of the core & 10$^{-5}$ K$^{-1}$ & \citet{labrosse_age_2001} \\
$\alpha_m$ & Modified Rayleigh number exponential & 0.4 & \citet{ChristensenAubert2006} \\
$B$ & Einstein coefficient (intrinsic viscosity) & 2.5 & \citet{costa_model_2009} \\
$B_{sol}$ & Solidus fit coefficient & 1.708$\times10^{-9}$ & \citet{elkins-tanton_linked_2008} \\
$C_{sol}$ & Solidus fit coefficient & -9.074$\times 10^{-3}$ & \citet{elkins-tanton_linked_2008} \\
$c_\mathrm{man}$ & Specific heat of the mantle & 1265 J/kg/K & \citet{elkins-tanton_linked_2008} \\
$c_c$ & Specific heat of the core & 840 & \citet{driscoll_thermal_2014} \\
$dV_\mathrm{liq}/dT_m$ & Change in mantle liquid volume with temp. & $8\times10^{17}$ m$^3$/K & \citet{driscoll_thermal_2014} \\
$D_i$ & Partitioning coefficient for radioisotope species $i$ & 0.002 & \citet{grott_volcanic_2011} \\
$D_{H_2O}$ & Partitioning coefficient for water & 0.01 & \citet{grott_volcanic_2011} \\
$D_{CO_2}$ & Partitioning coefficient for carbon dioxide & 0.01 & \citet{grott_volcanic_2011} \\
$D_\mathrm{Fe}$ & Length scale for Lindemann's Law & 7000 km & \citet{labrosse_age_2001} \\
$D_\mathrm{Core}$ & Length scale for core adiabat & 6340 km & \citet{labrosse_age_2001} \\
$D_{sol}$ & Solidus fit coefficient & 1.993$\times 10^4$ & \citet{elkins-tanton_linked_2008} \\
$d_\mathrm{H_2O}$ & Constant for solubility of $\mathrm{H_2O}$ in mafic melts & 2.3 & \citet{wogan_abundant_2020} \\
$\Delta T_\mathrm{sol,liq}$ & Mantle liquidus offset from solidus & 500 K & \citet{Andrault2011} \\
$\Delta \rho_\chi$ & Outer core comp. density difference & 700 kg/m$^3$ & \citet{driscoll_thermal_2014} \\
$\delta_\mathrm{ph}$ & Empirical viscosity parameter & 6.0 & \citet{costa_model_2009} \\
$E_\mathrm{A}$ & Viscosity activation energy & $3\times10^5$ J/mol & \citet{barnes_vplanet_2020} \\
$E_G$ & Gravitational energy released at ICB & 300 kJ/kg & \citet{driscoll_thermal_2014} \\
$\eta_{ic}$ & Average core to CMB adiabatic temp. jump & 0.8 & \citet{driscoll_thermal_2014} \\
$g$ & Surface gravity & 8.867 m/s$^2$ & Derived from $M_p$, $R_p$ \\
$g_\mathrm{cmb}$ & Gravity at core-mantle boundary & 10.27 m/s$^2$ & Derived from $\rho_c$, $R_c$ \\
$\gamma$ & Mantle adiabatic gradient & 0.5 K/km & \citet{driscoll_thermal_2014} \\
$\gamma_c$ & Core Gruneisen parameter & 1.3 & \citet{labrosse_age_2001} \\
$\gamma_{\mathrm{ph}}$ & Empirical viscosity parameter & 6.0 & \citet{costa_model_2009} \\
$k_{cr}$ & Thermal conductivity of the crust & 3 W/m/K & \citet{2001JGR...10616387A} \\
$k_\mathrm{UM}$ & Thermal conductivity of upper mantle & 4.2 W/m/K & \citet{driscoll_thermal_2014} \\
$\kappa$	&	Mantle thermal diffusivity &	$10^{-6}$	m$^2$ s$^{-1}$	& \cite{Hofmeister2015}\\
$L_\mathrm{Fe}$ & Latent energy of solidification at ICB & $750$ kJ/kg & \citet{driscoll_thermal_2014} \\
$L_\mathrm{melt}$ & Latent heat of melting & 320 kJ/kg & \citet{driscoll_thermal_2014} \\
$M_p$ & Mass of Venus & 4.867$\times 10^{24}$ kg & \citet{2017aeil.book.....C} \\
$M_E$ & Mass of Earth & 5.972$\times 10^{24}$ kg & \citet{prsa_nominal_2016} \\
$\mu_\mathrm{magma}$ & Molar mass of magma & 64.52 g/mol & \citet{wogan_abundant_2020} \\
$\mu_\mathrm{CO_2}$ & Molar mass of carbon dioxide & 18.02 g/mol & Given \\
$\mu_\mathrm{H_2O}$ & Molar mass of water & 44.01 g/mol & Given \\
$\Omega$ & Rotation rate of Venus & 2.993$\times 10^{-7}$ rad/s & Given \\
$\phi^*$ & Critical solid fraction & 0.8 & \citet{costa_model_2009} \\
$R$ & Universal gas constant & 8.314 J/mol/K & Given \\
$\mathrm{Ra}_\mathrm{crit}$ & Critical Rayleigh number & 660 & \citet{solomatov1995} \\
$R_c$ & Radius of the Venus' core & 3.087$\times 10^{6}$ m & Scaled from Earth\\
$Ro_0$ & Rossby coefficient & 0.85 & \citet{ChristensenAubert2006}  \\
$R_p$ & Radius of Venus & 6.0528$\times 10^{6}$ m & \citet{2017aeil.book.....C} \\
$R_E$ & Radius of Earth & 6.3781$\times 10^{6}$ m & \citet{prsa_nominal_2016} \\
$\rho_c$ & Density of the core & 10985 kg/m$^3$ & \citet{labrosse_age_2001} \\
$\rho_{cr}$ & Density of the crust & 2800 kg/m$^3$ & \citet{2002gedy.book.....T} \\
$\rho_\mathrm{ic}$ & Density of the inner core & 13000 kg/m$^3$ & \citet{labrosse_age_2001} \\
$\rho_m$ & Density of the mantle & 4800 kg/m$^3$ & \citet{2002gedy.book.....T} \\
$T_\mathrm{Fe,0}$ & Reference liquidus of iron core & 5451.6 K & \citet{driscoll_thermal_2014} \\
$\Theta$ & Frank-Kamenetskii parameter & 2.9 & \citet{reese_scaling_2005} \\
$\xi$ & Empirical viscosity parameter & $5\times10^{-4}$ & \citet{costa_model_2009} \\
\enddata
\end{deluxetable*}

We model the thermal evolution of the core both as a heat source to the mantle and to determine the strength of the magnetic field that arises from core convection.
In our model the core is heated primarily by radioactive decay ($Q_{i,\mathrm{core}}$) and secularly cools as:
\begin{equation}
Q_\mathrm{sec} = -\int_\mathrm{outer  core} \rho_c c_c \frac{\partial T_c(r)}{\partial t} \mathrm{d}V.
\label{eq:qseccore}
\end{equation}
The core also loses heat to the lower mantle through the core-mantle boundary ($Q_\mathrm{CMB}$ in both the mantle and core heat budgets).

Following \citet{labrosse_thermal_2015}, we assume an adiabat for the temperature distribution in the outer core given by:
\begin{equation} \label{eq:coreadiabatic}
    T_c(r) = T_L(r_\mathrm{IC}) \exp{(\frac{r_\mathrm{IC}^2 - r^2}{D_\mathrm{core}^2})},
\end{equation}
where $T_L(r_\mathrm{IC})$ is the core liquidus temperature at the surface of the inner core and $D_\mathrm{core}$ is a core adiabatic length scale.
When there is no inner core then $T_L(r_\mathrm{IC})$ in the above equation is replaced by $T_0$, the core central temperature.

Inner core growth also contributes to core heat flows, including the latent heating and gravitational energy change due to inner core growth, we obtain the following equation for the change in core central temperature over time:
\begin{equation} \label{eq:dotTCoreyesIC}
    \frac{\mathrm{d}T_0}{\mathrm{d}t} = \frac{-(Q_\mathrm{CMB} - Q_\mathrm{rad} - Q_\mathrm{Lat} - Q_\chi)}
    {4 \pi \rho c_c \int_{r_\mathrm{IC}}^{R_c} r^2 \exp \left(\frac{-r^2}{D_\mathrm{core}^2}\right) \, \mathrm{d}r },
\end{equation}
where $\rho_c$ and $c_c$ are the density and specific heat of the core, respectively. The integral in the denominator represents a volume integral over the core temperature distribution necessary for the secular cooling calculations.
This heat balance equation ignores secular cooling of the inner core, but including such a phenomenon results in only minor changes to the core's evolution \citep{labrosse_thermal_2015}. Note that when there is no inner core, the volume integral is taken over the whole core and $Q_\mathrm{Lat} = Q_\chi = r_\mathrm{IC} = 0$, recovering the formulation in \citet{driscoll_divergent_2013} and \citet{driscoll_thermal_2014}. Finally, we note that, following \cite{driscoll_divergent_2013}, we assume a modest amount of potassium heating in the core (up to 220 ppm), which can resolve the so-called new core paradox and thermal catastrophes implied by thermal models that do not include deep heating or an insulating layer \citep{Korenaga2006,LabrosseJaupart2007}. While this concentration of potassium may not be compatible with mineralogical experiments, it serves as an effective proxy for lower mantle insulation.

\subsection{Thermal Evolution of the Inner Core}

The core can solidify depending on its thermal history, chemical stratification, and the assumed liquidus temperature of iron.
Inner core solidification occurs where the adiabatic core temperature is less than the liquidus temperature of iron.
As the inner core grows, light elements are expelled into the liquid outer core, which acts to decrease the liquidus temperature in the outer core and slow inner core growth.
To model this influence on the inner core's growth, we follow \citet{labrosse_thermal_2015} and \citet{Bonati2021} to calculate $\frac{\mathrm{d}r_\mathrm{IC}}{\mathrm{d}t}$, the growth and shrinkage of the inner core over time, assuming a homogeneous core.

Following \citet{Labrosse2015}, we can rewrite the inner core-related heat flows in Eq. \eqref{eq:dotTCoreyesIC} in terms of $\frac{\mathrm{d}r_\mathrm{IC}}{\mathrm{d}t}$ by defining an expression that is purely a function of the inner core radius and has no time dependence, i.e. for $Q_\mathrm{sec}$, $Q_\mathrm{Lat}$, and $Q_\chi$:
\begin{equation}
    Q_i = P_i(r_\mathrm{IC}) \frac{\mathrm{d}r_\mathrm{IC}}{\mathrm{d}t}.
\end{equation}
We follow a similar procedure to \citet{Labrosse2015} and \citet{Bonati2021} to derive the $P_i$ functions, but in lieu of an equation of state, we assume a constant density in the core, $\rho_c$.
Using a constant density, we can analytically evaluate each integral in $Q_\mathrm{sec}$, $Q_\mathrm{Lat}$, and $Q_\chi$:
\begin{align}
    Q_\mathrm{sec} &= -\int_\mathrm{outer  core} \rho_c c_c \frac{\partial T_c(r)}{\partial t} \mathrm{d}V \label{QsecInt},\\
    Q_\mathrm{Lat} &= 4 \pi r_\mathrm{IC}^2 \rho_\mathrm{IC} T_L(r_\mathrm{IC}) \Delta S
    \frac{\mathrm{d}r_\mathrm{IC}}{\mathrm{d}t}, \\
    Q_\chi &= -\int_{V_\mathrm{OC}} \rho \mu' \frac{\partial \chi}{\partial t} \, dV,
\end{align}
where $\Delta S$ is the entropy of crystallization and is about 127 J/K/kg \citep{Hirose2013}.
$\mu(r)$ is the difference between the chemical potential at $r$ and the chemical potential at the inner-core boundary, given in \citet{Labrosse2015} as:
\begin{equation} \label{eq:lederiv}
    \frac{\partial \mu}{\partial r} = -\beta g,
\end{equation}
where $\beta = \frac{\Delta \rho_\chi}{\rho_{IC} \chi}$ and $\Delta\rho_\chi = 700$ kg/m$^3$ \citep{labrosse_age_2001}.
$\beta$ is a function of $r_\mathrm{IC}$ due to its dependence on $\chi$, the concentration of light elements in the outer core.
$\chi$ is a direct function of $r_\mathrm{IC}$:
\begin{equation} \label{eq:leconc}
    \chi(r_\mathrm{IC}) = \chi_0 \frac{M_c}{M_\mathrm{OC}} = \chi_0 \frac{M_c}{\frac{4}{3}\pi \rho_\mathrm{OC} (R_c^3 - r^3)},
\end{equation}
where $\chi_0$ is the initial light element concentration in the entire core and $\rho_\mathrm{OC}$ is the density of the outer core.

In our model, we assume a nearly pure liquid iron core (with some light elements present) so the iron solidus is approximately equivalent to the iron liquidus, which is given by Lindemann's Law.
To model the depression of the iron liquidus temperature by light elements, we introduce $\Delta T_\chi$ and a reference temperature, $\Delta T_{\chi,\mathrm{ref}}$ that increases with the inner core radius due to the partitioning of light elements into the outer core as the inner core solidifies so that the liquidus temperature is given by:
\begin{align}
    T_L(r) &= T_\mathrm{Fe,0} \mathrm{exp}\left(-2\left(1 - \frac{1}{3\gamma_c}\right)\frac{r^2}{D^2_\mathrm{Fe}}\right) - \Delta T_\chi(r),\\
    \Delta T_{\chi}(r_\mathrm{IC}) &= \chi(r_\mathrm{IC}) \frac{\Delta T_{\chi, \mathrm{ref}}}{\chi_\mathrm{OC,E}} \label{Tchieq},
\end{align}
where $T_\mathrm{Fe,0}$ is the reference liquidus temperature at the center of the core, $\gamma_c$ is the core Gruneisen parameter \citep{labrosse_age_2001}, and $D_\mathrm{Fe}$ is the length scale for the iron liquidus from Lindemann's Law (see Table \ref{consttable}).
We assume that no light elements are incorporated into the inner core and, following \citet{driscoll_thermal_2014}, that the fraction of light elements in the core $\chi_0$ is equal to the amount in Earth's outer core now, $\chi_\mathrm{OC,E} = 0.1$.

Using Eqs.~\eqref{QsecInt}-\eqref{Tchieq} we can derive the $P$ functions as such:
\begin{multline}
    P_\mathrm{Sec} = -4 \pi \rho_\mathrm{OC}
    \left( \frac{\mathrm{d}T_L(r_\mathrm{IC})}{\mathrm{d}r_\mathrm{IC}} 
    + \frac{2 T_L(r_\mathrm{IC})r_\mathrm{IC}}{D^2} \right)\\
    \left(\int_{r_\mathrm{IC}}^{R_c}
    r^2 \exp (\frac{r_\mathrm{IC}^2 - r^2}{D^2}) \,dr\right), 
\end{multline}
\begin{equation}
    P_\mathrm{Lat} = 4 \pi r_\mathrm{IC}^2 \rho_\mathrm{IC} T_L(r_\mathrm{IC}) \Delta S, 
\end{equation}
\begin{equation}   
    P_\chi = \left(\frac{9 \beta g_\mathrm{OC} \chi_0 M_c r_\mathrm{IC}^2}{(R_c^3 - r_\mathrm{IC}^3)^2}\right)
    \left(\int_{r_\mathrm{IC}}^{R_c} r^2(r - r_\mathrm{IC}) \, \mathrm{d}r \right).
\end{equation}
From these functions as well as the core heat balance equation, Eq.~\eqref{eq:dotTCoreyesIC}, we obtain the following equation for the change in the inner core radius over time:
\begin{equation} \label{eq:dRICDot}
    \frac{\mathrm{d}r_\mathrm{IC}}{\mathrm{d}t} = \frac{Q_\mathrm{CMB} - Q_\mathrm{rad}}{P_\mathrm{sec}(r_\mathrm{IC}) + P_\mathrm{Lat}(r_\mathrm{IC}) + P_\chi(r_\mathrm{IC})}.
\end{equation}

Finally, $Q_{CMB}$ is set by the thickness of the lower mantle thermal boundary layer, $\delta_{LM}$, i.e., a conducting shell around the core. We follow \citet{driscoll_thermal_2014} and define the boundary layer's thickness in terms of the mantle's critical Rayleigh number $Ra_c$ ($\approx 660$):
\begin{equation}
\delta_{LM}=\left( \frac{\kappa \nu_{LM}} {\alpha g \Delta T_{LM}} Ra_c \right)^{1/3},
\label{eq:delta_2} 
\end{equation}
where $\nu_{LM}$ is the lower mantle viscosity and $\Delta T_{LM}$ is the temperature difference between the deepest convecting layer of the mantle and the temperature at the CMB. The lower mantle and CMB temperatures, $T_{LM}$ and $T_{cmb}$, are extrapolations along the mantle and core adiabats: $T_{LM}=\epsilon_{LM}T_m$ and $T_{cmb}=\epsilon_c T_c$, where  $\epsilon_{LM}=\exp(-(R_{LM}-R_{m})\alpha g /c_p) \approx 1.3$
and $\epsilon_c\approx0.8$. In this case, the heat flow through the CMB is
\begin{equation}
Q_{cmb}=A_c k_{LM} \frac{\Delta T_{LM}}{\delta_{LM}},
\label{eq:q_cmb} 
\end{equation}
where $A_c$ is core surface area and $k_{LM}$ is lower mantle thermal conductivity.

\subsection{Mantle Convection Under a Stagnant Lid}

We model convective heat transfer using boundary layer theory as described in \citet{driscoll_thermal_2014}.
In this one-dimensional parameterization of convection, the heat transferred across the upper mantle layer that bounds the convective cell is proportional to the temperature jump across that boundary layer, $\Delta T_\mathrm{UM}$ and inversely proportional to the thickness of the boundary layer, $\delta_\mathrm{UM}$, such that
\begin{align} \label{umconv}
    Q_\mathrm{conv} &= A q_\mathrm{UM}, \\
    q_\mathrm{UM} &= k_\mathrm{UM} \frac{\Delta T_\mathrm{UM}}{\delta_\mathrm{UM}} = k_\mathrm{UM} \frac{T_\mathrm{UM} - T_\mathrm{Lid}}{\delta_\mathrm{UM}},
\end{align}
where $A$ is the surface area of the mantle below the upper mantle's thermal boundary layer, $k_\mathrm{UM}$ is the thermal conductivity of the upper mantle, and $\Delta T_\mathrm{UM}$ is the difference between the upper mantle temperature and the temperature at the base of the stagnant lid ($T_\mathrm{Lid}$). The upper mantle temperature, $T_{UM}$, is defined as the temperature at the base of the upper mantle thermal boundary layer, while the temperature at the base of the stagnant lid is the temperature at the top of the mantle thermal boundary layer.

Analogous to Eq.~(\ref{eq:delta_2}), the thickness of the upper mantle's thermal boundary layer $\delta_{UM}$ can be derived in terms of $Ra_c$: 
\begin{equation}
\delta_{UM}=D \left(Ra_c \frac{\nu_{UM}\kappa}{\alpha g \Delta T_{UM} D^3} \right)^{\beta}.
\label{eq:delta_1}
\end{equation}
where $D$ is the mantle thickness, $\nu_{UM}$ is the upper mantle viscosity, $\alpha$ is the thermal expansivity of the mantle, and $\kappa$ is the mantle's thermal diffusivity \citep{howard1966,solomatov1995,sotin1999,driscoll_thermal_2014}. Note that as the crust grows/shrinks, the thickness of the mantle changes, too.

When simulating convection in a mobile lid planet, $T_\mathrm{UM}$ represents the difference between the upper mantle temperature and the surface temperature as the cold upper boundary layer is involved in mantle convection for mobile lids.
However, we follow \citet{grasset_thermal_1998} to simulate mantle convection under a stagnant lid and define $\Delta T_\mathrm{UM}$ as the difference between the upper mantle temperature and the lower stagnant lid temperature.

The temperature at the base of the stagnant lid is defined as the temperature at which the mantle viscosity increases by an order of magnitude \citep{grasset_thermal_1998}, and is given by:
\begin{equation}
T_\mathrm{Lid} = T_\mathrm{UM} - \Theta \frac{R T_\mathrm{UM}^2}{E_A - \Delta E_\mathrm{H_2O} X^{\mathrm{H}_2\mathrm{O}}_\mathrm{man}},
\end{equation}
where $\Theta = 2.9$ accounts for spherical symmetry \citep{morschhauser_crustal_2011}, $R$ is the universal gas constant, $E_A$ is the activation energy of the upper mantle \citep{morschhauser_crustal_2011}, $\Delta E_\mathrm{H_2O}$ is a factor that depends on the assumed rock type and represents water's weakening of minerals that reduces the activation energy in this viscosity parameterization \citep{mcgovern_thermal_1989}, and $X^{\mathrm{H}_2\mathrm{O}}_\mathrm{man}$ is the weight fraction of water in the mantle.
Following \citet{mcgovern_thermal_1989}, we assume that the viscosity parameters for the mantle can be represented by dunite.

The viscosity of the upper mantle is a critical control on the convective heat transfer.
We start with the viscosity equation used in \citet{barnes_vplanet_2020}, which is a standard Arrhenius law that also incorporates the influence of the upper mantle melt fraction based on \citep{costa_model_2009}. 
We then decrease the viscosity activation energy in \citet{barnes_vplanet_2020} to incorporate dehydration stiffening due to water loss as prescribed by \citet{mcgovern_thermal_1989}.
With these modifications, the viscosity is calculated as:
\begin{equation} \label{eq:viscosity}
    \nu_\mathrm{UM} = \nu_0 \frac{ \exp( \frac{E_{A}}{RT_\mathrm{UM}} - \frac{\Delta E_\mathrm{H_2O} X^{\mathrm{H}_2\mathrm{O}}_\mathrm{man}}{RT_\mathrm{UM}} )}  {\epsilon \left(F_\mathrm{UM}\right)},
\end{equation}
where $\nu_0$ is the reference viscosity of the upper mantle, $E_{A}$ is the viscosity activation energy of the upper mantle, and $F_\mathrm{UM}$ is the upper mantle melt fraction calculated at the base of the upper mantle thermal boundary layer.
The mass fraction of water in the mantle ($X^{\mathrm{H}_2\mathrm{O}}_\mathrm{man}$) decreases the viscosity activation energy by an amount modulated by $\Delta E_\mathrm{H_2O}$, which is the change in activation energy due to the mantle water content.

Melting in the upper mantle decreases the viscosity, as shown in the experimental results of \citet{costa_model_2009}, and is represented by $\epsilon(F_\mathrm{UM})$ in the equation above.
\citet{costa_model_2009} provide the following relations for the role of melt on viscosity:
\begin{equation}
    \epsilon \left(F_{\mathrm{UM}}\right) = \frac{1 + \Phi^{\delta_{\mathrm{ph}}}} {\left( 1 - \mathcal{F}_\mathrm{phase}\right)^{B\phi^*} },
\end{equation}
where $\mathcal{F}_\mathrm{phase}$ is given by:
\begin{equation}
    \mathcal{F}_\mathrm{phase} = (1 - \xi)\mathrm{erf}\left( \frac{\sqrt{\pi}} {2(1-\xi)} \Phi(1+\Phi^{\gamma_\mathrm{ph}}) \right),
\end{equation}
which depends on the melt fraction in the upper mantle as $\Phi = F_{\mathrm{UM}}/\phi^*$. In this parameterization erf is the error function and $\delta_{\mathrm{ph}}, B, \phi^*, \xi$, and $\gamma_\mathrm{ph}$ are all empirical constants with values given in Table \ref{consttable}.

\subsection{Evolution of the Stagnant Lid}

The conventional treatment for modeling stagnant lid growth is to track the heat balance at the base of the stagnant lid \citep{spohn_mantle_1991}.
We can quantify this heat balance as:
\begin{equation}
    \rho_m c_m (T_{\mathrm{UM}} - T_L) \frac{dD_L}{dt} = -q_\mathrm{UM} - k_\mathrm{UM} \frac{\partial T}{\partial r} \bigg\rvert_{r = R_p - D_L} \label{staglid},
\end{equation}
where $T_\mathrm{UM}$ is the upper mantle temperature, $T_L$ is the temperature at the base of the stagnant lid (above the upper thermal boundary layer), $D_L$ is the thickness of the stagnant lid, and $q_\mathrm{UM}$ represents the heat flow from the mantle to the base of the stagnant lid.
Generally the right-hand side of Eq.~\eqref{staglid} is equivalent to the convective heat flow from the mantle, $q_\mathrm{UM}$, but it can also include a separate term that represents the assumption that melt produced in the mantle transmits heat to the base of the stagnant lid instead of the surface \citep{morschhauser_crustal_2011}.

In this work, we assume that melt produced in the mantle transmits heat away from the mantle, rather than the base of the stagnant lid to be consistent with our definition of eruption efficiency, which results in magma ending either in the crust or on the surface.
The second term on the right-hand side of Eq. \eqref{staglid} represents conduction of heat to the base of the stagnant lid through the boundary layer and is obtained by solving the steady-state heat equation
\begin{equation} 
    k_L \frac{\partial^2 T}{\partial r^2} = -H_L \label{thermgrad},
\end{equation}
which is piecewise with $H_L$, the volumetric radioactive heating rate due to all radioactive species, and $k_L$, the thermal conductivity. These parameters have different values depending on whether they are in the crust or mantle chemical component of the stagnant lid.

To solve Eq.~\eqref{thermgrad}, we take as our boundary conditions that the temperature at the surface of Venus is $T_s$ and the temperature at the base of the stagnant lid is $T_L$. We also enforce the continuity conditions that the temperature and heat flux are continuous at the crust-mantle chemical boundary. Lastly, we assume that the base of the stagnant lid is mantle material to obtain the solution of Eq. \eqref{thermgrad} at the base of the stagnant lid as:
\begin{align} \label{thermgradstaglidcalc}
    &\frac{\partial T}{\partial r} \bigg\rvert_{r = R_p - D_L} = -H_m \frac{R_p - D_L}{k_\mathrm{UM}} + \frac{(B_1 + B_2 + B_3)}{B_4} \\
    &B_1 = R_p D_{cr} H_m (k_\mathrm{UM} - k_{cr}), \\
    &B_2 = k_{cr} \left(\frac{D_L^2 H_{m}}{2} + R_p D_L H_{i,m} + k_\mathrm{UM} (T_L -  T_s) \right), \\
    &B_3 = D_{cr}^2 \left( k_\mathrm{UM} H_{m} + \frac{k_{cr} H_{m}}{2} + \frac{k_\mathrm{UM} H_{cr}}{2} \right), \\
    &B_4 = k_\mathrm{UM} (D_{cr} k_\mathrm{UM} - D_{cr} k_{cr} + D_L k_{cr}),
\end{align}
where $R_p$ is the radius of Venus, $D_L$ is the depth of the stagnant lid, $D_{cr}$ is the depth of the crust, $H_m$ and $H_{cr}$ are the volumetric mantle and crustal heating rate, respectively, due to all radioactive species, $k_\mathrm{UM}$ is the thermal conductivity of the upper mantle, and $k_{cr}$ is the thermal conductivity of the crust.

We assume that any crust that is produced by volcanism is buried by further volcanic eruptions akin to the heat pipe model of crustal burial \citep{oreilly_magma_1981,turcotte_heat_1989}.
Following \citet{foley_carbon_2018} we assume that the overlying pressure on buried crust increases until the crust experiences a change in phase from basalt to eclogite.
This phase transition results in a more dense parcel of rock that, being less buoyant than the overlying crust, delaminates from that crust and drips further down into the mantle \citep{morschhauser_crustal_2011,tosi_habitability_2017,foley_carbon_2018}.
\citet{foley_carbon_2018} provide the following equation
to model eclogite-driven crustal recycling behavior:
\begin{multline} \label{eq:dVCrustdt}
\frac{dV_{cr}}{dt} = \left(\frac{\dot{M}_\mathrm{melt}}{\rho_{cr}} - 4 \pi (R_p - D_{cr})^2 \mathrm{min}(0,\frac{dD_L}{dt})\right)\\\left(\tanh(20(D_{cr} - D_L)) + 1\right).
\end{multline}
This equation takes into account the creation of crust through eruptions as well, hence the $\dot{M}_\mathrm{melt}$ term.
The stagnant lid depth term represents the need for crust to return to the mantle if the stagnant lid as a whole is shrinking.
The tanh multiplier enforces the assumption that any crustal depth greater than the stagnant lid is recycled back into the mantle.

The stagnant lid tectonic regime, while not being as amenable to volatile recycling as plate tectonics, can still return incompatible elements (elements that preferentially partition from the solid mantle into the melt) and volatiles from the crust to the mantle through crustal recycling \citep{foley_carbon_2018}.
In our model any mass of crust that founders back into the mantle is assumed to transport a proportional mass of any incompatible element, $j$ (either a radioisotope or volatile species), to the mantle based on the mass fraction of each in the crust, $X^{j}_{cr}$.
Following \citet{foley_carbon_2018}, the change in number of radioisotopes and volatiles in the crust and mantle due to crustal recycling is then given by:
\begin{align}
    \frac{dN^j_{cr}}{dt} &= - \frac{N_{i,cr}}{V_{cr}} R_{cr,\mathrm{recycle}}\\
    \frac{dN^j_\mathrm{man}}{dt} &= \frac{N_{i,cr}}{V_{cr}} R_{cr,\mathrm{recycle}},
\end{align}
where $N^j_{cr}$ and $N^j_\mathrm{man}$ refers to the number of radioisotopes or volatile molecules (e.g., CO$_2$) in the crust or mantle reservoir and $R_{cr,\mathrm{recycle}}$ is the rate at which a unit volume of crust is returned to the mantle, which can be obtained from Eq. \eqref{eq:dVCrustdt}.

We can compare our model of stagnant lid evolution to \citet{foley_carbon_2018} by using their sample initial conditions and melt calculations to reproduce Figure 4 in their paper.
To make this comparison, we made temporary changes to our own model to match the formulation of mantle convection given in \citet{foley_carbon_2018}.
Primarily, we temporarily removed the core's heat flow from our mantle heat budget.
Additionally, we temporarily changed our solidus function from the third-order polynomial in \citet{elkins-tanton_linked_2008} to the linear model given in \citet{fraeman_influence_2010} used by \citet{foley_carbon_2018}.

\begin{figure*}[htb!]
    \includegraphics[width=\linewidth]{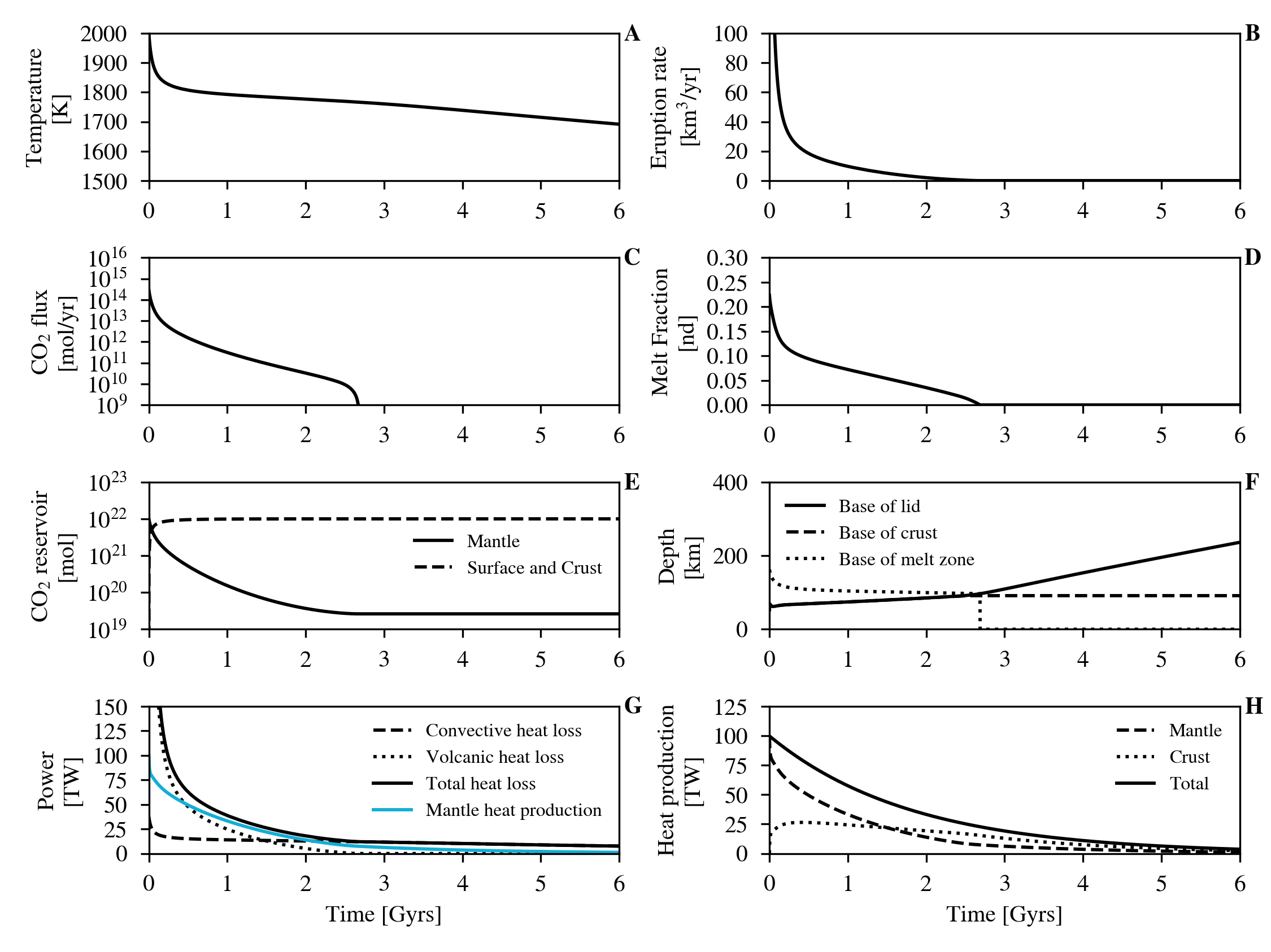}
    \centering
    \caption{A reproduction of Fig. 4 of \citet{foley_carbon_2018} but with a slightly modified version of our model. Most of the qualitative behavior in this figure is the same as in \citet{foley_carbon_2018} though the timing of certain events, such as the melt fraction reaching 0 are different due to a difference in choice of half-life and heat-producing elements. Where necessary, black lines representing different values are distinguished with solid, dashed, and dotted lines.}
    \label{fig:validatestagnantlid}
\end{figure*}

For our initial conditions to match \citet{foley_carbon_2018} we ran a simulation with an initial upper mantle temperature of 2000 K, an initial heating due to potassium-40 of 100 TW (neglecting all other heat-producing elements), an initial stagnant lid depth of 70 km, a reference kinematic viscosity of 9$\times$10$^6$ m$^2$/s, an initial mantle content of carbon dioxide of 10$^{22}$ mol and no influence of water on mantle viscosity.
Using these conditions, we were able to closely replicate Figure 4 in \citet{foley_carbon_2018} as can be seen in Fig. \ref{fig:validatestagnantlid}, which shows the time evolution of the same mantle properties that are found in Figure 4 of \citet{foley_carbon_2018}.

While our mantle properties evolve with the same trend and have similar values with those in \citet{foley_carbon_2018}, the timing of key events such as the melt fraction zeroing out or the crossover between convective and volcanic heat loss is different between our model and that of \citet{foley_carbon_2018}.
This difference is likely due to our differing approaches to heat-producing radioactive elements, while we model radiogenic heat using $^{40}$K, $^{232}$Th, $^{238}$U, and $^{235}$U each with their own half-lives, \citet{foley_carbon_2018} model radiogenic heat using a hypothetical element with a half-life of 2.94 Gyrs representing a composite of $^{238}$U, $^{235}$U, $^{232}$Th, and $^{40}$K.

\subsection{Melt Transport of Radioisotopes and Volatiles}
Modeling the thermally driven transport of radioisotopes and water is required for self-consistency as these elements play important roles in the thermal evolution of the mantle.
These incompatible elements are transported to the crust because they are preferentially incorporated into the partial melt in the mantle.
As melt from the mantle reaches the crust and surface, the mantle becomes relatively depleted and the crust becomes relatively enriched in these elements \citep{taylor_geochemical_1995,morschhauser_crustal_2011,noack_coupling_2012}.

Melting occurs in our model where the temperature in the mantle is higher than the solidus temperature of KLB-1 peridotite given in \citet{elkins-tanton_linked_2008}.
The depth of the melt region, $D_\mathrm{melt}$ is defined by the intersection of the solidus and the mantle geotherm, $T(z)$.
Following \citet{godolt_habitability_2019}, we ensure that $D_\mathrm{melt}$ cannot be greater than a depth corresponding to 8 GPa as magma generated below this depth will not be buoyant enough to rise to the surface.
We benchmark our model to \citet{godolt_habitability_2019} to set the groundwork for future exoplanet studies but caution that a 12 GPa cutoff is more appropriate for Venus. 
We found that using 12 GPa for our models did not yield qualitative differences in evolutionary paths.
We derive our solidus equation by generalizing the fit for Earth given in \citet{elkins-tanton_linked_2008} and \citet{barnes_vplanet_2020} and changing the variables such that the solidus is a cubic function of depth instead of radius, thus yielding:
\begin{align}
    T_\mathrm{sol}(z) &= A z^3 + B z^2 + C z + D \\
    A &= -A_\mathrm{sol} \\
    B &= 3A_\mathrm{sol} R_E + B_\mathrm{sol} \\
    C &= -3 A_\mathrm{sol} R_E^2 - 2 B_\mathrm{sol} R_E - C_\mathrm{sol} \\
    D &= A_\mathrm{sol} R_E^3 + B_\mathrm{sol} R_E^2 + C_\mathrm{sol} R_E + D_\mathrm{sol},
\end{align}
where $z$ is the depth, $R_E$ is the radius of Earth, and $A_{sol}$, $B_\mathrm{sol}$, $C_\mathrm{sol}$, and $D_\mathrm{sol}$ are the original solidus fit values given in \citet{elkins-tanton_linked_2008} and \citet{barnes_vplanet_2020} and listed in Table \ref{consttable}.

We evaluate the upper mantle melt fraction at the point just below the thermal upper boundary layer, which is where $T_\mathrm{UM}$ is defined as well, such that:
\begin{equation} \label{meltequation}
    F_\mathrm{UM} = \frac{T_\mathrm{UM} - T_\mathrm{sol}(D_L+\delta_\mathrm{UM})}{T_\mathrm{sol}(D_L+\delta_\mathrm{UM}) - T_\mathrm{liq}(D_L+\delta_\mathrm{UM})},
\end{equation}
where $D_L$ is the thickness of the stagnant lid and $\delta_\mathrm{UM}$ is the thickness of the thermal upper boundary layer.
We consider the liquidus temperature, $T_\mathrm{liq}$ to be greater than the solidus temperature by a constant, $\Delta T_\mathrm{sol,liq} = 500$ K \citep{driscoll_thermal_2014,barnes_vplanet_2020}.
Following \citet{foley_carbon_2018} we calculate the melt production rate by assuming a cylindrical upwelling of melt that moves at the convective velocity, $v_\mathrm{conv}$, such that:
\begin{align}
    \dot{M}_\mathrm{melt} &= 17.8 \pi R_p v_\mathrm{conv}(D_\mathrm{melt} - D_L) F_\mathrm{UM} \rho_m \label{eqmeltmassfluxman}, \\
    v_\mathrm{conv} &= \frac{c_2 \kappa}{R_p - R_c} \left(\frac{\mathrm{Ra}_i}{\Theta}\right)^{2/3},
\end{align}
where the coefficient 17.8 is derived from assuming Earth-like convection cell length scales of about 0.45 $R_p$ \citep{foley_carbon_2018}. Note that $\dot{M}_\mathrm{melt}$ is directly related to $Q_\mathrm{melt}$ as:
\begin{equation}
    Q_\mathrm{melt} = \epsilon_\mathrm{erupt} \dot{M}_\mathrm{melt} c_\mathrm{man} \Delta T_\mathrm{melt}.
\end{equation}
The melt eruption temperature, is found by extrapolating the temperature of the melt at the melt depth along the mantle adiabat:
\begin{equation}
    \Delta T_\mathrm{melt} = T_\mathrm{melt} - \gamma D_\mathrm{melt} - T_\mathrm{surf},
\end{equation}
where $\gamma$ represents the mantle adiabatic gradient, i.e., the increase in mantle temperature with depth due to convection.
Our model holds $T_\mathrm{surf}$ constant at 700 K.

We treat the incorporation of incompatible elements from the mantle into the melt by assuming an aggregate liquid fractional melting model \citep{white2020geochemistry}.
With this model we can calculate the weight fraction of any element $j$ in the melt as:
\begin{equation}
    X^j_\mathrm{melt} = X^j_\mathrm{man} \frac{1 - (1-F_\mathrm{UM})^{1/D_j}}{F_\mathrm{UM}} \label{fracmelt},
\end{equation}
where $X^j_\mathrm{man}$ is the weight fraction of that element in the mantle and $D_j$ is the partitioning coefficient for the element.
Following \citet{foley_carbon_2018}, we use a value of 0.002 for the partitioning coefficient for all radioisotopes.
For the partitioning coefficient of water and carbon dioxide, we follow \citet{grott_volcanic_2011} and use a value of 0.01.
At these values the amount of each element that is incorporated into the melt from the mantle is roughly inversely proportional to the melt fraction such that the melt is relatively rich in incompatible elements at lower melt fractions.

Elements and volatiles in the melt are then transported out of the mantle with a mass flux given by
\begin{equation}
    \frac{dM^j_\mathrm{man}}{dt} = \dot{M}_\mathrm{melt} X^j_\mathrm{melt}, \label{manfluxout}
\end{equation}
where $\dot{M}_\mathrm{melt}$ is the mass of melt leaving the mantle per unit time calculated from Eq. \eqref{eqmeltmassfluxman}.
All of the radioisotopes that are erupted from the mantle enter the crust.
Only some of the volatiles that are erupted from the mantle are outgassed to the atmosphere; the rest are emplaced in the crust due to either being trapped in intrusive melt or high solubilities resulting in the volatiles exsolving from extrusive melt.
The maximum amount of volatiles that can be outgassed to the atmosphere (assuming any amount of volatiles in extrusive melt are outgassed to the atmosphere) is given by
\begin{equation}
    \frac{dM^j_\mathrm{atm}}{dt} = \dot{M}_{melt} X^j_\mathrm{melt} \epsilon_\mathrm{erupt},
\end{equation}
where $\epsilon_{erupt}$ represents the fraction of melt that is erupted to the surface. 
The actual outgassing rate is typically lower due to solubility effects described in the next section.

\subsection{Developing a Secondary Atmosphere from Mantle Outgassing} \label{outgassing}

Outgassing, the bubbling out of gases from melt, is the key process in our model that connects the planet's interior evolution to its atmospheric evolution.
Some of the gases dissolved in the melt in the mantle exsolve from the melt as it rises to the surface due to the decreasing pressure.
The gases that do not leave the melt contribute to the increasing volatile content in the crust and may be recycled back into the mantle in the future.
While the amount of volatiles outgassed depends strongly on the weight fraction of that volatile in the erupted melt, the process is complicated by a variety of factors, including the composition of the melt, the structure of the overlying crust, and the dynamics in the conduit \citep{oppenheimer_44_2014}.

The pressure, $P$, above the gas bubbles leaving the magma sets the partial pressures of gases inside a melt bubble ($p\mathrm{CO_2},p\mathrm{H_2O}$), which ultimately dictates how much of each volatile is outgassed.
To take into account pressure effects on volatile solubilities we apply a simplified version of the \volcgases\footnote{Publicly available at https://github.com/Nicholaswogan/VolcGases} model as described in \citet{wogan_abundant_2020} that adapted the solubility relations for water and carbon dioxide found in \citet{iacono-marziano_new_2012}.
We follow the C-O-H system as described in \citet{wogan_abundant_2020}, but neglect carbon and hydrogen speciation into more reducing gases as we assume an oxidizing mantle, which simplifies the pressure inside the melt bubble as due only to water and carbon dioxide such that $P = p\mathrm{H_2O} + p\mathrm{CO_2}$.

The particular species of carbon- and hydrogen-bearing gases that exsolve out from the melt and into the atmosphere depends on the oxidation state of the source region of that melt: the upper mantle \citep{gaillard_theoretical_2014,sossi_redox_2020,ortenzi_mantle_2020}.
In this work, we assume that the oxidation state of the upper mantle of Venus is similar to Earth’s, which is relatively oxidized at a value about equal to that of the Fayalite-Magnetite-Quartz (FMQ) buffer \citep{stagno_oxidation_2013,armstrong_deep_2019}.
This assumption allows us to neglect the outgassing of more reduced carbon- and hydrogen-bearing species such as methane and hydrogen gas and consider only the outgassing of carbon dioxide and water vapor.

By neglecting other carbon-bearing species (especially carbon monoxide), our model overestimates the amount of carbon that speciates to carbon dioxide; similarly our model also overestimates the production of hydrogen that speciates to water by neglecting contributions in the melt bubble due to hydrogen gas.
At reducing mantle conditions we would also expect that carbon overwhelmingly partitions into graphite instead of carbonates and decreases the amount of carbon dioxide outgassed \citep{hirschmann_ventilation_2008}.
\citet{krissansen-totton_was_2021} point out that these two effects of a reducing mantle partially offset each other when it comes to the outgassing of carbon dioxide and so our assumptions are likely reasonable for probing Venus' history.

Earth's oxidizing mantle is thought to result from a variety of geochemical processes that occur during differentiation and core formation including disproportionation \citep{wood_accretion_2006} and magma ocean crystallization \citep{deng_magma_2020}. As these processes are likely experienced by any rocky planet with a similar size and composition to Earth such as Venus \citep{deng_magma_2020} we assume an oxidizing mantle such that carbon partitions into carbon dioxide and hydrogen partitions into water.
While \citet{krissansen-totton_was_2021} and \citet{warren_narrow_2023} demonstrate that both reducing and oxidizing mantle assumptions can reproduce modern-day Venus, our work makes no assumptions on the former habitability of Venus (a point that helps distinguish between their oxidizing and reducing evolutions) and so our assumption of an oxidizing mantle is consistent with their results. 

Given water and carbon dioxide in surface melt, the fraction of each volatile that exsolves out from the melt into the atmosphere depends on the solubility of each volatile in that melt \citep{grott_volcanic_2011}.
The two primary influences on solubility are the thermomechanical environment of the melt (temperature and pressure) and the chemical properties (i.e., composition) of the melt itself \citep{iacono-marziano_new_2012}.
With many empirical models that  measure the solubility of water and carbon dioxide in silicate melts available \citep{papale_modeling_1997,papale_modeling_1999,newman_volatilecalc:_2002,hirschmann_ventilation_2008,iacono-marziano_new_2012}, we use the model described in \citet{iacono-marziano_new_2012} as it has an open-source implementation described in \citet{wogan_abundant_2020}.

Following the water and carbon dioxide solubility calculations in \citet{wogan_abundant_2020} and matching observations of the crustal composition of Venus from the Venera mission \citep{smrekar_venus_2018}, we take our magma composition to be mafic with the chemical properties of the alkali-basaltic Mt. Etna sample given in \citet{iacono-marziano_new_2012}.
Assuming other basaltic compositions, such as a Mauna Loa composition \citep{Powers1955} resulted in some variations in the amount of carbon dioxide and water in Venus' atmosphere but did not yield qualitative differences in evolutionary paths. 
We did not consider an ultramafic melt composition as this is outside the limits of the model calibrated in \citet{iacono-marziano_new_2012}. For example, applying a KLB-1 peridotite composition \citep{Davis2009} to this solubility model resulted in carbon dioxide outgassing rates multiple orders of magnitude higher than any mafic melt composition.

To calculate the amount of carbon dioxide and water outgassed, we start with their solubilities.
The solubility of carbon dioxide and water is given in \citet{iacono-marziano_new_2012} as well as Equations 1 and 2 in \citet{wogan_abundant_2020}) as:
\begin{align} 
    \ln(x^\mathrm{post}_\mathrm{CO_2}) &= x^\mathrm{post}_\mathrm{H_2O} d_\mathrm{H_2O} + a_\mathrm{CO_2} \ln(p\mathrm{CO_2}) + S_1 \label{solequationco2}, \\
    \ln(x^\mathrm{post}_\mathrm{H_2O}) &= a_\mathrm{H_2O} \ln(p\mathrm{H_2O}) + S_2 \label{solequationh2o},
\end{align}
where $x^\mathrm{post}_\mathrm{H_2O}$ and $x^\mathrm{post}_\mathrm{CO_2}$ are the mole fractions of water and carbon dioxide remaining in the melt after outgassing, while $a_\mathrm{CO_2},d_\mathrm{H_2O},S_1,$ and $S_2$ are all constants related to the solubility of carbon dioxide and water in a Mt. Etna melt from \citet{iacono-marziano_new_2012}.
$a_\mathrm{CO_2}$ and $d_\mathrm{H_2O}$ are solubility coefficients for water and carbon dioxide, and $S_1$ and $S_2$ are coefficients related to the composition of the melt, which we assume to be constant.
We pair Eqs. \eqref{solequationco2} and \eqref{solequationh2o} with the following atom conservation equations
\begin{align}
    \frac{X^\mathrm{CO_2}_\mathrm{melt} \mu_\mathrm{magma}}{\mu_\mathrm{CO_2}} &= \frac{p\mathrm{CO_2}}{P}\alpha_\mathrm{gas} + (1-\alpha_\mathrm{gas})x^\mathrm{post}_\mathrm{CO_2} \label{conservationco2},\\
    \frac{X^\mathrm{H_2O}_\mathrm{melt} \mu_\mathrm{magma}}{\mu_\mathrm{H_2O}} &= \frac{p\mathrm{H_2O}}{P}\alpha_\mathrm{gas} + (1-\alpha_\mathrm{gas})x^\mathrm{post}_\mathrm{H_2O} \label{conservationh2o},
\end{align}
to relate the mass fraction of carbon dioxide and water initially in the melt to the moles of carbon dioxide and water that exsolve from the melt and remain in the melt. $\alpha_\mathrm{gas}$ refers to the mole fraction in the gas phase.

We use Eqs.\eqref{solequationco2}--\eqref{conservationh2o} and the bisection method to solve for the partial pressure of carbon dioxide in the melt bubble, $p\mathrm{CO_2}$.
We then use $p\mathrm{CO_2}$ to calculate $\alpha_\mathrm{gas}$, which enables the calculation of the outgassing mass fluxes for carbon dioxide ($R_\mathrm{CO_2}$) and water ($R_\mathrm{H_2O}$) as
\begin{align}
    R_\mathrm{CO_2} &= \frac{\alpha_\mathrm{gas}\mu_\mathrm{CO_2}}{\mu_\mathrm{magma} (1-\alpha_\mathrm{gas})} \frac{p\mathrm{CO_2}}{P} \dot{M}_\mathrm{melt} \epsilon_\mathrm{erupt}, \\
    R_\mathrm{H_2O} &= \frac{\alpha_\mathrm{gas}\mu_\mathrm{H_2O}}{\mu_\mathrm{magma} (1-\alpha_\mathrm{gas})} \frac{p\mathrm{H_2O}}{P} \dot{M}_\mathrm{melt} \epsilon_\mathrm{erupt}.
\end{align}
Subtracting these outgassing fluxes from Eq. \eqref{manfluxout} gives the flux of water and carbon dioxide into the crust.

\begin{figure*}[htb!]
    \includegraphics[width=\linewidth]{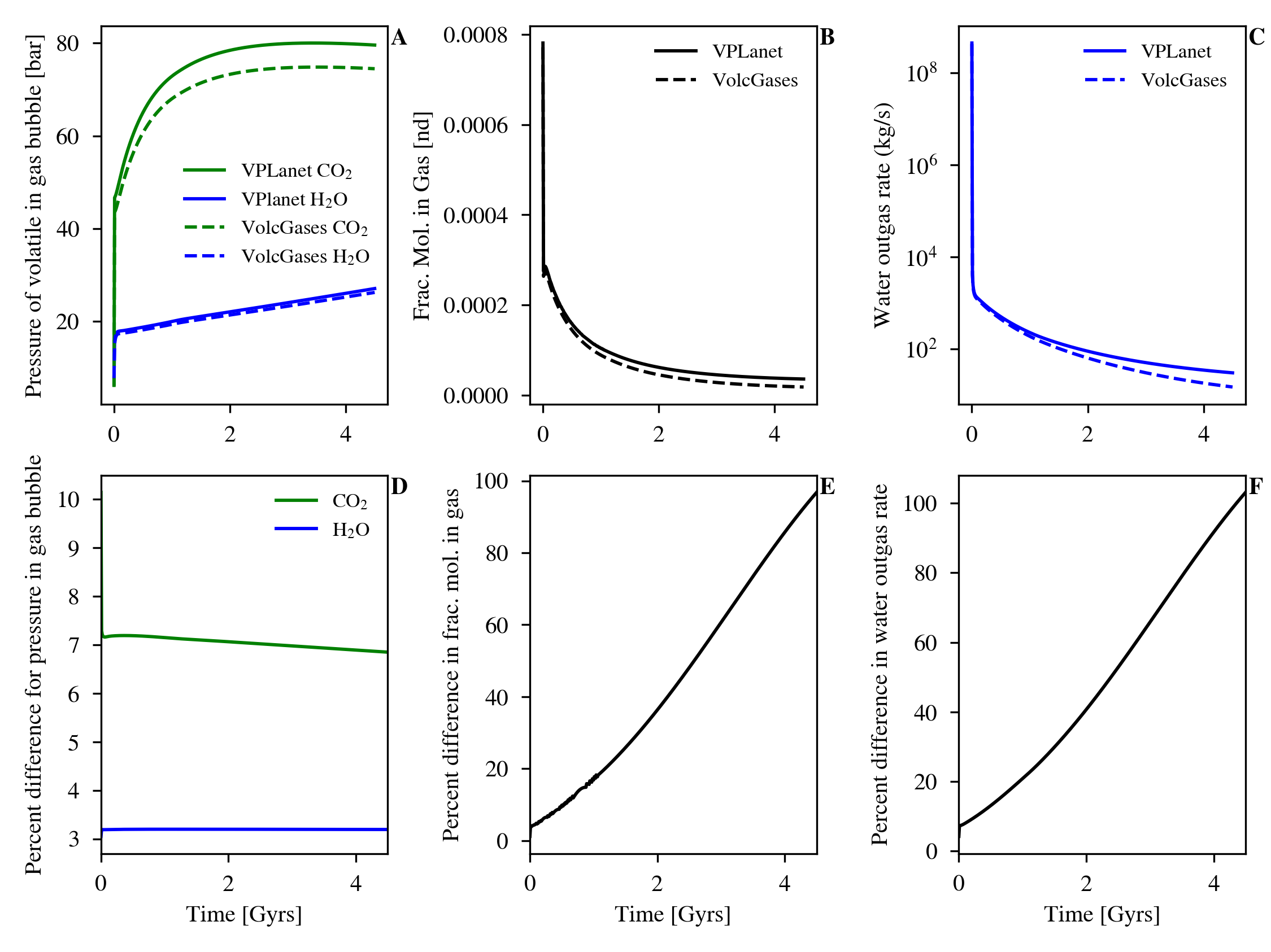}
    \centering
    \caption{Comparison between the outgassing model in \vplanet (solid lines) and in \volcgases (dotted lines). The top row displays true values while the bottom row displays the percent difference between each model. The first column illustrates the difference in the partial pressure of CO$_2$ (green) and H$_2$O (blue) in the gas bubble. The second column illustrates the difference in the fraction of moles in the gas phase vs. the melt phase, $\alpha_\mathrm{gas}$. The third column illustrates the difference in the water outgassing rate, calculated from $P_\mathrm{H_2O}$ and $\alpha_\mathrm{gas}$.}
    \label{fig:validateoutgassing}
\end{figure*}

We incorporated this simplified model into \vplanet and validated it by comparing results from a single \vplanet simulation to \volcgases \citep{wogan_abundant_2020}.
To make this comparison, we take the surface pressure, eruption temperature, $X_\mathrm{melt}^\mathrm{CO_2}$, and $X_\mathrm{melt}^\mathrm{H_2O}$ values from the \vplanet simulation as a function of time and feed those as inputs to \volcgases to obtain $p$CO$_2$, $p$H$_2$O, and $\alpha_\mathrm{gas}$.
We then compare the \volcgases-calculated values to those same outputs as in \vplanet, as shown in Fig. \ref{fig:validateoutgassing}.
As expected, our simplified model overestimates the amount of CO$_2$ exsolved by about 7\% and the amount of H$_2$O exsolved by about 3\% compared to \volcgases, as we neglect the contribution of more reduced gases such as CO and H$_2$ to the total pressure in the melt bubble.

The difference in outgassing predictions between \vplanet and \volcgases grows over time as the amount of CO$_2$ in the atmosphere grows.
The outgassing difference is driven by a growing difference in $\alpha_\mathrm{gas}$ between the two models as can be seen in Fig. \ref{fig:validateoutgassing}.
Initially the $\alpha_\mathrm{gas}$ difference is low, but as the simulation evolves the surface pressure increases and the gas pressure in the melt bubble grows to equilibrate with the increasing surface pressure.
The \vplanet model assumes that this increasing gas pressure is due solely to CO$_2$ and H$_2$O, while the \volcgases model includes the contributions of other H- and C-bearing species, most critically, CO and H$_2$.
This increasingly larger value of $x^\mathrm{post}_\mathrm{CO_2}$ in the \vplanet model drives the greater value of $\alpha_\mathrm{gas}$, which in turn drives greater outgassing fluxes.

\subsection{Atmospheric Escape}

Atmospheric escape in our model is driven by the photolysis of water due to the XUV radiation from the Sun.
We assume that the hydrogen produced from photolysis escapes into space based on the hydrodynamic escape mechanism described in \citet{barnes_vplanet_2020}; see also \citet{watson_dynamics_1981} and \citet{hunten_mass_1987}.
Hydrogen initially escapes in the energy-limited regime, which can carry away the oxygen produced by photodissociation depending on the amount of energy present \citep{hunten_mass_1987}.
Following \citet{luger_extreme_2015}, we assume that due to the lack of oxygen in the modern Venus atmosphere, there must be some efficient sink (that we do not explicitly model) and thus any oxygen produced from photolysis does not remain in the atmosphere, i.e., it is immediately absorbed by surface sinks such as a magma ocean.

Because carbon dioxide builds up quickly in all of our Venus simulations we also calculate when the hydrogen escape is diffusion-limited due to the substantial amount of carbon dioxide in the atmosphere.
When the atmosphere consists of more than 60\% carbon dioxide, we switch from energy-limited to diffusion-limited escape, where water must diffuse through carbon dioxide to reach the base of the thermosphere. 
Our formalism for diffusion-limited escape through carbon dioxide is based on Eq. (13) in \citet{luger_extreme_2015} with the binary diffusion coefficient for carbon dioxide given by \citet{zahnle_mass_1986}.

\subsection{Stellar Evolution}

\begin{figure*}[htp!]
    \centering
    \includegraphics[width=\linewidth]{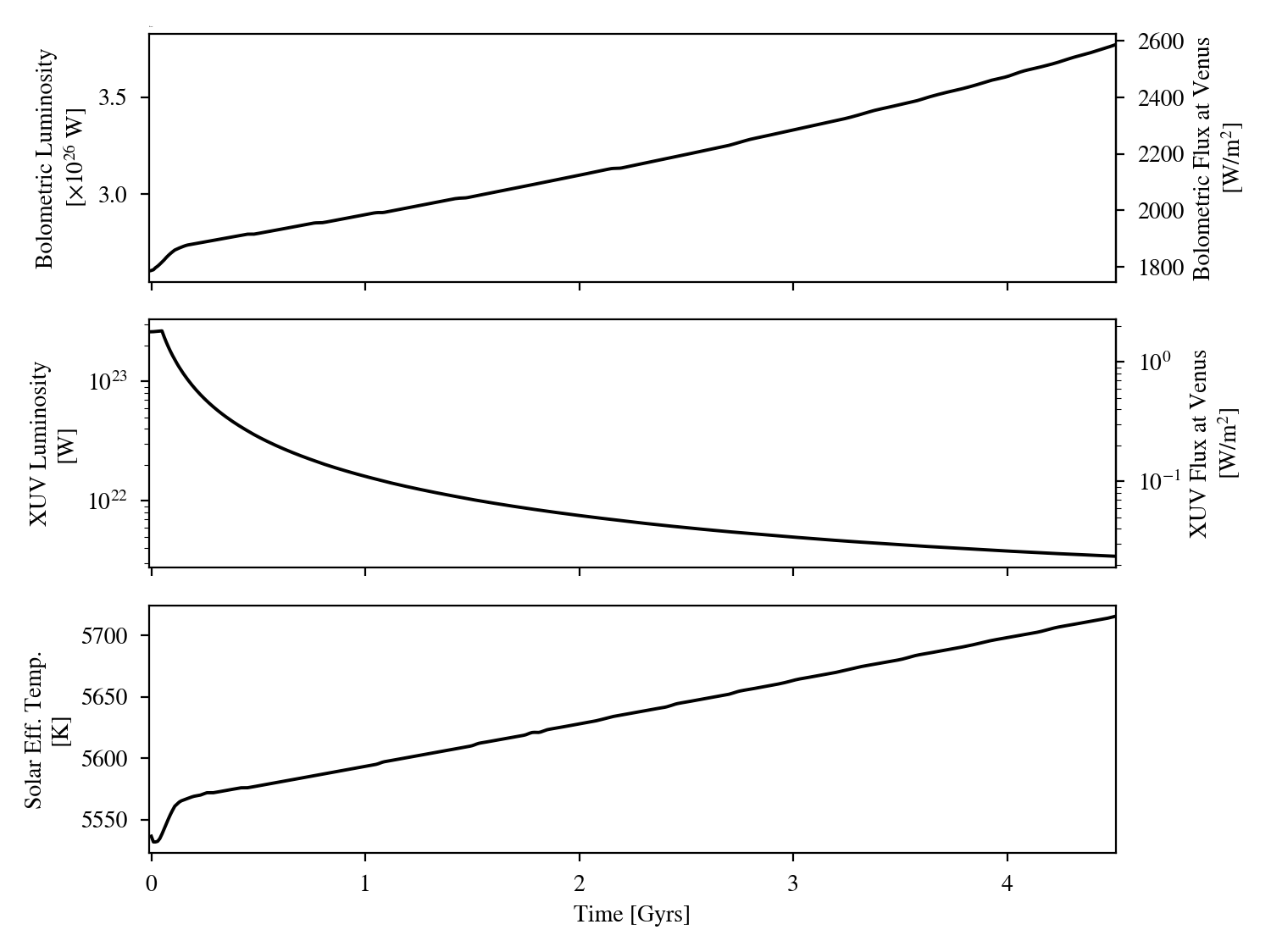}
    \caption{The top panel shows the evolution of the Sun's bolometric luminosity over time both in Watts and in terms of the incident flux at Venus. The middle panel shows the evolution of the Sun's XUV luminosity over time again in Watts and in terms of the incident flux at Venus. Luminosity over time is calculated based on the interpolated \citet{baraffe_new_2015} model. The evolution of the effective temperature of the Sun is shown in the bottom panel.}
    \label{fig:SunEvolvePlot}
\end{figure*}

The XUV environment of Venus changes over time with the evolution of the Sun, we model this evolution using the \stellar module in \vplanet.
The \stellar module uses a bicubic spline interpolation of the stellar evolution models in \citet{baraffe_new_2015} to calculate the evolution of a star's radius, temperature, and luminosity over time.
From the star's luminosity, we can calculate its XUV luminosity, $L_\mathrm{XUV}$ based on the empirical broken power law in \citet{ribas_2005},
\begin{equation}
    \frac{L_\mathrm{XUV}}{L_\mathrm{bol}} = 
        \begin{cases}
            f_\mathrm{sat} & \text{if } t \leq t_\mathrm{sat} \\
            f_\mathrm{sat} \left(\frac{t}{t_\mathrm{sat}}\right)^{-\beta_\mathrm{XUV}} & \text{if } t > t_\mathrm{sat},
        \end{cases}
\end{equation}
where $L_\mathrm{bol}$ is the bolometric (across all wavelengths) luminosity, $f_\mathrm{sat}$ is the initial ratio of XUV to bolometric luminosity, $t_\mathrm{sat}$ is the saturation timescale and $\beta_\mathrm{XUV}$ is the empirically derived power law exponent.
Based on \citet{jackson2012} and \citet{ribas_2005}, we use $f_\mathrm{sat} = 10^{-3}$, $t_\mathrm{sat} = 100$ Myrs, and $\beta_\mathrm{XUV} = 1.23$ to model the Sun.
The changing bolometric and XUV luminosity and incident flux on Venus are shown in Fig.~\ref{fig:SunEvolvePlot}.

\subsection{Dynamo Evolution of the Core}
Multiple methods have been proposed to calculate the strength of a dynamo, and here we consider two: buoyancy and the magnetic Reynolds number, $Rm$. Buoyancy in the core may be due to both thermal convection and compositional convection driven by solidification of an inner core.
Here, we reproduce the equations for these buoyancy fluxes from \citet{barnes_vplanet_2020} for completeness:
\begin{align} \label{eq:buoyancyeqs}
    \mathcal{F}_\mathrm{th} &= \frac{\alpha_c g_c}{\rho_c c_c}q_{c,\mathrm{conv}}, \\
    \mathcal{F}_\chi &= g_{ic} \frac{\Delta \rho_\chi}{\rho_c}\left(\frac{R_\mathrm{ic}}{R_c}\right)\dot{R}_\mathrm{ic},
\end{align}
where the subscript $c$ refers to bulk core properties, $q_{c,\mathrm{conv}}$ represents the core convective heat flux, which is the difference between total and conductive heat flux at the core-mantle boundary, $g_\mathrm{ic}$ is the gravity at the inner-core boundary, and $\Delta \rho_\chi$ is the outer core compositional density difference due to light element production at the inner core boundary.
Numerical studies \citep[e.g.][]{Christensen2009Nature} have found that the magnetic dipole moment is proportional to the cube root of the net buoyancy flux, so Venus will have a magnetic field whenever
\begin{equation} \label{dynamobuoyancycondition}
    \mathcal{F}_\mathrm{th} + \mathcal{F}_\chi > 0.
\end{equation}

The thermal buoyancy flux, $\mathcal{F}_\mathrm{th}$, depends on the efficiency of convective heat transfer at the core-mantle boundary compared to the efficiency of conduction up the core-mantle boundary.
This relationship is expressed by
\begin{equation}
    q_{c,conv} = q_{cmb} - q_{c,ad} = k_\mathrm{LM}\frac{T_\mathrm{CMB} - T_\mathrm{LM}}{\delta_\mathrm{LM}} - \frac{k_\mathrm{OC}T_\mathrm{CMB}R_C}{D_\mathrm{core}^2}.
\label{eq:cmb_heat_flux}
\end{equation} 
The compositional buoyancy flux, $\mathcal{F}_\chi$ depends on the efficiency of core cooling: if the core cools too quickly it will completely solidify and there can be no dynamo, but if the core doesn't cool at all, the inner core will never nucleate and the compositional buoyancy flux will remain zero. Given the thermal cooling rate of the core, the magnetic dipole moment $\mathcal{M}$ is estimated from the empirical scaling law,
\begin{equation} 
\mathcal{M} = 4\pi R_c^3 \gamma_d \sqrt{\rho_c/2\mu_0} \left( (\mathcal{F}_\mathrm{th} + \mathcal{F}_\chi) D_c \right)^{1/3}, 
\label{eq:magmom} 
\end{equation}
where $\gamma_d$ is the saturation constant for fast rotating dipolar dynamos, $\mu_0=4\pi \times 10^{-7}~\mbox{H~m}^{-1}$ is magnetic permeability, $D_c=R_c-R_{ic}$ is the dynamo region shell thickness, and $R_{ic}$ is the inner core radius \citep{olson_christenen_2006}.

An alternative method to determine the presence of a dynamo is to compare the magnetic Reynolds number to the critical value for magnetic field generation. This quantity is calculated as:
\begin{equation}
    R_m = \frac{v_\mathrm{conv} D}{\lambda},
\end{equation}
where $v_\mathrm{conv}$ is the convective velocity of the fluid, $D$ is some fluid length-scale and $\lambda$ is the magnetic diffusivity of the fluid, which is inversely proportional to the electric conductivity of the core \citep{olson_christenen_2006}.
Following \citet{olson_christenen_2006} we define a critical magnetic Reynolds number of 40, above which we assume that Venus maintains a dynamo-generated magnetic field.
To calculate the magnetic Reynolds number, we follow \citet{driscoll_geodynamo_2019} and calculate the convective velocity as
\begin{equation}
    v_\mathrm{conv} = Ro_0 \Omega D \left(\frac{r_c^2 (\mathcal{F}_\mathrm{th} + \mathcal{F}_\chi)}{D^4\Omega^4}\right)^{\alpha_m},
\end{equation}
where $Ro_0 = 0.85$ and $\alpha_m = 0.4$ are nominal values used in \citet{driscoll_geodynamo_2019} and \citet{ChristensenAubert2006}, $\Omega$ is the angular rotation rate of Venus (corresponding to a rotational period of 243 days), and $D = R_c - R_\mathrm{IC}$ is the length scale of the liquid outer core.

In our simulations we find that both methods yield nearly identical results with the only difference being small timing changes for when the geodynamo is quenched and/or restarted. Even when using a much faster rotation rate, such as a full rotation in 8 hours, we find that the magnetic Reynolds number results are nearly identical to our buoyancy flux approach results. This similarity arises because the convective velocity is essentially a reparameterization of the total buoyancy flux. Thus, when the buoyancy is greater than 0, the corresponding magnetic Reynolds number is greater than 40. We therefore arbitrarily choose to calculate our magnetic dipole moment using Eq.~(\ref{eq:magmom}) rather than via the magnetic Reynolds number since the two methods predict such similar results.

Regardless of how the magnetic field is calculated, the above discussion elucidates why a coupled whole-planet modeling approach is necessary to accurately model the core dynamo. The lower mantle temperatures and viscosities act as boundary conditions for the core's thermal history, and hence geodynamo generation.
Similarly, for Venus, the lack of an observable magnetic field is an important constraint that can only be matched with a self-consistent core thermal history model.

\section{Numerical Methods} \label{sec:methods}

The derivatives described in Sec. \ref{sec:modeldesc} above are solved using a modified version of the \vplanet software that is described in detail in \citet{barnes_vplanet_2020}.
In short, \vplanet solves a large ``matrix" of coupled differential equations using the fourth-order Runge-Kutta method with variable timestepping, enforcing physical limitations such as a non-negative amount of water in any planetary reservoir, at each timestep.
\vplanet is publicly available\footnote{\url{https://github.com/VirtualPlanetaryLaboratory/vplanet}}.

\subsection{Parameter Selection}
To simulate a variety of potential Venus evolutions, we varied many parameters that correspond to both initial conditions and physical assumptions in the model.
Many of these parameters are poorly constrained for Venus, so we varied them across reasonable ranges based on values derived for Earth and used in previous models.
Below we describe the justification and values of these parameter ranges, which are also listed in Table \ref{rangevary}.

For the initial thermal state of Venus, we varied the average mantle temperature ($T_\mathrm{m}$) and the temperature jump across the core-mantle boundary ($\Delta T_\mathrm{CMB}$).
For these ranges we explored values similar to the thermal states simulated in \citet{driscoll_thermal_2014}, with our upper mantle temperatures ranging from 1750 to 2100 K and the temperature at the CMB ranging from 3250 to 4800 K.

We followed \citet{barnes_vplanet_2020} in determining the amounts of radioactive $^{40}\mathrm{K}$ in the mantle ($^{40}\mathrm{K}_\mathrm{m}$), ranging from 0.5 to 1.5 times the total amount predicted for Earth.
In determining how much $^{40}\mathrm{K}$ is in the core ($^{40}\mathrm{K}_\mathrm{c}$), we took a fraction of $^{40}\mathrm{K}_\mathrm{m}$ ($\eta_\mathrm{K}$) ranging from $0.01-1$, where $1$ means equal partitioning between the mantle and core.
These values correspond to a current amount of potassium in the core that ranges from 0.5 ppm to 200 ppm.
This range is large, but designed to accommodate both the smaller $^{40}\mathrm{K}_\mathrm{c}$ values derived from partitioning experiments \citep{hirose_composition_2013} and the larger $^{40}\mathrm{K}_\mathrm{c}$ values argued for by models trying to explain the thermal evolution of the core \citep{nimmo_influence_2004}.

We used the estimates of the amount of water that Earth formed with (1-10 terrestrial oceans, TO) in \citet{raymond_making_2004} as our range of initial values of total water mass ($M^\mathrm{H_2O}_p$).
In setting up our simulations, we partitioned this water between the interior and the surface such that the initial amount of water on the surface of the planet was some fraction ($\eta_\mathrm{H_2O}$) of the total water the planet formed with.
We varied $\eta_\mathrm{H_2O}$ between 0.1 and 0.5. For CO$_2$, we followed \citet{driscoll_divergent_2013} and placed all of it in the mantle initially. We show below that this assumption should not affect our results significantly.

While some parameter choices represent initial conditions, others represent different physical assumptions about the dynamics of Venus' mantle and core.
We modeled a variety of rock rheologies given by the reference kinematic viscosity of the mantle (the $\nu_0$ in Eq. \ref{eq:viscosity}), the viscosity jump from the upper to lower mantle ($\nu_\mathrm{jump}$ in Eq. \ref{eq:viscjump}), and the depression of the Arrhenius activation energy due to water ($\Delta E_\mathrm{H_2O}$ also in Eq. \ref{eq:viscosity}).
Note that the range of reference viscosities we sampled are given in kinematic mantle viscosity units of $m^2$/s, this corresponds to a dynamic mantle viscosity range of 4.8$\times 10^{10}$ to 2.4$\times 10^{13}$ Pascal-seconds.
We followed \citet{driscoll_thermal_2014} in determining our ranges for the basic Arrhenius equation values ($\nu_0$ and $\nu_\mathrm{jump}$).
Because \citet{driscoll_thermal_2014} do not provide a formulation of how water influences viscosity, we follow the assumptions made in \citet{mcgovern_thermal_1989} that the activation energy depression ($\Delta E_\mathrm{H_2O}$) ranges from $8\times10^5$ to $6\times10^6$ K/wt. fraction, representing the range between Anita Bay and \AA heim dunites.

The amount of melt that erupts to the surface, the eruption efficiency ($\epsilon_\mathrm{erupt}$), is also a key control on the thermal evolution of a planet \citep{driscoll_thermal_2014,lourenco_plutonic-squishy_2020}.
For the lower limit of our $\epsilon_\mathrm{erupt}$ values, we chose a relatively low value of 0.01 so that we could fully explore the effects of intrusive melt-dominated dynamics, such as the squishy lid described in \citet{lourenco_plutonic-squishy_2020}.
Our upper limit of 0.2 was chosen based on values corresponding to a shut-off of the Venusian dynamo found in simulations in \citet{driscoll_thermal_2014}. 

As the potential for a dynamo in the core is also a key constraint on our simulations, we chose a variety of values for the core reference liquidus depression due to light elements ($\Delta T_{\chi,\mathrm{ref}}$), and the thermal conductivity of the outer core ($k_\mathrm{OC}$).
\citet{hirose_composition_2013} cites a depression up to about 700 K at core-relevant pressures, and so we varied $\Delta T_{\chi,\mathrm{ref}}$ between 0 and 700 K.
We vary $k_\mathrm{OC}$ based on the range of modern estimates explored by \citet{orourke_prospects_2018}, with a lower limit of 40 W/m/K and an upper limit of 200 W/m/K that represents higher values derived from experiments and a hotter CMB temperature.

\begin{deluxetable*}{lllll}
\tablecaption{Varied parameters in study \label{rangevary}} 
\tablehead{\colhead{Notation} & \colhead{Description} & \colhead{Range} & \colhead{Units} & \colhead{Distribution}}
\startdata
$T_m$ & Initial average mantle temp. & [2500, 3000] & K & Uniform \\
$\Delta T_\mathrm{CMB}$ & Initial temp. jump across CMB & [0, 900] & K & Uniform \\
$\nu_0$ & Reference kinematic mantle viscosity & [10$^7$, 5$\times$10$^9$] & m$^2$/s & Log-Uniform \\
$\nu_\mathrm{jump}$ & Viscosity jump from upper to lower mantle & [2, 10] & nd & Uniform \\
$\Delta T_{\chi,\mathrm{ref}}$ & Core reference liquidus depression & [0, 700] & K & Uniform \\
$^{40}\mathrm{K}_\mathrm{man}$ & Initial amount of $^{40}\mathrm{K}$ in the mantle & [0.5, 1.5] & Relative to Earth & Uniform \\
$\eta_K$ & Initial fraction of $^{40}\mathrm{K}$ in the core & [0.01, 1] & nd & Uniform \\
$\mathrm{k}_\mathrm{OC}$ & Thermal conductivity of outer core & [40, 200] & W/m/K & Uniform \\
$\epsilon_\mathrm{erupt}$ & Fraction of extrusive volcanism & [10$^{-2}$, 2$\times10^{-1}$] & nd & Log-Uniform \\
$\Delta E_\mathrm{H_2O}$ & Depression of $E_A$ due to water & [$8 \times 10^5$, $6 \times 10^6$] & K/wt. frac. & Uniform \\
$M^\mathrm{H_2O}_p$ & Initial amount of water on planet & [1, 10] & Earth oceans & Uniform \\
$\eta_\mathrm{H_2O}$ & Initial fraction of water on surface of planet & [0.1, 0.5] & nd & Uniform \\
$M^\mathrm{CO_2}_p$ & Initial amount of carbon dioxide in mantle & [91.9, 121.9] & bars & Uniform \\
\enddata
\end{deluxetable*}

\subsection{Generating Simulations that Match Modern-day Venus}
To create our suite of input files with these parameter choices, we used the publicly available sampling software, \vsp\footnote{\url{https://github.com/VirtualPlanetaryLaboratory/vspace}}.
With \vplanet, we created 234,000 different input files, each representing a set of parameter choices.
We than ran our simulations with \vplanet in parallel on 40 cores across 6 nodes on a supercomputer.
We ran our simulations with a dynamic timestepping value of 0.01 from 0 to 4.5 Gyrs with outputs every 1 Gyr.
Running this many simulations across 6 nodes with 40 cores each took a total of about 2 wallclock days, or about 30 seconds per simulation.

\section{Results} \label{sec:results}

From our set of 234,000 we identified 808 simulations (0.35\% of all simulations) that matched key observables of modern-day Venus to within five standard deviations (see Table \ref{constraint_table}).
For each successful trial, we then looked at the time evolution of values fundamental to the thermal evolution of Venus, the mantle temperature, core temperature, magnetic moment, radius of the inner core, amount of water in the atmosphere, and upper mantle melt fraction, with the goal of identifying qualitative differences in their evolutions.
Our results that match Venus' present-day state help constrain the initial conditions of Venus.
The summary statistics for our varied parameters and initial conditions for our simulations that reproduced Venus are shown in Table \ref{tab:initialvaluestats}.

\begin{deluxetable*}{lllll}
\tablecaption{Summary statistics for varied parameters and initial conditions \label{tab:initialvaluestats}} 
\tablehead{\colhead{Parameter Name} & \colhead{Min.} & \colhead{Max.} & \colhead{Mean} & \colhead{Stdev.}}
\startdata
Init. avg. mantle temp. [K] & 2500 & 3000 & 2724 & 145 \\
Init. temp. jump across CMB [K] & 10 & 892 & 392 & 249 \\
Reference kinematic mantle viscosity [log$_{10}$(m$^2$/s)] & 7.0 & 9.70 & 8.44 & 0.77 \\
Visc. jump from upper to lower mantle & 1.00 & 99.8 & 46 & 28.7 \\
Core reference liquidus depression [K] & 5.2 & 698 & 231 & 199 \\
Lindemann reference temperature [K] & 5602 & 6999 & 6429 & 390 \\
Initial amount of $^{40}$K in mantle [Rel. to Earth] & 0.50 & 1.50 & 1.00 & 0.287 \\
Initial fraction of $^{40}$K in core & 0.01 & 1.0 & 0.36 & 0.26 \\
Thermal conductivity of outer core [W/m/K] & 41 & 203 & 145 & 41 \\
Fraction of extrusive volcanism & 0.01 & 0.20 & 0.064 & 0.050 \\
Depression of $E_A$ due to water [$\times 10^6$ K/wt. frac.] & 0.80 & 6.00 & 2.99 & 1.46 \\
Initial amount of water on planet [TO] & 1.38 & 10.0 & 5.95 & 1.96 \\
Initial fraction of water on surface of planet & 0.10 & 0.50 & 0.31 & 0.11 \\
Initial amount of CO$_2$ in mantle [bars] & 90.8 & 129 & 110 & 9.7 \\
\enddata
\end{deluxetable*}

Our results provide evidence for a wide variety of potential evolutionary histories that explain Venus.
The summary statistics for some key mantle and core values across all our evolutions are shown in Table \ref{tab:finalvaluestats}.
We can see that there is quite a variety for the final thermal state of the planet, with a wide range of mantle and core temperatures as well as upper mantle melt fractions.
The amount of water left in the planet's mantle is also greatly variable, with some simulations losing quite a bit of water and others containing a substantial water reservoir in the mantle.
While the water outgassing rate is constrained by the escape rate and has a narrow acceptable range, the carbon dioxide outgassing rate is more variable and highlights the variety of acceptable eruption rates at 4.5 Gyr.
Overall the high standard deviation and minimum-maximum spread across these parameters provides evidence for a multimodal distribution of results that represents our qualitatively different evolutionary scenarios.

\begin{deluxetable*}{lllll}
\tablecaption{Statistics for key internal variables that reproduce modern Venus  \label{tab:finalvaluestats}} 
\tablehead{\colhead{Final Value} & \colhead{Min.} & \colhead{Max.} & \colhead{Mean} & \colhead{Stdev.}}
\startdata
Avg. Mantle Temp. [K] & 2259 & 2708 & 2456 & 96 \\
Avg. Core Temp. [K] & 4137 & 5986 & 4738 & 278 \\
Crust Depth [km] & 21 & 140 & 77 & 26 \\
Stagnant Lid Depth [km] & 21 & 140 & 78 & 26 \\
H$_2$O Outgassing Rate [$\times 10^{11}$ mol/yr] & 4.5 & 14.8 & 8.65 & 2.7 \\
CO$_2$ Outgassing Rate [$\times 10^{11}$ mol/yr] & 0.0016 & 5.94 & 2.08 & 1.31 \\
Upper Mantle Melt Fraction & 0.00025 & 0.49 & 0.16 & 0.11 \\
Water in Mantle [TO] & 0.40 & 5.93 & 2.49 & 1.08 \\
Eruption Rate [km$^3$/yr] & 0.00060 & 6.89 & 1.67 & 1.32 \\
Upper Mantle Viscosity [log$_{10}$(Pa$\cdot$s)] & 19.2 & 21.1 & 20.1 & 0.37 \\
\enddata
\end{deluxetable*}

\subsection{Venus' distinct evolutionary scenarios}

\begin{figure*}[htp!]
    \centering
    \includegraphics[width=\linewidth]{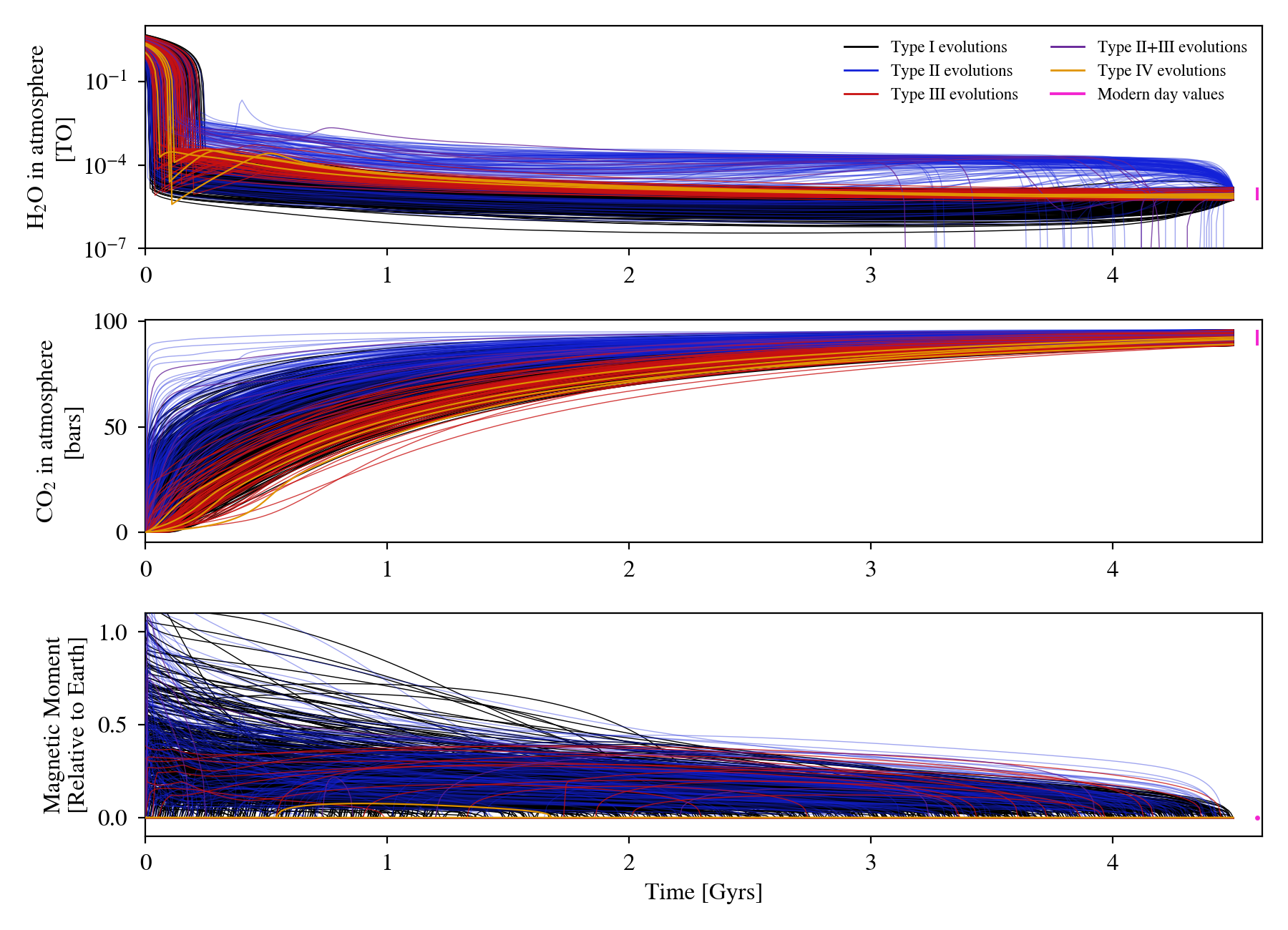}
    \caption{The evolution of all of our simulations that matched modern-day Venus constraints (shown offset from 4.5 Gyrs in pink). Our simulations are color-coded by the different types of evolutionary scenarios described in the text. In the top panel we show that each evolution converges to the 2.868 $\pm$ 0.2868 mbars of H$_2$O in Venus' atmosphere \citep{arney_2014}. While Type I and Type III evolutions reach that value similarly, Type II evolutions tend to either outgas more water over time, leading to higher amounts in the atmosphere, or cease melting for a time, leading to epochs with no water in the atmosphere. Note that Type II+III evolutions exhibit outgassing rates more in line with Type II than Type III evolutions. The oscillations characteristic of the Type IV evolutions can best be seen in this panel. Note that in the top panel, every case in which the atmospheric water drops below $10^{-7}$ TO, later epochs of outgassing replenish atmospheric water to the observed value. The middle panel shows the evolutions converging to the 92.254 $\pm$ 0.765 bar of CO$_2$ in Venus' atmosphere \citep{2017aeil.book.....C}. Again we can see that Type II evolutions have higher outgassing rates and thus outgas CO$_2$ more quickly than the other evolutions. In the bottom panel, we show how each evolution leads to a magnetic moment of $<10^{-5}\mathcal{M}_\oplus$~\citep{phillips_russell_1987}. Note that many cases have transient magnetic fields.}
    \label{fig:modern_values}
\end{figure*}

\begin{figure*}[htp!]
    \centering
    \includegraphics[width=\linewidth]{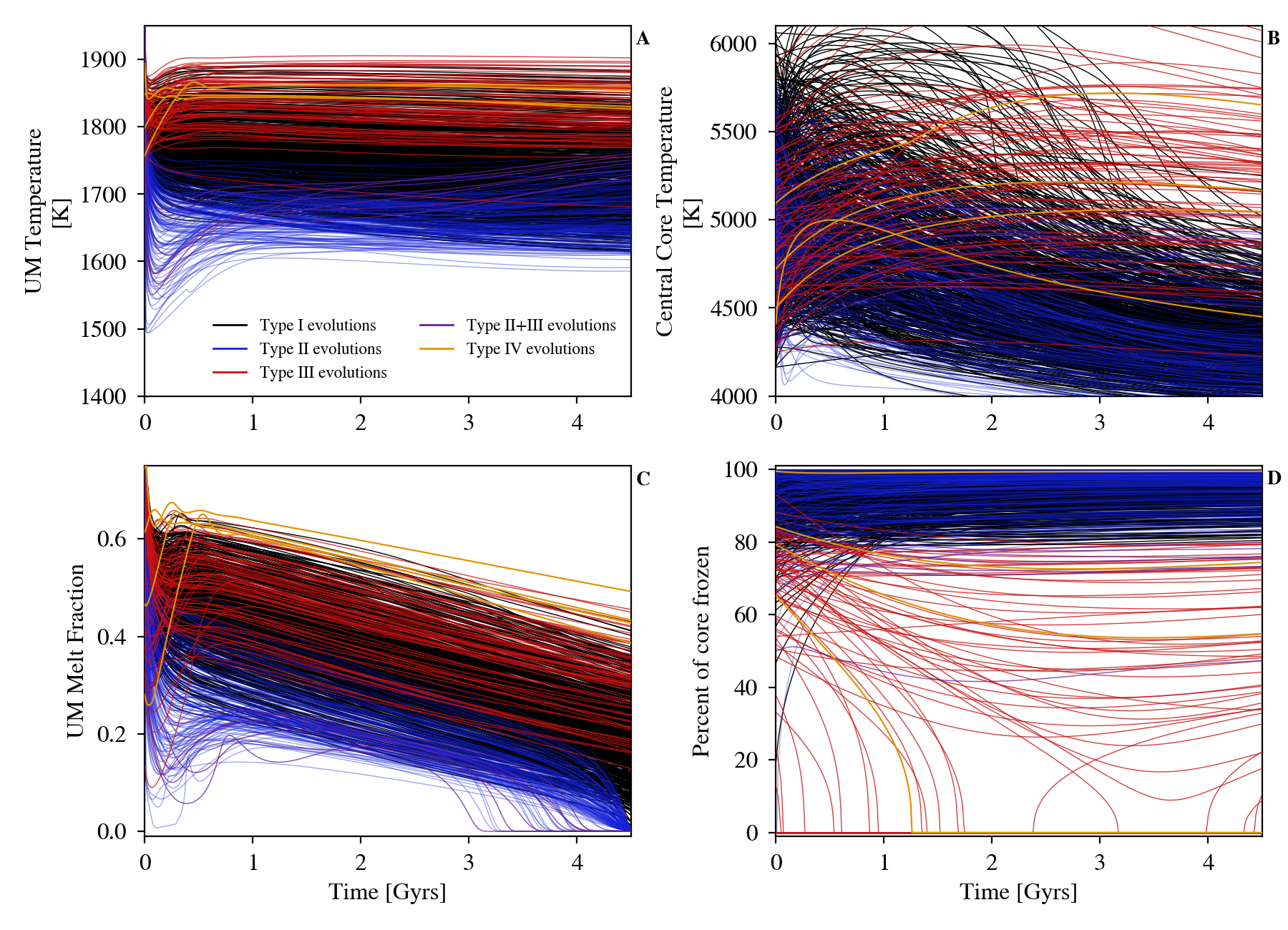}
    \caption{The evolution of a few key mantle and core variables for all our simulations, color-coded by the different evolutionary types. The upper left and upper right panels show the evolution of the upper mantle (UM) temperature and core temperature, respectively. Notice that in Type II evolutions the mantle temperature tends to increase more rapidly than the other types, while Type III evolutions tend to maintain a higher mantle and core temperature than the other types. The bottom left panel shows the upper mantle melt fraction over time, which highlights the characteristic behavior of the Type II evolutions: ending with a low melt fraction. The bottom right panel shows the fraction of the core that is solid over time. Note that Type III and Type IV evolutions both end with an inner core radius fraction less than (and for most simulations, much less than) 0.8. No simulation results in a completely solid core. The Type II+III behavior can be seen across the panels, where they behave more like Type II evolutions for mantle-relevant outputs (like the mantle temperature), but behave more like Type III evolutions for core-relevant outputs (as can be seen in the evolution of the inner core over time).}
    \label{fig:evolve_plots}
\end{figure*}

\begin{figure*}
    \includegraphics[width = \linewidth]{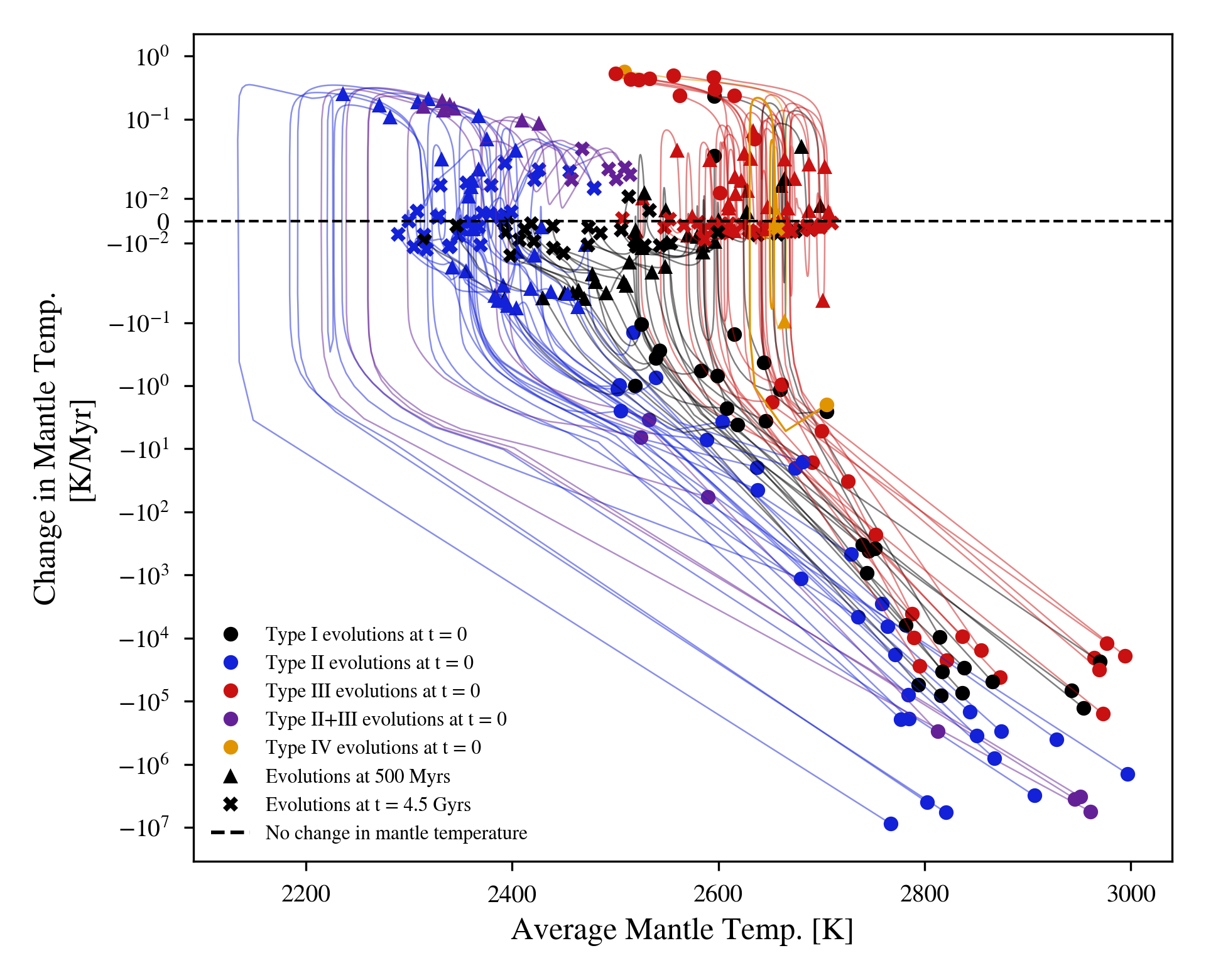}
    \caption{The evolution of a random subset of 30 simulations for each evolutionary type in the mantle phase space. Note that the $y$-axis is plotted as a symmetric logarithm: the values are plotted on a logarithmic scale except in the range $-10^{-2}$-$10^{-2}$, which is plotted on a linear scale, allowing for the plotting of positive and negative values on a logarithmic scale. The filled circles represent the initial state of each simulation, the triangles represent the value after 500 Myr, and the filled x represents the final state of each simulation and color encodes the evolutionary type. Most of the plotted simulations cool quickly and then stabilize, except for Type II evolutions, which heat up due to dehydration stiffening, which slows convection. Note that while some simulations start by heating up, no Type II evolution does.}
    \label{fig:mantlephaseplots}
\end{figure*}

The results of the simulations described in the previous section can be organized into types based on the mantle melt fraction evolution and core solidification history.
Our defined scenarios are based on the behavior after the first 50 Myrs, because before then the evolutions tend to exhibit transient behavior that is sensitive to initial thermal conditions.
The four distinct evolutionary scenarios are described below, with a short name given to each type that represents a key feature of that scenario:  
\begin{enumerate}
    \item Type I (Conventional): The mantle and core temperature, as well as the amount of water in the atmosphere and the melt fraction evolve relatively smoothly, with little noteworthy changes in behavior over time. These cases represent about 72\% of all simulations that reproduce Venus. Because these are similar to many parameterized mantle convection models, we call these conventional evolutions.
    \item Type II (Low Melt): The melt fraction becomes very small (below 0.01) during the evolution. Some of these scenarios include a brief shutdown of all upper mantle melting. These solutions show a brief, non-monotonic increase or decrease in the amount of water in the atmosphere. These solutions also lose a large amount of water from the mantle over time due to a breakdown in crustal recycling. These cases represent about 18\% of all simulations that reproduce Venus.
    \item Type III (Smaller Inner Core): The inner core radius never reaches 80\% of the total core radius. The mantle temperature, melt fraction, and amount of water in the atmosphere tend to evolve relatively smoothly. These cases represent about 10\% of all simulations that reproduce Venus.
    \item Type IV (Oscillatory): Four evolutions that appear to have damped oscillatory behavior in the evolution of many parameters such as the mantle temperature for the first 500 Myrs. While some of these evolutions also end with a smaller inner core, their oscillatory behavior is unique and so they are defined as their own evolutionary type. While 4 examples is a small sample, we did discover many thermal histories that were incompatible with modern Venus showed similar oscillations, so this type of behavior might be more prevalent with different assumptions and/or be relevant for terrestrial exoplanets.
\end{enumerate}

We also found that there was enough significant overlap between some Type II evolutions and some Type III evolutions to call out these simulations as Type II+III evolutions.
These do not represent their own unique type of evolution, but do exhibit characteristics of both Type II and Type III evolutions.
The Type II+III evolutions represent about 0.9\% of all simulations.
The time evolutions of the observed properties from every simulation that reproduce the modern-day evolution of Venus, see Table \ref{constraint_table}, are shown in Fig.~\ref{fig:modern_values}, color-coded by type.

We next show the evolution of the mantle and core temperatures as well as the upper mantle melt fraction and percent of the core that's frozen in Fig. \ref{fig:evolve_plots}.
We used these parameters to separate evolutions into the four evolutionary scenarios described above.
Similarly, we also provide a mantle temperature phase curve for a selection of 30 simulations across all of our identified evolutionary types in Fig. \ref{fig:mantlephaseplots}. Note that most of the movement in this parameter space occurs within the first 500 Myr.

Fig. \ref{fig:mantlephaseplots} also provides a way to distinguish between our evolutionary scenarios, with the mantles of Type II evolutions consistently heating up after transient cooling due to increasing mantle viscosity caused by dehydration stiffening.
Most Type I and Type III evolutions end with the mantle cooling, though some appear to be heating towards the end of the evolution, potentially due to higher viscosities and lower eruption efficiencies resulting in less efficient cooling from convection and eruptions.

\begin{figure*}
    \includegraphics[width = \linewidth,trim={1.5cm 0 2cm 0},clip]{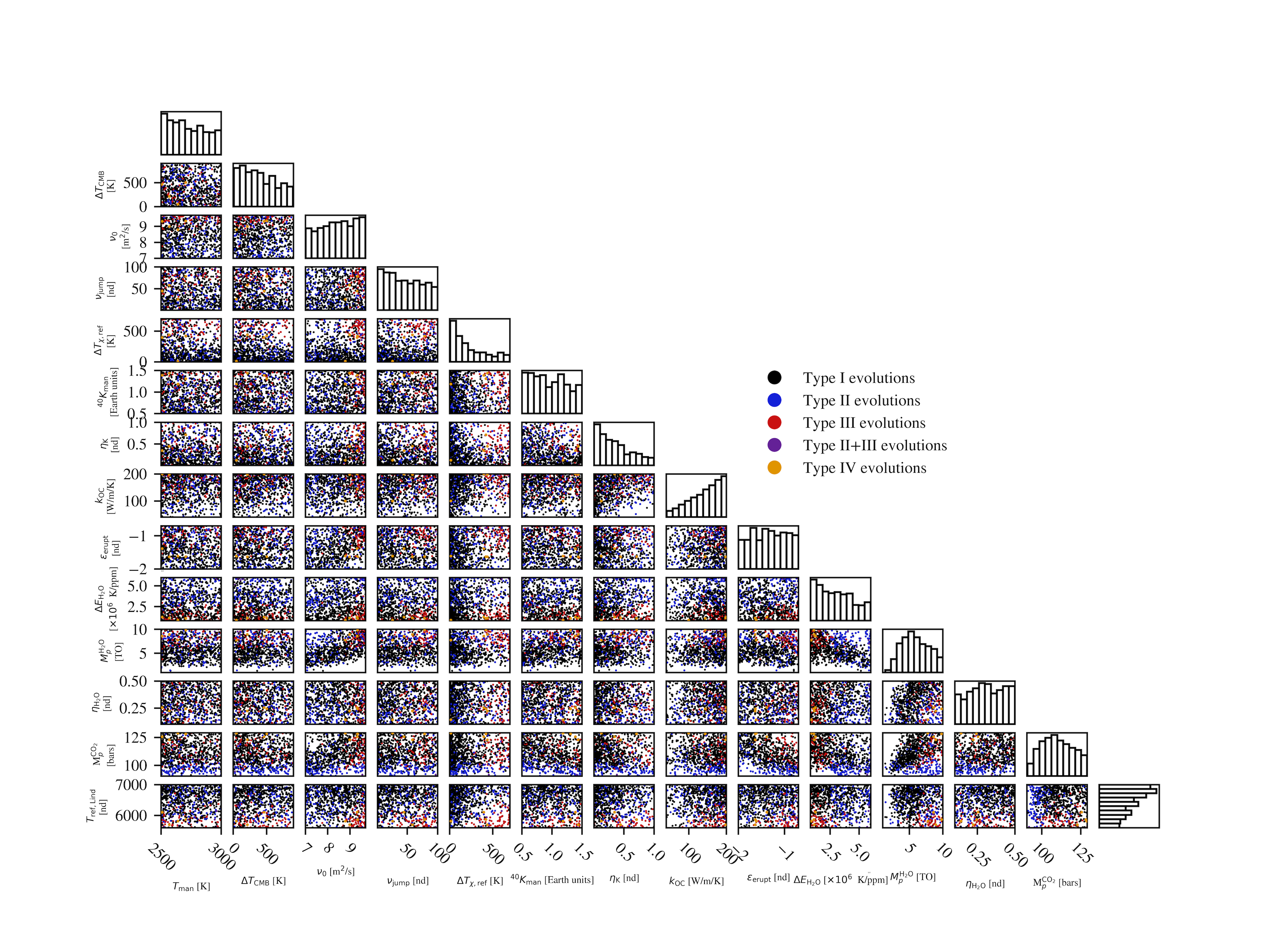}
    \caption{Initial parameter choices for all evolutions that reproduce Venus in a space defined by all of our varied parameters: the temperature jump across the core-mantle boundary ($\Delta T_\mathrm{CMB}$), the reference viscosity ($\nu_\mathrm{ref}$), the jump in viscosity from the upper to the lower mantle ($\nu_\mathrm{jump}$), the core liquidus depression ($\Delta T_{\chi,\mathrm{ref}}$), the amount of potassium-40 in the mantle ($^{40}$K$_\mathrm{man}$), the amount of potassium-40 in the core expressed as a fraction of how much is in the mantle ($\eta_\mathrm{K}$), the thermal conductivity of the outer core ($k_\mathrm{OC}$), the eruption efficiency ($\epsilon_\mathrm{erupt}$), the depression of the viscosity activation energy due to water ($\Delta E_\mathrm{H_2O}$), the amount of water the planet formed with ($M^\mathrm{H_2O}_p$), the fraction of that water that is initially in the planet's atmosphere ($\eta_\mathrm{H_2O}$), the initial amount of carbon dioxide in the planet's mantle ($M^\mathrm{CO_2}_p$), the reference core liquidus temperature ($T_\mathrm{ref,Lind}$), and the initial average temperature of the mantle ($T_m$). Each simulation is color-coded based on the identified evolutionary type described in the text. The histograms show the distribution of all simulations marginalized over all variables except for each variable below them.}
    \label{fig:options_triangle}
\end{figure*}

We show in Fig. \ref{fig:options_triangle} where each successful simulation lies in a subset of parameter space (see Table \ref{rangevary}).
Fig. \ref{fig:options_triangle} consists of two-dimensional slices of the full twelve-dimensional parameter space, with each point representing a single simulation's parameter values, color-coded by the type of evolutionary scenario.
The histograms show the distribution of all simulations marginalized over all variables except for each variable below them (or to the left in the case of $M^\mathrm{CO_2}_p$).
Because each parameter we chose was sampled over a uniform (or log-uniform) distribution, any deviance in the histograms away from uniformity indicates parameter values that are less likely to reproduce modern Venus.

We can see here that to match Venus requires a generally lower mantle temperature and a core that is not superheated compared to the mantle, as well as a relatively low core liquidus depression (likely to prevent the core from remaining liquid and powering a magnetic dynamo today).
Simulations with high amounts of potassium in the core ($>$150 ppm) are also disfavored.
We will defer an analysis of the distribution of evolutionary scenarios in Fig. \ref{fig:options_triangle} to the following subsections, where we analyze those distributions in lower-dimensional subspaces.

\subsection{Type I (Conventional) Evolutions}

\begin{figure*}[htp!]
    \includegraphics[width=\linewidth]{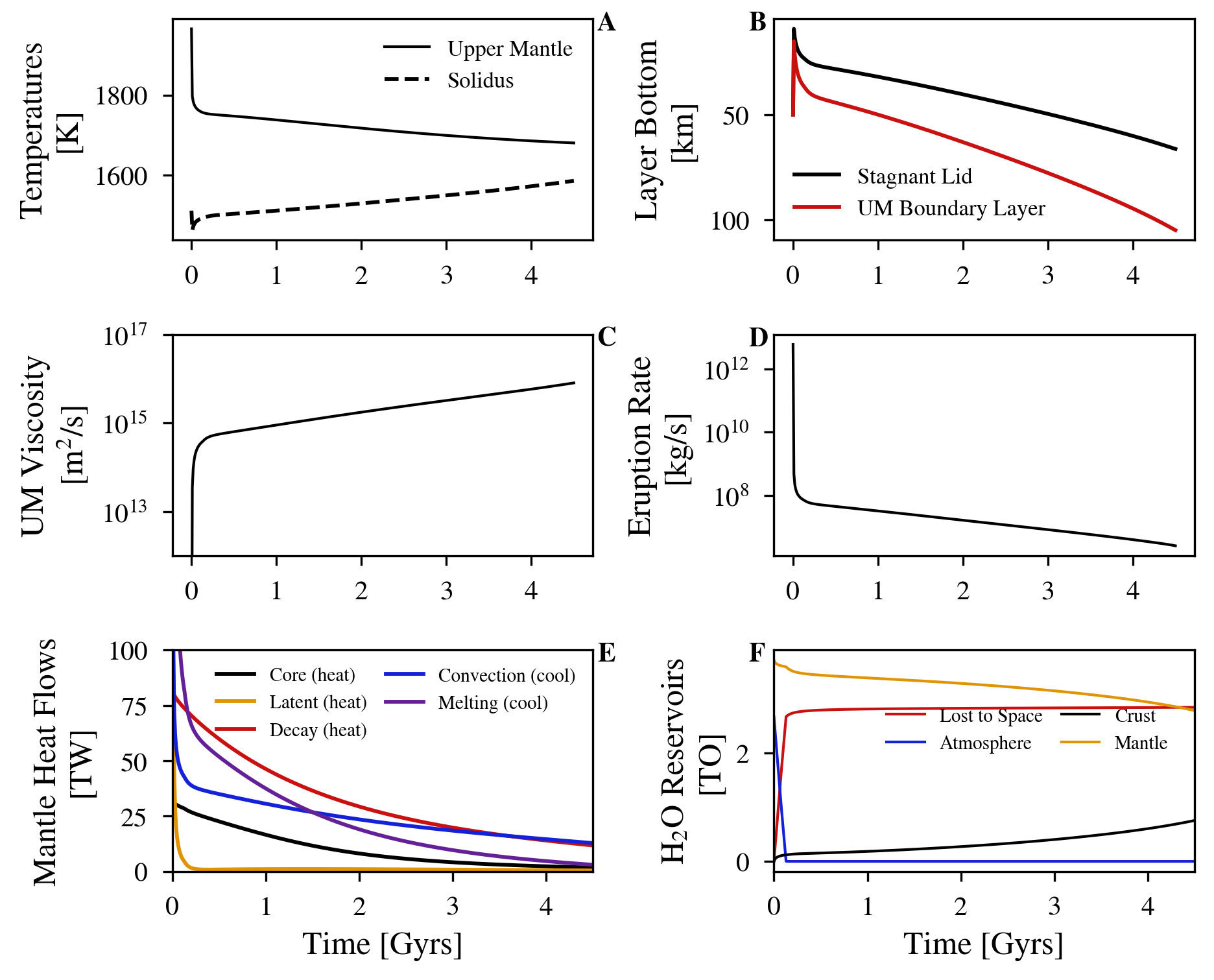}
    \caption{Time evolution of select key mantle variables for a sample evolution with Type I behavior. Panel \textbf{A} plots the upper mantle and solidus temperatures over time; panel \textbf{B} shows the depth to the bottom of the stagnant lid layer and the upper mantle thermal boundary layer; panel \textbf{C} shows the upper mantle viscosity; panel \textbf{D} shows the eruption rate; panel \textbf{E} shows the different heat sources and sinks (cooling mechanisms) for the mantle, and panel \textbf{F} shows the evolution of the water reservoirs in the simulation. Note that all of these variables evolve rather stably after short transient behavior at the beginning of the simulation. As the planet cools, the stagnant lid and thermal boundary layer slowly increase, which causes the solidus temperature and eruption rate to slowly decrease. Most of the water is lost to space during the early transient stage, with the rest sequestered in the crust and mantle. Parameter choices for this evolution can be found in Table \ref{TypeIEvolParameters}. \label{fig:TypeISampleEvol}}
\end{figure*}

The Type I evolution makes up a majority of our simulations that match Venus and are primarily defined by the differences from Type II, III, and IV evolutions rather than by their own characteristics.
These evolutions behave like typical parameterized convection models, for example those found in \citet{mcgovern_thermal_1989,foley_carbon_2018} and \citet{driscoll_thermal_2014}.
The behavior of key mantle properties such as the temperature, eruption rate, outgassing rates, and the various heating and cooling rates tend to smoothly decline after some transient behavior at the beginning of the simulations, as seen in Fig. \ref{fig:TypeISampleEvol}.
This behavior is consistent with previous parameterized cooling models where the temperature decreases slowly due to the temperature-viscosity feedback.
 
\begin{deluxetable}{ll}
    \tablewidth{\linewidth}
    \tablecaption{Parameters for Type I sample evolution shown in Fig. \ref{fig:TypeISampleEvol} \label{TypeIEvolParameters}}
    \tablehead{\colhead{Parameter} & \colhead{Values}}
    \startdata
    $T_m$ [K] & 2808 \\
    $\Delta T_\mathrm{CMB}$ [K] & 698 \\
    log$_{10}(\nu_0$ [m$^2$/s]) & 8.02 \\
    $\nu_\mathrm{jump}$ [nd] & 16.7 \\
    $\Delta T_{\chi,\mathrm{ref}}$ [K] & 349 \\
    $T_\mathrm{ref,Lind}$ [K] & 6328 \\
    $^{40}$K$_m$ [Earth] & 1.20 \\
    $\eta_K$ [nd] & 0.26 \\
    k$_\mathrm{OC}$ [W/m/K] & 192 \\
    $\epsilon_\mathrm{erupt}$ [nd] & 0.020 \\
    $\Delta E_\mathrm{H_2O}$ [K/wt. frac.] & 3.13$\times10^6$ \\
    $M^{H_2O}_p$ [TO] & 6.38 \\
    $\eta_\mathrm{H_2O}$ [nd] & 0.42 \\
    $M^\mathrm{CO_2}_p$ [bar] & 102.7 \\
    \enddata
\end{deluxetable}

The generally constant cooling evolution characteristic of most Type I mantles is also present in their cores.
In differentiating Type I evolutions from Type III evolutions, we can see from Fig. \ref{fig:evolve_plots} that most Type I evolutions tend to have a decreasing core temperature, whereas Type III evolutions tend to have an increasing core temperature. A small fraction of Type I cases, however, experienced initially rising core temperatures for up to 2 Gyr, followed by cooling. For these Type I cases, this initial heating, powered by radioactive decay and latent heat from solidification, does not significantly affect mantle water content nor the size of the the inner core, so we classify them as Type I. Note that we find that core heating can occur for most values of the initial $^{40}$K abundance in the core.

Many Type I evolutions also exhibit a magnetic field in the past.
The initial magnetic field strength of most Type I evolutions is comparable to Earth's (see Fig. \ref{fig:modern_values}), with many of these fields weakening, but persisting over billions of years of Venusian history.
These magnetic fields are strong and recent enough to potentially be detectable through remnant magnetization in the Venusian crust \citep{orourke_2019}.

\begin{figure}[]
    \includegraphics[width=\linewidth]{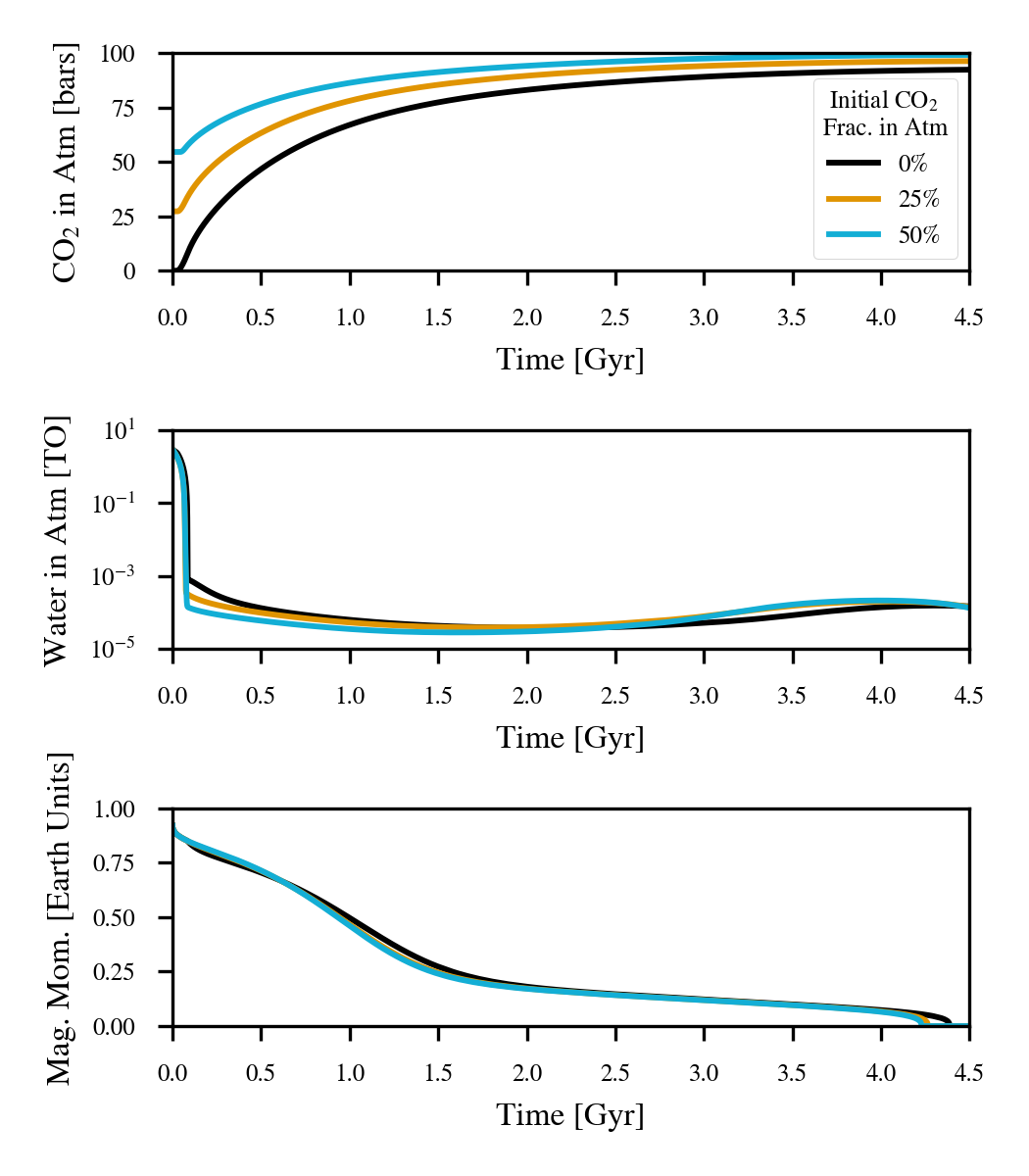}
    \caption{Comparison of Type I evolutions with different initial fractions of CO$_2$ in the atmosphere (black = 0\%; orange = 25\%; blue = 50\%).  The top panel shows that changing the initial atmospheric abundance by 50\% only results in a 7\% change in CO$_2$ partial pressure at 4.5 Gyr. The middle and bottom panels show that atmospheric water mass and magnetic moment are negigibly impacted by the initial CO$_2$ partitioning.\label{fig:CheckPartitioning}}
\end{figure}

Finally, we examine our decision to initially place all CO$_2$ in the mantle by relaxing that requirement for one Type I simulation and quantifying the differences. Holding everything else constant, we placed 25\% and 50\% of the CO$_2$ in the atmosphere initially and simulated to 4.5 Gyr. In Fig.~\ref{fig:CheckPartitioning}, we see that this assumption has a negligible impact on the final water abundance and magnetic field strength, but that there is a noticeable difference in atmospheric CO$_2$. However, even for the 50/50 partitioning, the final CO$_2$ pressure is only 99 bars, or about 7\% larger than the nominal case. We therefore conclude that initially sequestering all CO$_2$ in the mantle does not qualitatively affect our results.

\subsection{Type II (Low Melt) Evolutions}
Type II evolutions represent about 18.4\% of all of our successful simulations.
An example of the volatile and mantle evolution for a Type II evolution is shown in Fig. \ref{fig:LowMeltSampleEvol}.
Many characteristics of Type II evolutions can be seen in Fig. \ref{fig:LowMeltSampleEvol}, such as the loss of a substantial amount of water from the mantle as well as a dramatic increase in the upper mantle viscosity, upper mantle solidus temperature (calculated at the base of the upper mantle thermal boundary layer), stagnant lid depth, and upper mantle thermal boundary layer thickness that corresponds to a decrease in the melt fraction.

\subsubsection{Characteristics of Type II Behavior}

The substantial loss of water from the mantle over the course of Type II evolution drives many other behaviors in the mantle.
The loss of water in the mantle leads to significant dehydration stiffening, increasing the mantle's viscosity.
For water to recycle back into the mantle, there must both be a high enough eruption rate to bury the old crust deeper into the stagnant lid and a thin enough stagnant lid so that the crust can reach the bottom of the stagnant lid and delaminate back to the mantle.
The eruption rates decreases and the stagnant lid thickens with increasing viscosity, so water loss from the mantle both increases viscosity of the mantle and decreases the rate at which water recycles from the crust back into the mantle.
The shutdown of mantle recycling also results in Type II evolutions having a significantly lower amount of carbon dioxide retained in the mantle.
While much water is lost from the mantle from this process, note that no Type II evolution ends up with a completely devolatilized mantle.

\begin{figure*}[htp!]
    \includegraphics[width=\linewidth]{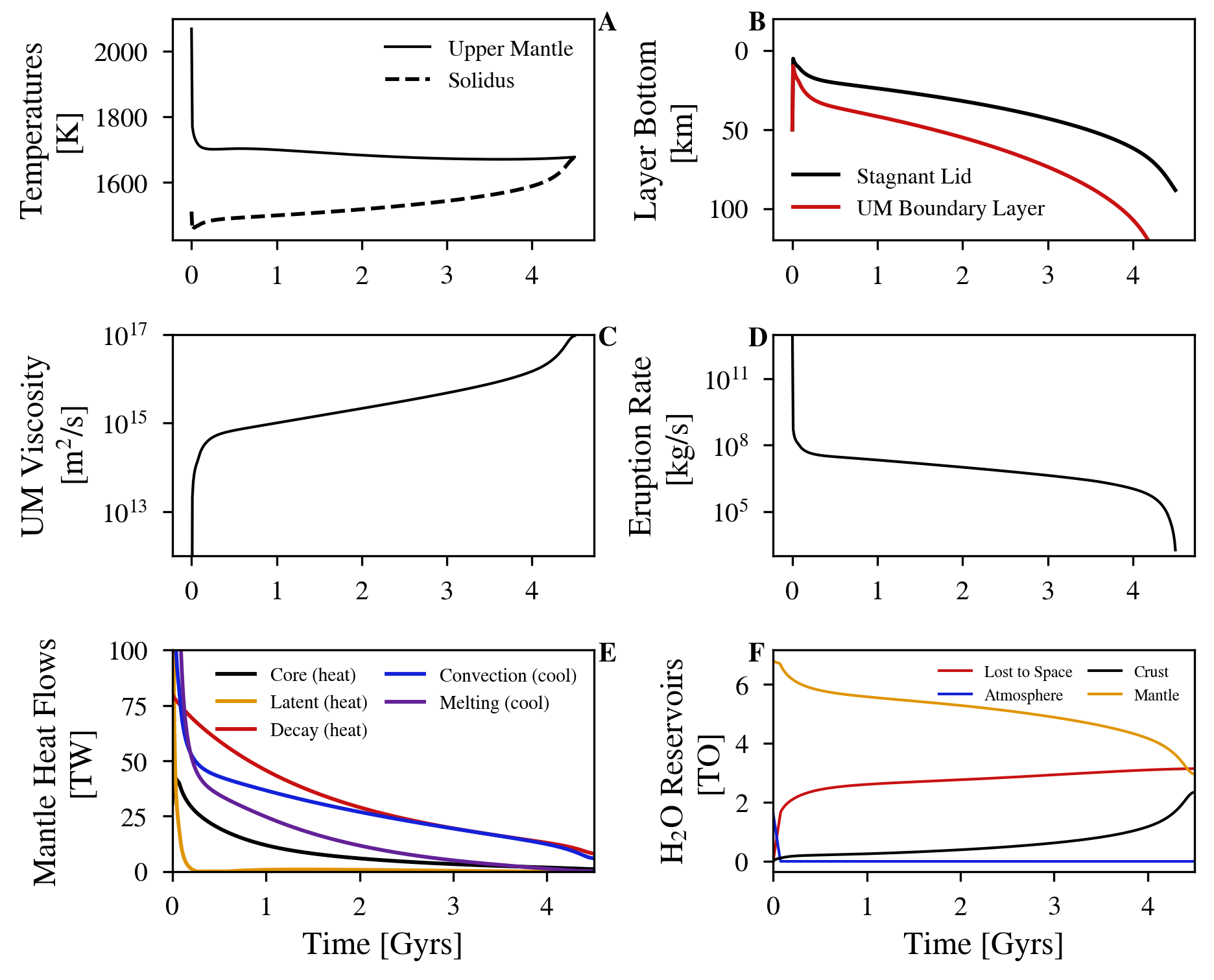}
    \caption{Time evolution of select key mantle variables for a sample evolution with Type II behavior. Panel \textbf{A} plots the upper mantle and solidus temperatures over time; panel \textbf{B} shows the depth to the bottom of the stagnant lid layer and the upper mantle thermal boundary layer; panel \textbf{C} shows the upper mantle viscosity; panel \textbf{D} shows the eruption rate; panel \textbf{E} shows the different heat sources and sinks (cooling mechanisms) for the mantle, and panel \textbf{F} shows the evolution of the water reservoirs in the simulation. Note that the Type II behavior can be seen as the upper mantle viscosity rises towards the end of the simulation, causing the melt zone (beneath the thermal boundary layer and stagnant lid) to be pushed deeper to higher pressures causing the solidus temperature to sharply rise. This causes the eruption rate and thus the steady state amount of water in the atmosphere to drop as well. Note that the mantle has two periods of drastic water loss, the first at the beginning to replenish water lost to space from the atmosphere, and the second driven by the increase in viscosity. Parameter choices for this evolution can be found in Table \ref{lowmeltsampleparameters}. \label{fig:LowMeltSampleEvol}}
\end{figure*}

Type II evolutions are also characterized by a low upper mantle melt fraction (less than 1\%) at the end of the evolution with a correspondingly low eruption rate.
These low melt fractions cannot be attributed to lower mantle temperatures because Type II evolutions end with a significantly lower melt fraction than non-Type II evolutions with the same mantle temperature.
Fig. \ref{fig:LowMeltSampleEvol} instead shows that the decrease in the melt fraction is due to a drastic increase in the solidus temperature at about 4.2 Gyrs, matching the sudden rise in the upper mantle viscosity and the stagnant lid and thermal boundary layer depths.
As dehydration stiffening in Type II evolutions increases the mantle's viscosity, the stagnant lid and upper mantle thermal boundary layer are pushed deeper into the mantle.
While the mantle temperature is higher at greater depths, the effect of greater pressure increases the solidus temperature in the melting zone (defined as the part of the mantle immediately beneath the stagnant lid and upper mantle thermal boundary layer), which results in the characteristic low melt fractions.

Fig. \ref{fig:LowMeltSampleEvol}F shows that, by the end of the evolution, the amount of water in the atmosphere is trending downward.
Such behavior is characteristic of the Type II evolution, in contrast with non-Type II Venus histories that have a relatively steady amount of water in their atmosphere at 4.5 Gyrs. 
Type II evolutions only match modern-day Venus values of water exactly at 4.5 Gyrs but lose all the water in their atmosphere shortly after 4.5 Gyrs. We discuss this point more below. 
The amount of water in the atmosphere decreases as Type II evolutions are in the process of losing all mantle melting soon after 4.5 Gyrs resulting in a lack of outgassing to replenish the water loss in the atmosphere.
Thus, Type II evolutions represent a Venus that is in the process of melt and outgassing cessation, i.e., a Type II Venus is magmatically dying and, if Venus is Type II, then we happen to be catching Venus at this transitional epoch.

\begin{deluxetable}{ll}
    \tablewidth{\linewidth}
    \tablecaption{Parameters for sample Type II evolution shown in Fig. \ref{fig:LowMeltSampleEvol} \label{lowmeltsampleparameters}}
    \tablehead{\colhead{Parameter} & \colhead{Values}}
    \startdata
    $T_m$ [K] & 2955 \\
    $\Delta T_\mathrm{CMB}$ [K] & 490 \\
    log$_{10}(\nu_0$ [m$^2$/s]) & 7.47 \\
    $\nu_\mathrm{jump}$ [nd] & 9.22 \\
    $\Delta T_{\chi,\mathrm{ref}}$ [K] & 139 \\
    $T_\mathrm{ref,Lind}$ [K] & 6185 \\
    $^{40}$K$_m$ [Earth] & 1.15 \\
    $\eta_K$ [nd] & 0.37 \\
    k$_\mathrm{OC}$ [W/m/K] & 68.8 \\
    $\epsilon_\mathrm{erupt}$ [nd] & 0.014 \\
    $\Delta E_\mathrm{H_2O}$ [K/wt. frac.] & 3.83$\times10^6$ \\
    $M^\mathrm{H_2O}_p$ [TO] & 8.4 \\
    $\eta_\mathrm{H_2O}$ [nd] & 0.19 \\
    $M^\mathrm{CO_2}_p$ [bar] & 107.9 \\
    \enddata
\end{deluxetable}

\subsubsection{Important Parameters and Mechanisms for Type II Behavior}

To better understand what initial conditions and dynamics lead to Type II evolutions, we trained a random forest classifier \citep{Breiman_2001} on our data to identify the parameters that most strongly lead to a Type II evolution.
A random forest classifier estimates how data points in a parameter space would be labeled based on a training set of already-labeled data points.
The random forest classifier accomplishes this goal non-linearly by creating an ensemble of decision trees that use the value of each initial condition for a simulation and the label of that simulation to predict a label for a putative initial condition.
From the random forest classifier, we can obtain the relative importance of each parameter in determining whether a certain simulation setup will result in a Type II evolution based on how important that parameter is in the ensemble of decision trees.

\begin{figure*}[htp!]
    \includegraphics[width=\linewidth]{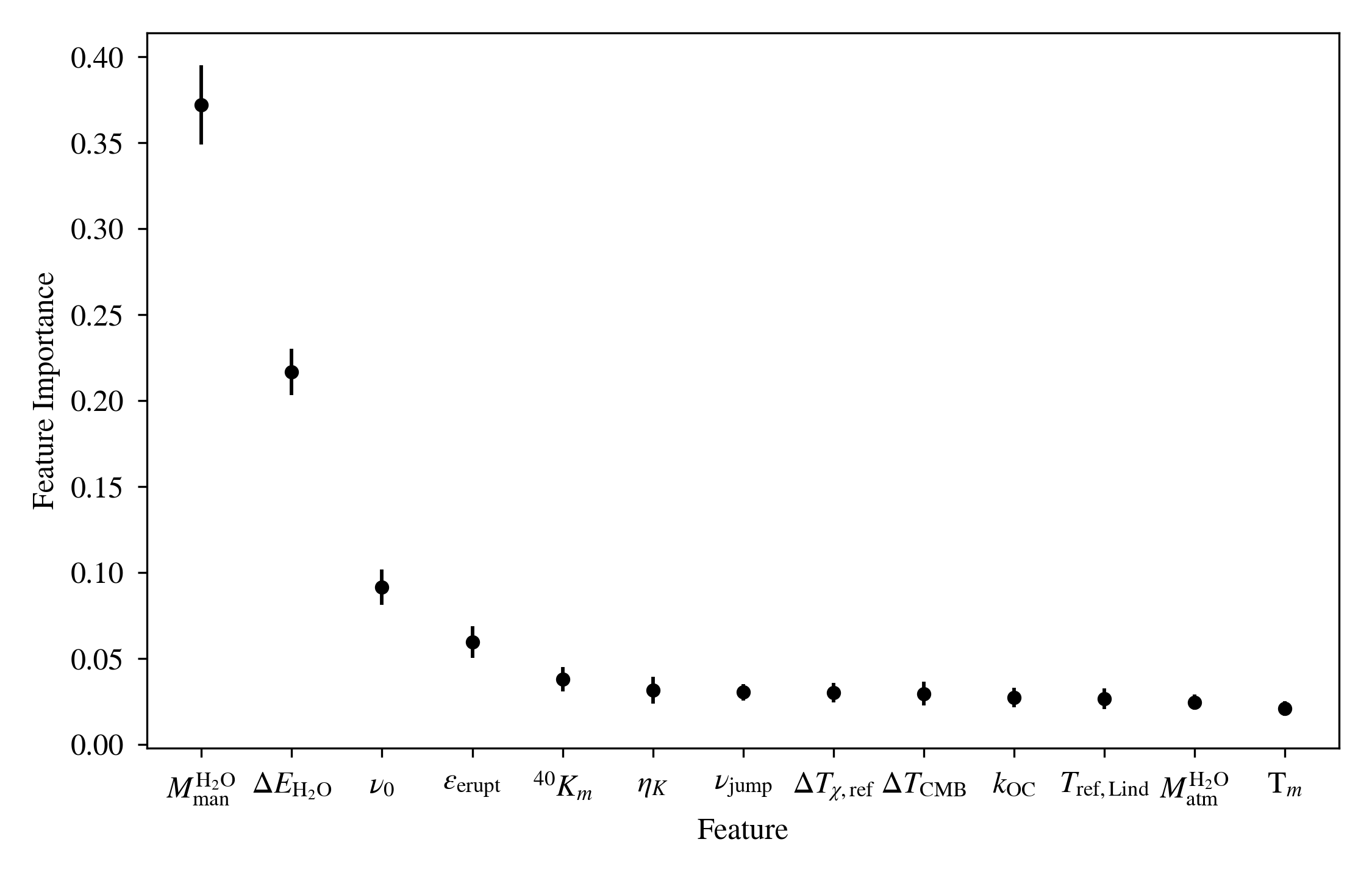}
    \caption{A feature's relative importance in determining whether a simulation will follow the Type II evolutionary scenario. The feature variables are described in Table \ref{rangevary}. The error bars represent the standard deviation for the feature importance as calculated from 50 different random forest models trained on the same data. Note that even within error, $M^{H_2O}_m$ (initial water in the mantle) is by far the most important variable, with $\Delta E_\mathrm{H_2O}$ (the coupling strength between mantle water content and viscosity) next. \label{fig:LowMeltFeatureImportance}}
\end{figure*}

\begin{figure*}
    \includegraphics[width=\linewidth]{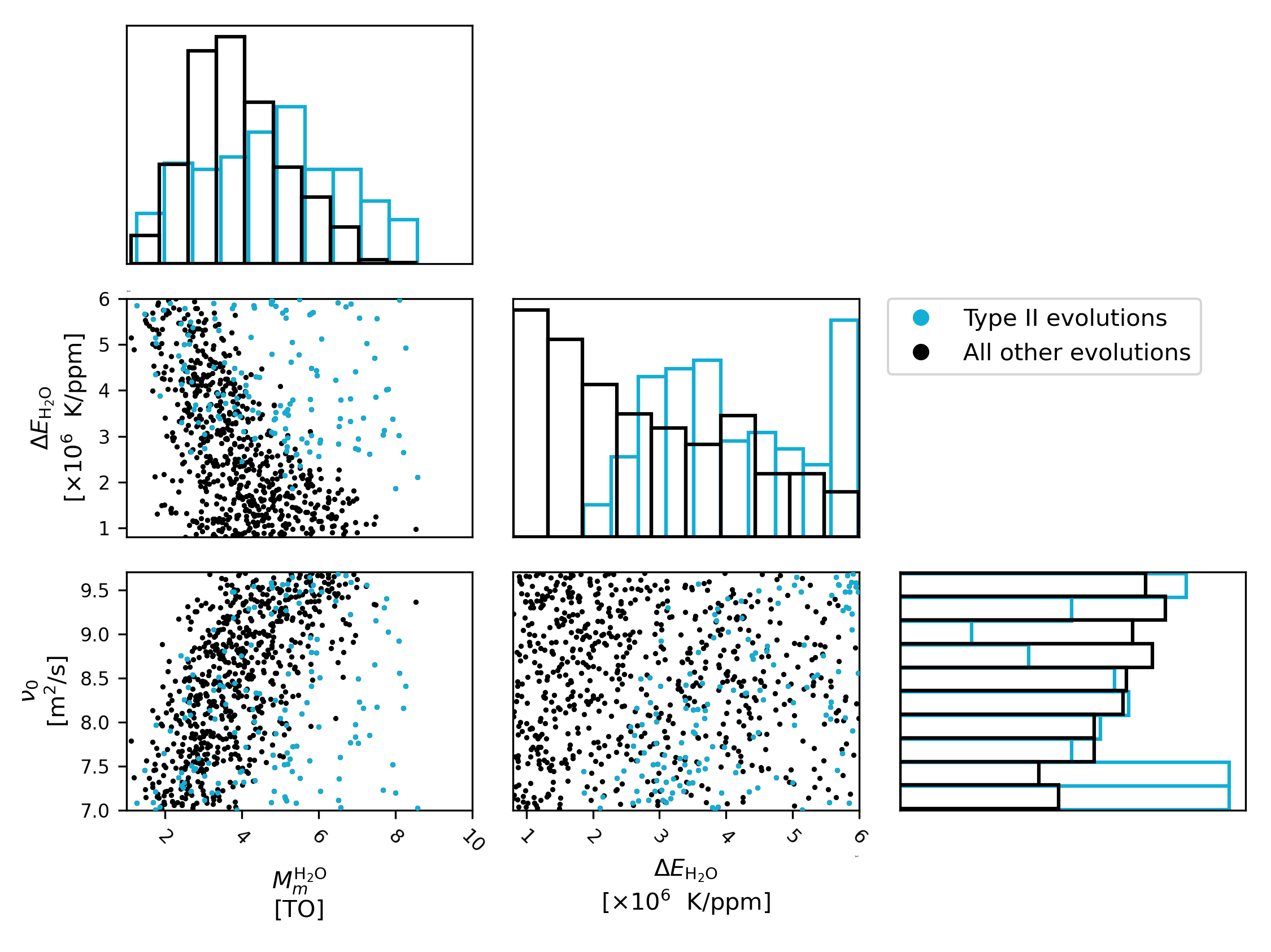}
    \caption{The scatter plots show where every simulation lies in a subset of the full parameter space defined by three variables: the initial amount of water in the mantle ($M^{H_2O}_m$), the strength of coupling between mantle water content and viscosity ($\Delta E_\mathrm{H_2O}$), and the reference viscosity ($\nu_0$). The histograms show the distribution of Type II evolutions (blue) and non-Type II evolutions (black) marginalized over all parameters except for each parameter below them (or to the left in the case of the eruption efficiency). Note that many of the simulations that exhibit Type II behavior (the blue points) carve out a region in the $\Delta E_\mathrm{H_2O}$-$M^{H_2O}_m$ space that is mostly distinct, despite some overlap, with a Type I band. While $M^\mathrm{H_2O}_m$ has a higher feature importance score, $\Delta E_\mathrm{H_2O}$ has a histogram distribution for Type II evolutions that is more distinguishable from that for Type I evolutions, highlighting the water-mantle viscosity coupling as a key driver of Type II behavior.\label{fig:LowMeltFeatureTriangle}}
\end{figure*}

Using the random forest classifier in the \textit{Scikit-learn} Python package \citep{scikit-learn} with the default value of 100 estimators and defining all of our varied parameters as features for the classifier, we can obtain the importance of each parameter in determining whether a simulation will yield a Type II evolution as seen in Fig. \ref{fig:LowMeltFeatureImportance}.
The values in Fig.~\ref{fig:LowMeltFeatureImportance} were obtained by training 50 different random forest classifiers and averaging the resulting feature importance values, with the error bars given by the standard deviation of the learned feature importance values.
We found that the mean feature importance value stayed relatively constant after averaging around 30 different random forest classifiers.
Physically, we can interpret the most important parameters in Fig.~\ref{fig:LowMeltFeatureImportance} as those that drive the interior dynamics into a Type II evolution.

Fig.~\ref{fig:LowMeltFeatureImportance} illustrates that only a few parameters have a high feature importance, which provides a good start for prioritizing which parameters to investigate when trying to identify the driving mechanisms for Type II evolutions.
The most important parameters are thus the initial amount of water in the mantle, the strength of the coupling of viscosity to water in the mantle, and the eruption efficiency. 
We can then look for trends in a three-dimensional subset of the whole parameter space defined by just these parameters in Fig. \ref{fig:LowMeltFeatureTriangle}.

The strongest distinction between Type II and all other evolutions seen in Fig. \ref{fig:LowMeltFeatureTriangle} is the strong positive correlation between the water-viscosity coupling strength and the initial mantle water content.
The simulations with a high initial mantle water content and a high water-viscosity coupling make up the majority of Type II evolutions.
Following the influence of both of these parameters on the viscosity (the $\Delta E_\mathrm{H_2O}X^\mathrm{H_2O}_\mathrm{man}$ term in Eq. [\ref{eq:viscosity}]), we can see that initially Type II evolutions will have a very low viscosity that then increases greatly. As Type II evolutions lose water, the influence of dehydration stiffening on their viscosity becomes stronger than that of a non-Type II evolution with a lower $\Delta E_\mathrm{H_2O}$ value.

The rapid increase in viscosity over time in Fig.~\ref{fig:LowMeltSampleEvol} occurs because of a feedback between viscosity and water content in the mantle.
As the viscosity increases, the vigor of convection decreases and the stagnant lid grows more quickly. Conversely, the crust grows more slowly as a lower convective velocity results in melt being transported to the surface more slowly.
Recycling of volatiles in the mantle only occurs when the crust grows larger than the stagnant lid, and so if the crust grows more slowly (though not necessarily slower than the stagnant lid) and the stagnant lid grows more quickly, the rate of recycling will net decrease with an increase in viscosity.
As recycling decreases, water begins to be lost from the mantle at a greater rate, which further increases the viscosity through dehydration stiffening, which in turn results in more water being lost.
This feedback is amplified for Type II evolutions and results in a substantial loss of water from their mantles, and ultimately from the planet as a whole when the water is lost to space.

\begin{figure*}[htp!]
    \includegraphics[width=\linewidth]{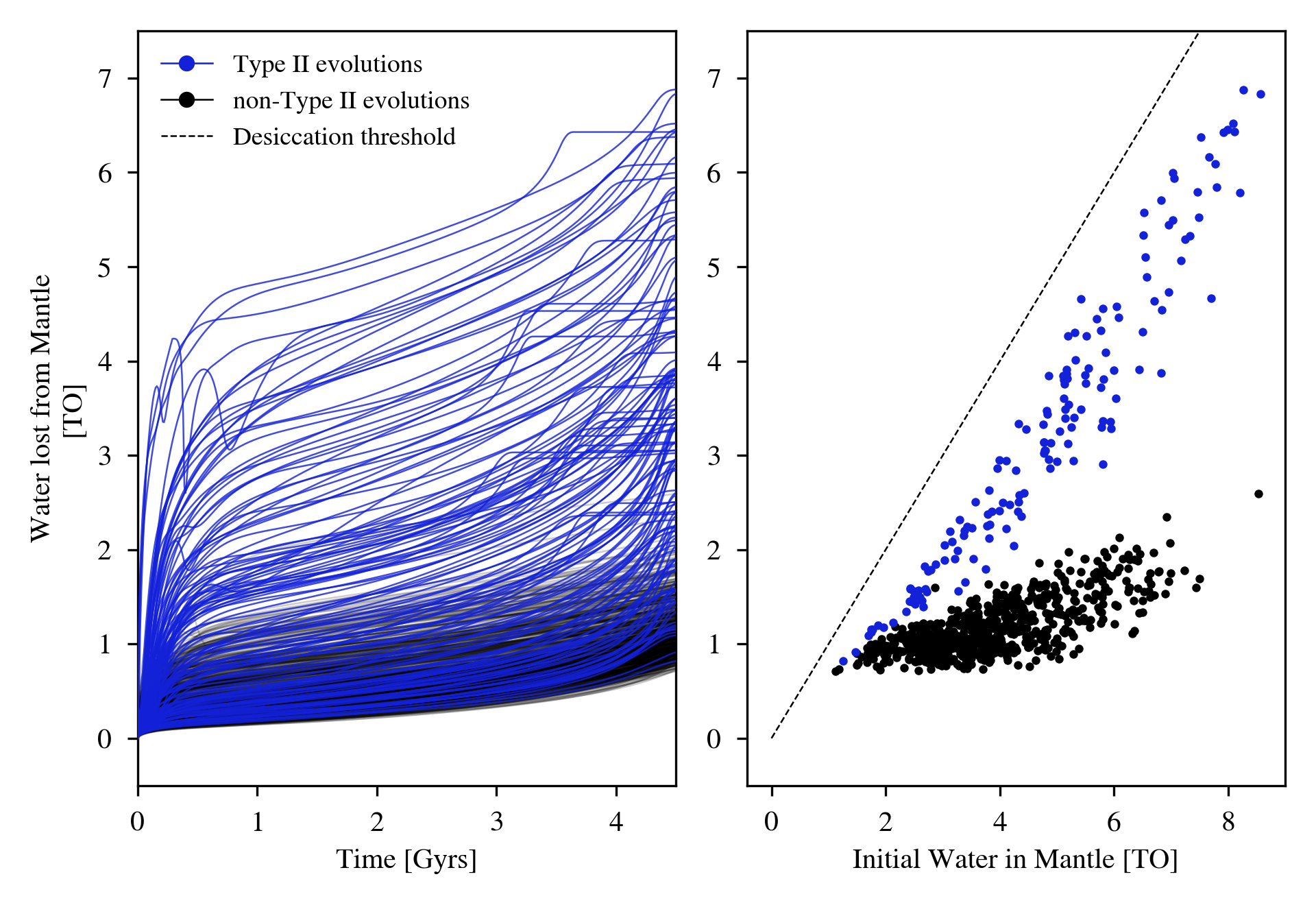}
    \caption{The evolution of the water lost from the mantle (left) and the net amount of water lost from the mantle (right). Occasional decreases in the amount of water lost from the mantle over time are due to large ingassing events. The simulations that exhibit Type II behavior are colored blue, while all other simulations are colored black. The simulations that exhibit Type II behavior end with more net water lost from their mantle for the same initial amount of water in their mantle than all other simulations. Note that no simulations lose their entire initial inventory of water from their mantles. \label{fig:LowMeltLowRecycling}}
\end{figure*}

The substantial loss of water in Type II evolutions is plainly seen in Fig. \ref{fig:LowMeltLowRecycling}, which shows the water loss from the mantle for Type II evolutions in blue and all other evolutions in black.
The right panel shows a clear separation between the Type II cases and all other cases: by the end of the simulation, the Type II evolutions have lost more water from their mantle than a non-Type II case with the same initial amount of water in the mantle. The clear separation between the two types in this parameter space is due to the Type IIs not trapping and/or recycling water in the crust as much as the other Types.

The left panel shows that there are two major water loss periods for low melt fraction evolutions.
The first is during the beginning of the evolution when the initial viscosities are very low due to the high $\Delta E_\mathrm{H_2O} X^\mathrm{H_2O}_\mathrm{man}$ value. 
The low initial viscosity allows for significantly more melt and water carried by that melt to reach the surface.
The second major water loss period is towards the end of the evolution when the viscosity drastically increases due to dehydration stiffening, which decreases the amount of water recycled back into the mantle.

While Type II evolutions lose a substantial amount of water from their mantle, they never completely desiccate.
Desiccation does not occur because of the dependence on the melt fraction in our melting model, Eq. \eqref{fracmelt}.
The amount of water leaving the mantle is given by Eq. \eqref{manfluxout}, where the melt production rate, $\dot{M}_\mathrm{melt}$ is inversely proportional to the upper mantle melt fraction, while the amount of water in the melt is proportional to the upper mantle melt fraction for high enough values of the melt fraction.
While the amount of water dissolved into the melt increases with decreasing melt fraction, the amount of melt erupted similarly decreases, so these two processes work against each other.
Thus, with all other variables and parameters equal, we can then write the dependence on the melt fraction for the rate at which water is lost from the mantle as:
\begin{equation} \label{waterlossratemelt}
    R^\mathrm{H_2O}_\mathrm{out,man} \propto 1 - (1 - F_\mathrm{UM})^{100},
\end{equation}
where the exponent of 100 comes from $1/D_{H_2O}$ in Eq. \eqref{fracmelt} and $D_{H_2O} = 0.01$.

\begin{figure*}[htp!]
    \centering
    \includegraphics[width=\linewidth]{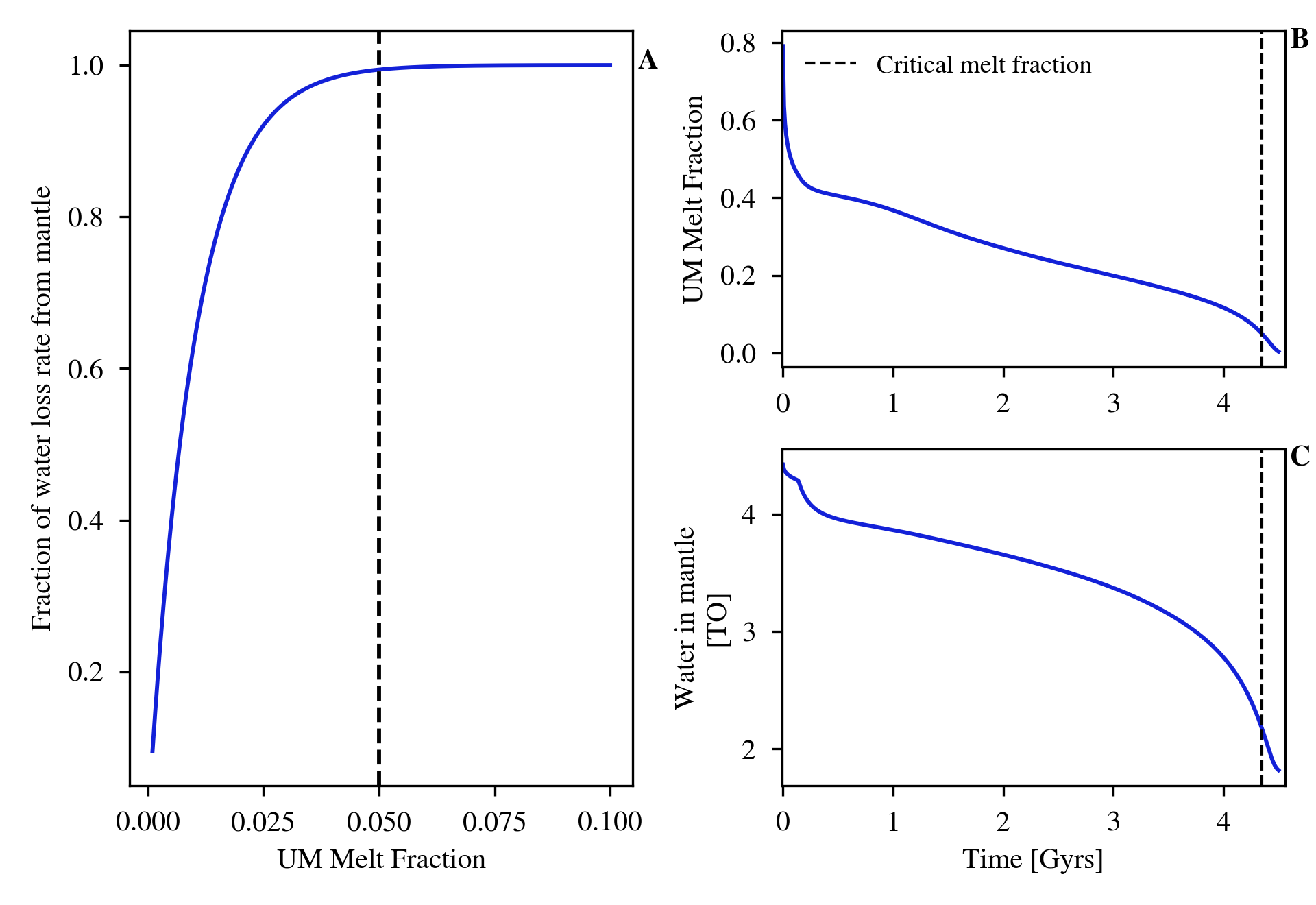}
    \caption{The fraction of the water loss rate when changing upper mantle melt fraction, \textit{ceteris paribus} (left panel) and the evolution of the upper mantle melt fraction and amount of water in the mantle for a Type II evolution (right column). The vertical black dashed line in the left panel shows the time when the melt fraction drops below the estimated critical value of 0.05. Below a melt fraction of 0.05 the water loss rate from the mantle begins to decrease with the melt fraction.\label{fig:MeltFractionCutoff}}
    \label{fig:my_label}
\end{figure*}

Eq.~\eqref{waterlossratemelt} is plotted in Fig.~\ref{fig:MeltFractionCutoff}, where it can be seen that the rate at which water is lost from the mantle does not depend on the melt fraction until the melt fraction declines to a critical value of about 0.05.
This critical value increases with the partitioning coefficient such that the critical value is about five times the value of the partitioning coefficient.
As the melt fraction drops below that critical value, so does the rate at which water is lost from the mantle.
We can see from the right panels of Fig. \ref{fig:MeltFractionCutoff} that when the melt fraction drops below 0.05, the evolution of water in the mantle for that simulation reaches an inflection point and slowly stops declining.
The evolution shown in Fig. \ref{fig:MeltFractionCutoff} is representative of all Type II evolutions: While the planet loses a large amount of water due to a weak volatile recycling process, it still does not fully desiccate because at very low melt fractions less water enters the melt from the mantle.

\begin{figure*}[htp!]
    \includegraphics[width=\linewidth]{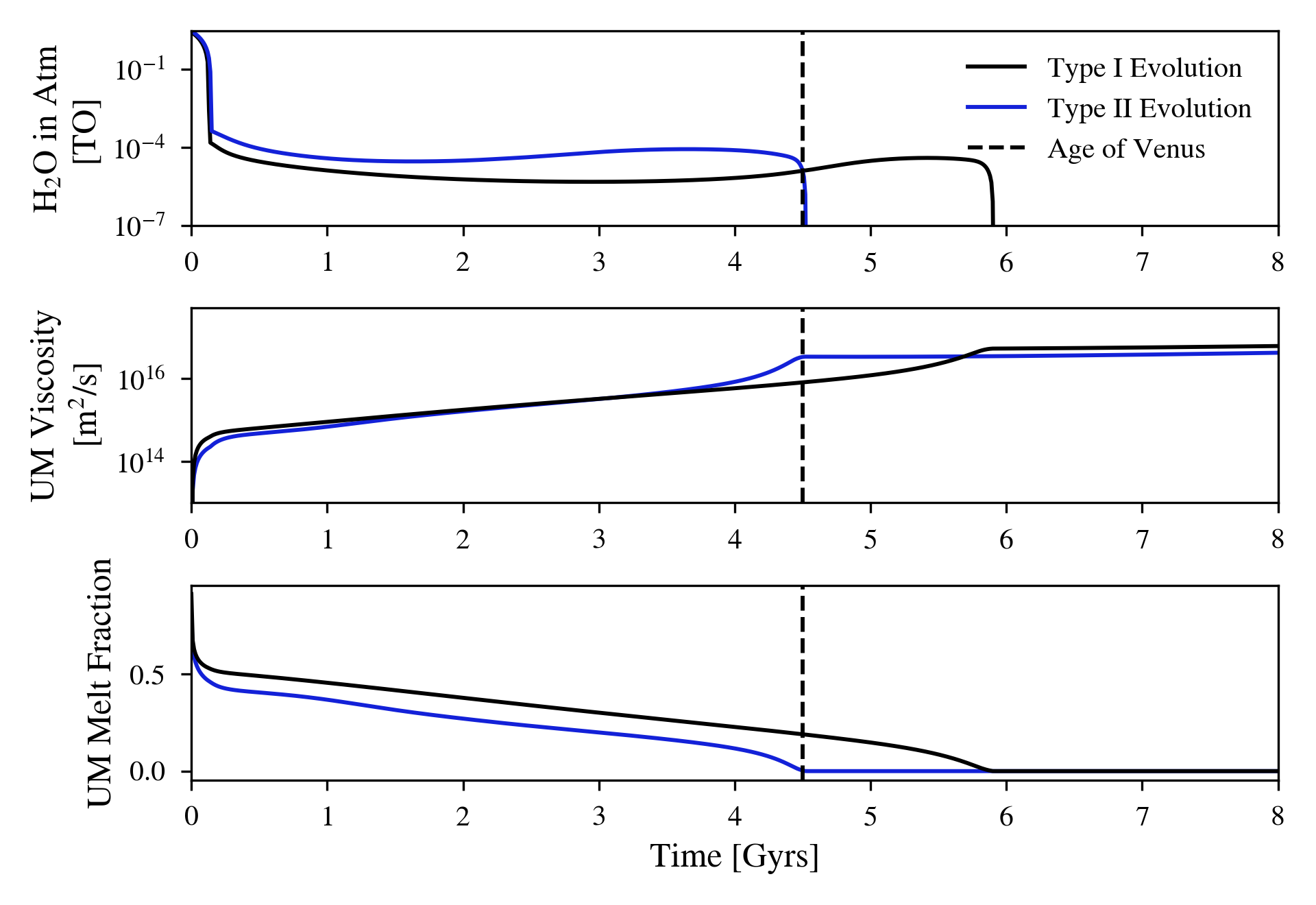}
    \caption{Evolution of key mantle parameters for a Type I evolution (black) and a Type II evolution (blue). Venus' age is demarcated with a dashed vertical black line. Note that after about 5.5 Gyrs the Type I evolution begins to exhibit Type II-like behavior (a ramp-up in viscosity, a ramp-down to 0 in the upper melt fraction, and a subsequent total loss of water in the atmosphere). Parameter choices for these evolutions can be found in Table \ref{tab:TimeComparison}. \label{fig:TimeComparison}}
\end{figure*}

\begin{deluxetable*}{lll}
    \tablewidth{\linewidth}
    \tablecaption{Parameters for sample Type I and Type II evolutions shown in Fig. \ref{fig:TimeComparison} \label{tab:TimeComparison}}
    \tablehead{\colhead{Parameter} & \colhead{Type I Evolution} & \colhead{Type II Evolution}}
    \startdata
    $T_m$ [K] & 2809 & 2723 \\
    $\Delta T_\mathrm{CMB}$ [K] & 698 & 839\\
    log$_{10}(\nu_0$ [m$^2$/s]) & 8.02 & 7.89\\
    $\nu_\mathrm{jump}$ [nd] & 17 & 22\\
    $\Delta T_{\chi,\mathrm{ref}}$ [K] & 349 & 57\\
    $T_\mathrm{ref,Lind}$ [K] & 6328 & 5864\\
    $^{40}$K$_m$ [Earth] & 1.20 & 0.66\\
    $\eta_K$ [nd] & 0.26 & 0.93\\
    k$_\mathrm{OC}$ [W/m/K] & 192 & 94\\
    $\epsilon_\mathrm{erupt}$ [nd] & 0.02 & 0.017\\
    $\Delta E_\mathrm{H_2O}$ [K/wt. frac.] & 3.13$\times10^6$ & 3.51$\times10^6$\\
    $M^\mathrm{H_2O}_p$ [TO] & 6.4 & 7.4\\
    $\eta_\mathrm{H_2O}$ [nd] & 0.42 & 0.40\\
    $M^\mathrm{CO_2}_p$ [bar] & 103 & 109\\
    \enddata
\end{deluxetable*}

While in some of our simulations, melting may restart for a brief million year-long period, in general we found that the process of melt cessation is irreversible.
While the loss of melting heats the mantle due to the loss of melt cooling, dehydration stiffening increases the viscosity more and leads to a deepening stagnant lid and thermal boundary layer that inhibits melting by increasing the solidus temperature.
While we do not investigate the effects of water on the solidus temperature in this paper, such a mechanism would even further inhibit melting as the drastic loss of water from the mantle that is characteristic of Type II evolutions would lead to higher solidus temperatures with no water to depress them \citep{Katz2003}.
So even if the mantle temperature increases after melt cessation, the loss of water in a Type II evolution ensures that the mantle will not return to a state conducive to melt generation.
If Venus has followed a Type II evolution, then it may be in the process of melt cessation with low levels of volcanism that will halt after a few more million years.

The dehydration stiffening and subsequent melt cessation in Type II evolutions may represent a general trend for planets that are magmatically dying.
In Fig. \ref{fig:TimeComparison} we show a Type I evolution and a Type II evolution evolving up to 8 Gyrs.
While these two evolutions appear to have little in common, at around 5 Gyrs the Type I evolution begins to exhibit characteristic Type II behaviors: a rise in upper mantle viscosity, a loss in mantle water, and eventually a decrease and cessation in upper mantle melting.
We found such Type II-like behavior in many of our Type I evolutions when extended beyond the lifetime of Venus.
This result provides evidence that the Type II evolution may not be just a particular type of evolution for Venus, but could represent a general evolutionary path that planets take when their interior melting begins to cease.

\subsection{Type III (Smaller Inner Core) Evolutions}

Type III evolutions, representing about 10.0\% of all simulations, are unique in that they end with a smaller, sometimes nonexistent, inner core.
The cutoff for a Type III evolution is a simulation that ends with an inner core that is less than 80\% the radius of the core.
This threshold is chosen roughly as the break in the population distribution of simulations based on their final inner core radius, as shown in Sec. \ref{sec:discussion}.
This cutoff also matches the ``intermediate case'' of a partially liquid core in \citet{dumoulin_tidal_2017}, whereas above this value we asymptotically approach their completely solid core model.
While some Type III simulations end with an inner core close to 80\% the radius of the core, most end with a much smaller inner core radius.
The dynamics of Type III evolutions depend primarily on the thermal evolution of the core, but also on the thermal evolution of the mantle, which acts as a boundary condition for core cooling/heating in our whole-planet model.
Thus, by better understanding this potential evolutionary scenario for Venus' core, we can better understand the evolution of the planet as a whole.

\begin{figure*}[htp!]
    \includegraphics[width=\linewidth]{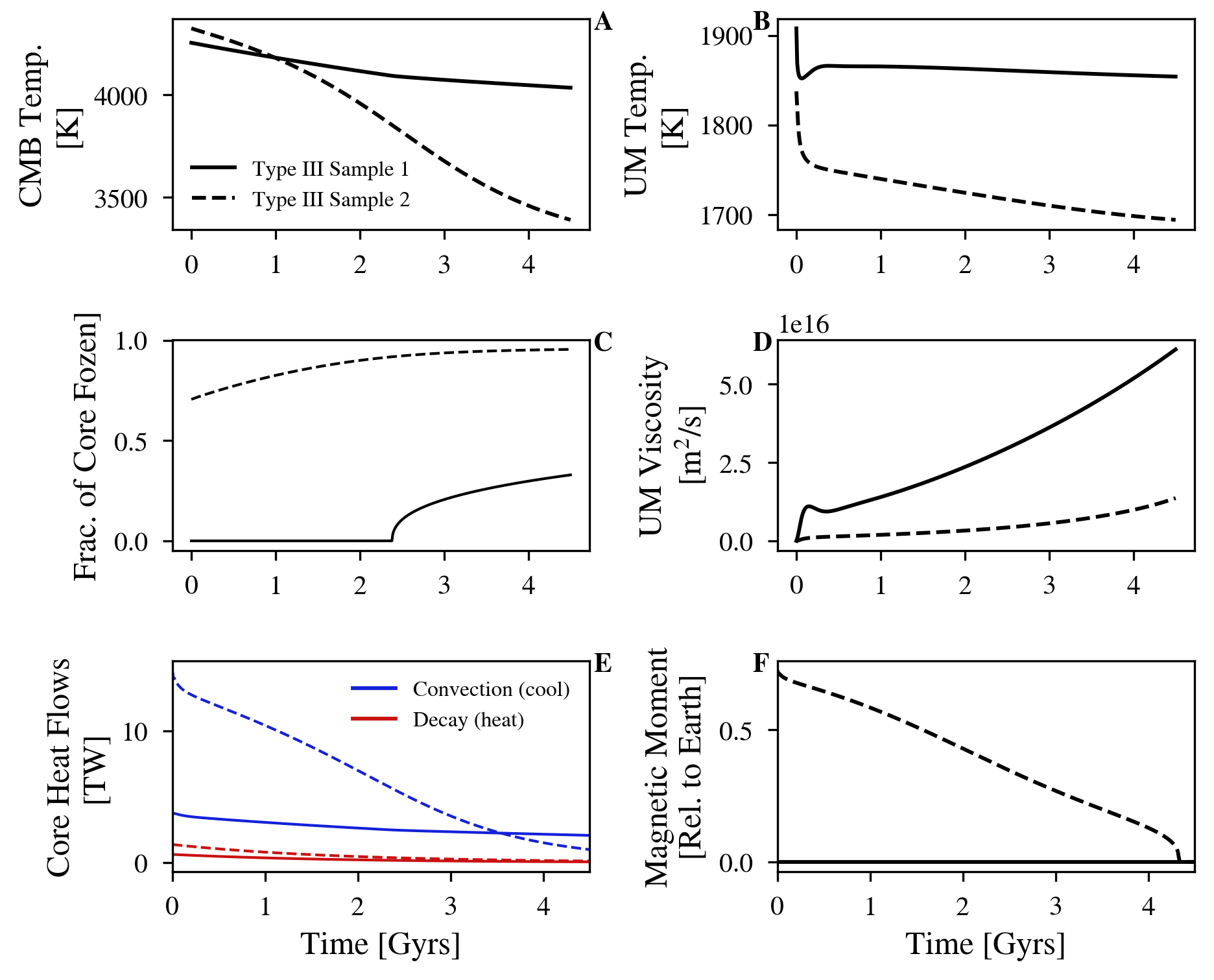}
    \caption{Time evolution of select key core and mantle variables for two sample Type III evolutions. Variables for sample evolution 1 are always plotted with a solid line, variables for sample evolution 2 are always plotted with a dashed line. Panel \textbf{A} plots the temperature at the core-mantle boundary over time; Panel \textbf{B} plots the upper mantle temperatures over time; panel \textbf{C} shows the fraction of the core frozen over time; panel \textbf{D} shows the upper mantle viscosity; panel \textbf{E} shows the heat source (radioactive decay in red) and heat sink (convective cooling in blue) for the core of each sample evolution; panel \textbf{F} shows the magnetic field strength relative to present-day Earth's over time. Parameter choices for this evolution can be found in Table \ref{TypeIIIEvolParameters}. \label{fig:TypeIIISampleEvol}}
\end{figure*}

\begin{deluxetable}{lll}
    \tablewidth{\linewidth}
    \tablecaption{Parameters for sample Type III evolutions shown in Fig. \ref{fig:TypeIIISampleEvol} \label{TypeIIIEvolParameters}}
    \tablehead{\colhead{Parameter} & \colhead{Sample 1} & \colhead{Sample 2}}
    \startdata
    $T_m$ [K] & 2725 & 2625 \\
    $\Delta T_\mathrm{CMB}$ [K] & 652 & 852 \\
    log$_{10}(\nu_0$ [m$^2$/s]) & 9.49 & 8.08 \\
    $\nu_\mathrm{jump}$ [nd] & 57 & 71 \\
    $\Delta T_{\chi,\mathrm{ref}}$ [K] & 630 & 134 \\
    $T_\mathrm{ref,Lind}$ [K] & 5746 & 5786 \\ 
    $^{40}$K$_m$ [Earth] & 1.45 & 1.30 \\
    $\eta_K$ [nd] & 0.013 & 0.03 \\
    k$_\mathrm{OC}$ [W/m/K] & 105 & 198 \\
    $\epsilon_\mathrm{erupt}$ [nd] & 0.12 & 0.04 \\
    $\Delta E_\mathrm{H_2O}$ [K/wt. frac.] & 1.35$\times10^6$ & 2.85$\times10^6$ \\
    $M^\mathrm{H_2O}_p$ [TO] & 6.5 & 4.1 \\
    $\eta_\mathrm{H_2O}$ [nd] & 0.13 & 0.14 \\
    $M^\mathrm{CO_2}_p$ [bar] & 108.3 & 99.7 \\
    \enddata
\end{deluxetable}

There is a wide diversity of behaviors within the Type III evolutions. Fig. \ref{fig:TypeIIISampleEvol} illustrates this point by plotting two sample Type III evolutions, whose initial conditions are shown in Table \ref{TypeIIIEvolParameters}. While both samples have a growing inner core and a cooling core temperature over time, Sample 1 ends with a final inner core radius that is much lower than that of Sample 2.
Sample 2 also has a much lower mantle viscosity, which enables convective cooling of the core more than in Sample 1.
Sample 2's strong convective cooling helps maintain a magnetic field in its past, whereas Sample 1 does not exhibit any magnetic field throughout its history (despite having a positive compositional core buoyancy flux due to the growing inner core).
While some Type III evolutions exhibit a past magnetic field, the majority of Type III evolutions do not exhibit a past magnetic field.

\begin{figure*}[htp!]
    \includegraphics[width=\textwidth]{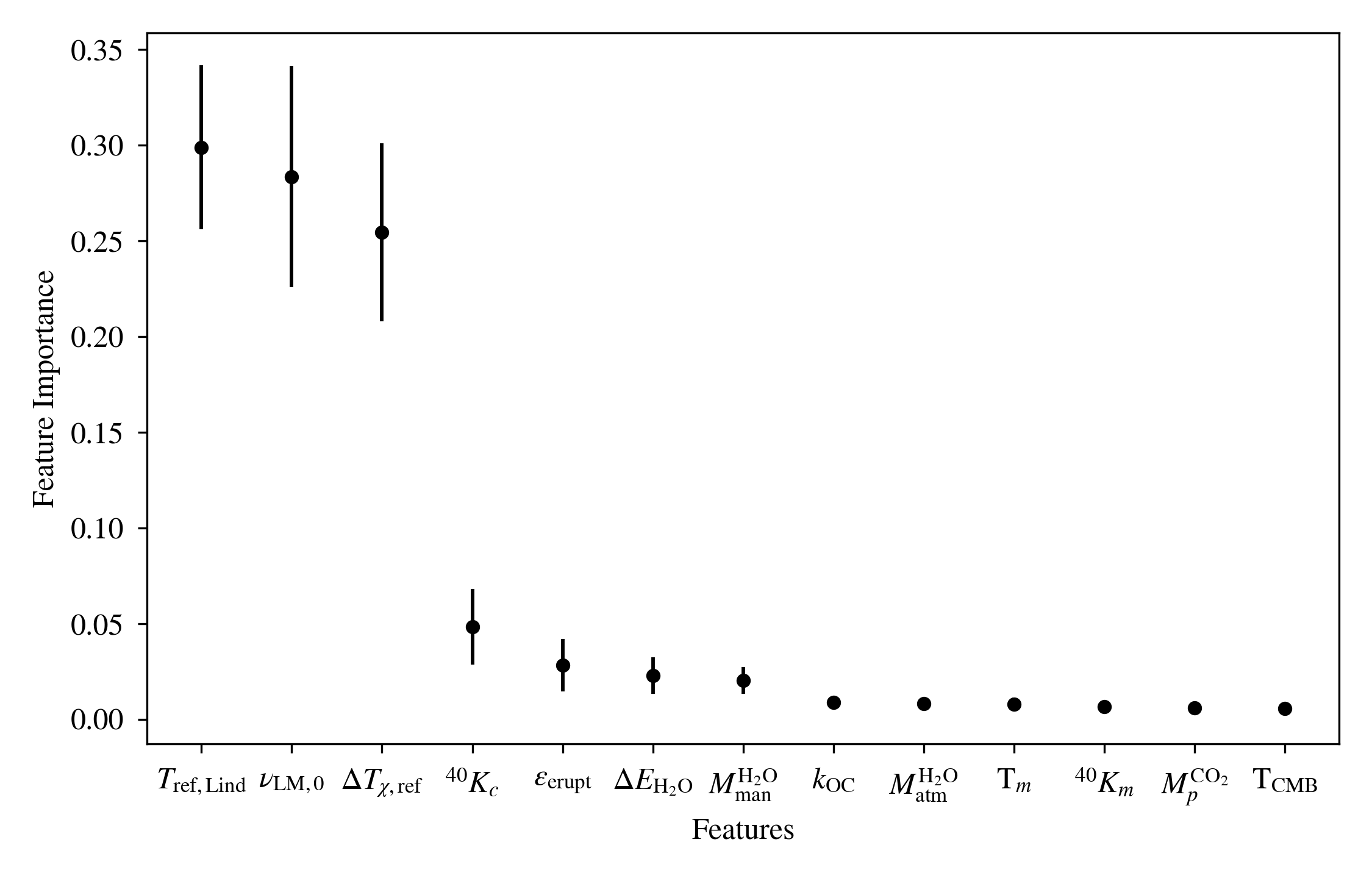}
    \caption{Each feature's relative importance in determining whether a simulation will end as a Type III simulation. The feature variables are described in Table \ref{rangevary}. The error bars represent the standard deviation for the feature importance as calculated from 50 different random forest models trained on the same data. $T_\mathrm{ref,Lind}$ and $\Delta T_{\chi,\mathrm{ref}}$ set the threshold for inner core nucleation, while $\nu_\mathrm{LM,0}$ sets the CMB heat flow and thus whether a core will evolve to a state that promotes inner core nucleation.\label{fig:CoreNotFrozenImportantFeatures}}
\end{figure*}

In order to identify the influence of each varied parameter on the dynamics that lead to Type III behavior, we again derived feature importance values from a random forest classifier.
This method is identical to that described earlier for identifying important features of Type II evolutions, with the only difference being an introduction of four new features, the temperature at the core-mantle boundary ($T_\mathrm{CMB}$) instead of $\Delta T_\mathrm{CMB}$, the amount of potassium in the core ($^{40}$K$_c$) instead of $\eta_\mathrm{K}$, the reference Lindemann temperature $T_\mathrm{ref,Lind}$, and the reference viscosity of the lower mantle, $\nu_\mathrm{LM,0}$, which we define as the upper mantle reference viscosity multiplied by a jump parameter that represents the increase in viscosity with temperature and pressure, so:
\begin{equation}
    \nu_\mathrm{LM,0} = \nu_0 \nu_\mathrm{jump}.
\end{equation}
The derived feature importance values are shown in Fig. \ref{fig:CoreNotFrozenImportantFeatures}. 
The primary importance of the reference Lindemann temperature, $T_\mathrm{ref,Lind}$, and the core liquidus depression value, $\Delta T_{\chi,\mathrm{ref}}$, makes sense as these parameters define the threshold temperature for the evolution to reach in order for its core to freeze.
Whether the core evolves to meet that temperature threshold depends on its thermal evolution, which is primarily dictated by $\nu_\mathrm{LM,0}$, the third important feature in our analysis. Type III evolutions tend to have much higher mantle viscosities and thus struggle with extracting heat from their cores.

\begin{figure*}[htp!]
    \includegraphics[width=\linewidth]{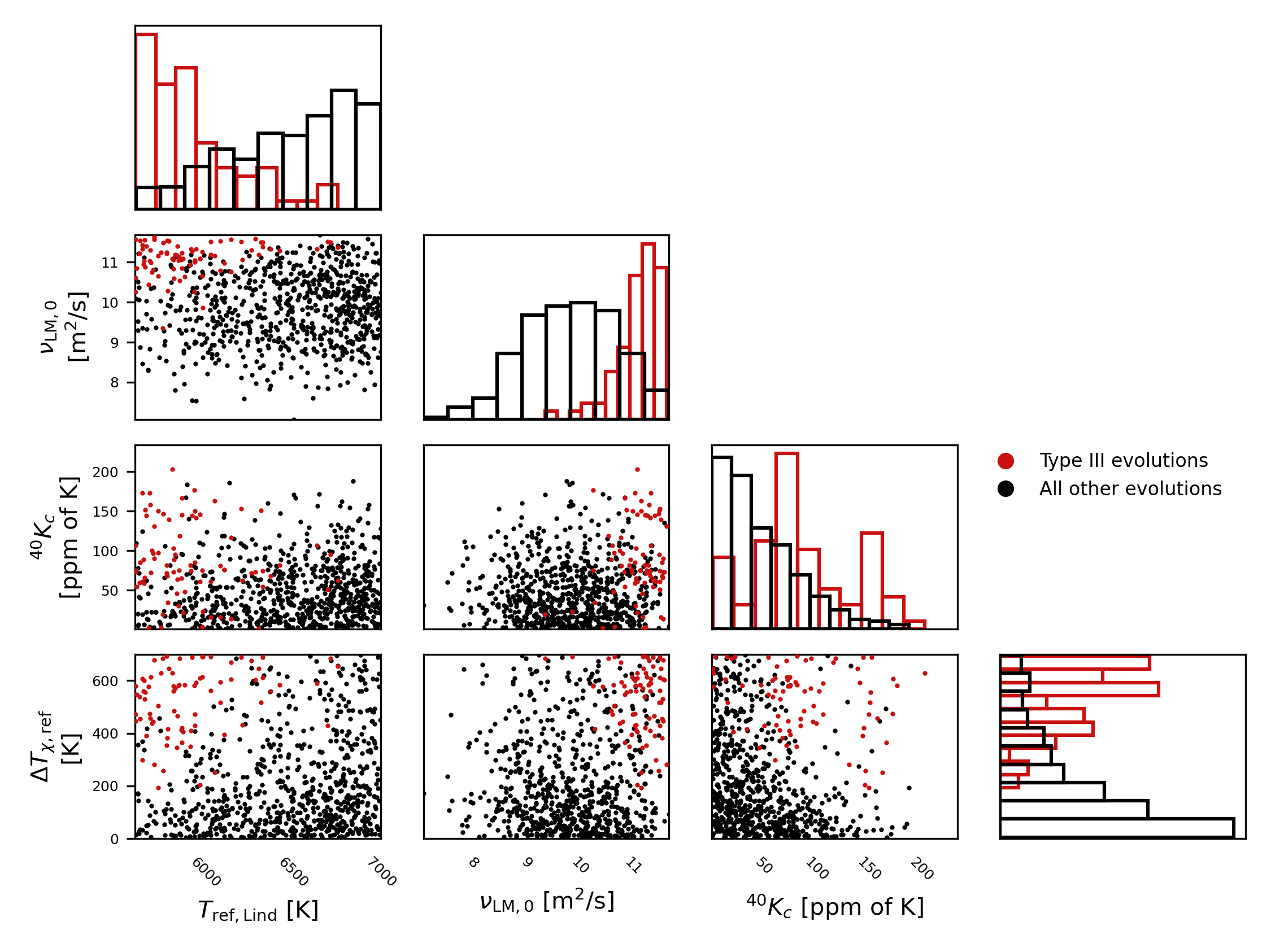}
    \caption{The scatter plots in this corner plot show where every simulation lies in a subset of the full parameter space defined by four variables: the reference viscosity for the lower mantle ($\nu_\mathrm{LM,0}$), the amount of $^{40}K$ in the core ($^{40}K_c$ in units of ppm of K in the core), the liquidus depression ($\Delta T_{\chi,\mathrm{ref}}$), and the reference core liquidus temperature ($T_\mathrm{ref,Lind}$). The histograms show the distribution of Type III evolutions (red) and non-Type III evolutions (black) marginalized over all parameters except for each parameter below them (or to the left in the case of the initial mantle temperature). Note that the lower mantle reference viscosity, the liquidus depression, and the reference liquidus temperature appear to exert the strongest control on whether a simulation will end with a mostly frozen core.\label{fig:CoreNotFrozenFeatureTriangle}}
\end{figure*}

To better understand the interplay of these important parameters and their influences on core dynamics, we plot in Fig. \ref{fig:CoreNotFrozenFeatureTriangle} each of our simulations in a four-dimensional subset of the whole parameter space defined by the four most important Type III-determining parameters as well as the amount of potassium-40 in the core, $^{40}K_c$, which acts as a heat source to the core.
In agreement with Fig. \ref{fig:CoreNotFrozenImportantFeatures}, we can see that Type III evolutions cluster separately from non-Type III evolutions rather well in the $\Delta T_{\chi,\mathrm{ref}}$-$T_\mathrm{ref,Lind}$-$\nu_\mathrm{LM,0}$ space.
While the amount of potassium-40 in the core does indeed play a role in determining the thermal evolution of the core, it does not appear to exert a primary control in whether a simulation will follow Type III evolutionary behavior or not.
Type III evolutions tend to have higher lower mantle reference viscosities; higher viscosities make it harder for the cores of Type III evolutions to lose heat and thus helps them preserve a partially liquid core.

\begin{figure*}[htp!]
    \includegraphics[width=\linewidth]{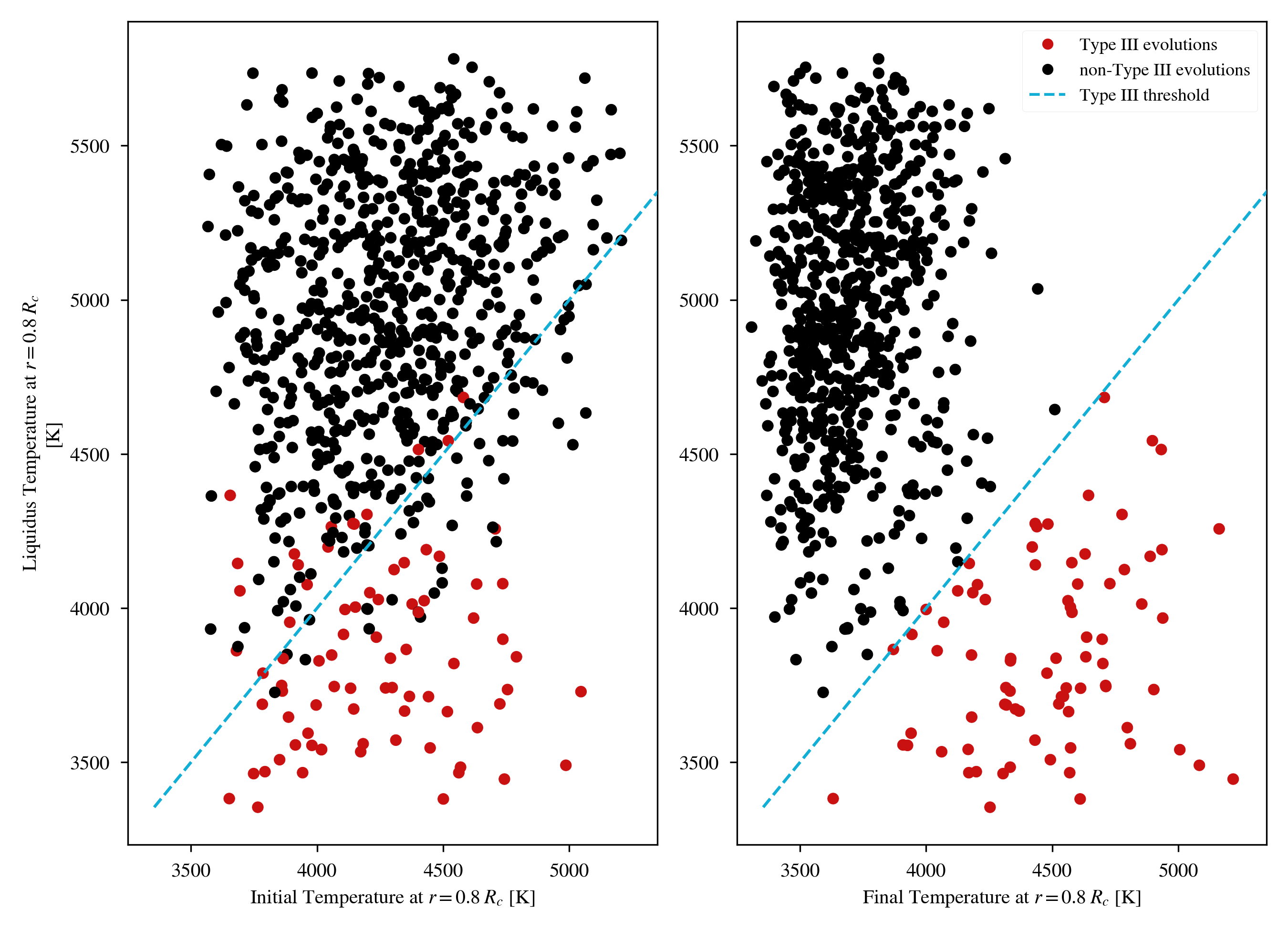}
    \caption{These two panels show the thermal state of the core for Type III (red) and non-Type III (black) at the beginning (left) and end (right) of the evolution. Each simulation is shown in a space defined by the liquidus temperature at 80\% of the core's radius and the actual temperature at 80\% of the core's radius. The cyan dashed line represents the threshold for a simulation to be considered Type III, ending with an inner core that is less than 80\% of the core's radius. \label{fig:FinalRICState}}
\end{figure*}

Type III evolutions tend to start with a small inner core and either heat up or do not cool such that their inner core remains small.
This thermal evolution pattern can be seen in Fig. \ref{fig:FinalRICState}.
Panel A shows the liquidus temperature at 80\% of the core radius, and the core temperature at 80\% of the core radius for each simulation at our first time-step.
Panel B shows the same values for each simulation at the final time-step (4.5 Gyrs).
The non-Type III simulations tend to cool over time and end to the left of the cyan line (the threshold for being a Type III evolution), whereas Type III simulations tend to either heat up or not have a substantial change in their core temperature over time.
There are only a few Type III evolutions that would not be Type III evolutions  if the liquidus temperature at 80\% of the core was increased.
Instead, the majority of Type III evolutions become Type III evolutions primarily due to their thermal evolution, meaning that the lower mantle reference viscosity exerts the strongest influence on the state of Venus' core, demonstrating the importance of modeling the mantle evolution to self-consistently assess the core's evolution and state.

Type III evolutions that start to the left of the dashed line have a high viscosity, but also have a higher amount of potassium in the core, enabling these simulations to heat up enough to cross the dashed line and melt their inner cores.
Non-Type III evolutions that start to the right of the dashed line, conversely, have a low viscosity, which allows these simulations to quickly cool and cross over the dashed line as their cores completely freeze over.
Neglecting potassium in the core would still likely result in some Type III evolutions, as many simulations would still struggle to remove the heat of formation to the mantle through convection.

\begin{figure*}[htp!]
    \includegraphics[width=\linewidth]{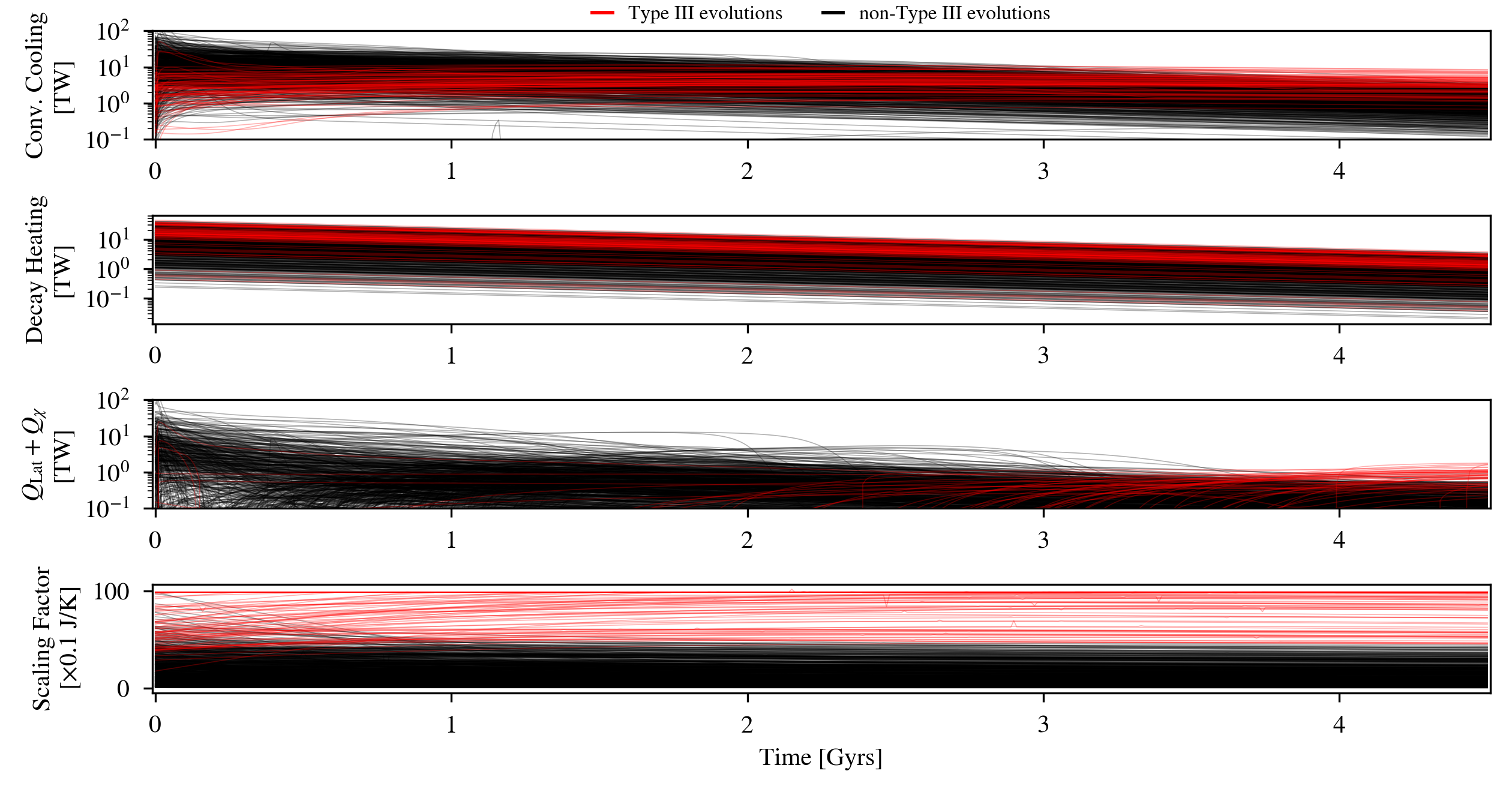}
    \caption{The evolution of the different factors in the core temperature evolution (Eq.~\ref{eq:dotTCoreyesIC}). The upper panel shows the amount of cooling due to convective loss to the mantle ($Q_\mathrm{CMB}$), the second panel shows the amount of heating due to decay of radioactive elements ($Q_{i,\mathrm{core}}$), the third panel shows the sum of core heating/cooling effects (latent, gravitational, and compositional energy), and the bottom panel shows the scaling factor due to the size of the inner core and its influence on secular cooling, i.e., the denominator in Eq.~(\ref{eq:dotTCoreyesIC}). Note that Type III evolutions (red) have a higher scaling factor and thus will cool or heat less from the same amount of energy as non-Type III evolutions. \label{fig:TDotCorePartsPlot}}
\end{figure*}

By plotting each of the key contributors to the core's thermal evolution (the cooling due to convection, the heating due to radioactive decay, cooling/heating due to inner core melting/solidification, and, in the denominator, the scaling of the heat capacity of the core based on the presence of an inner core) in Fig. \ref{fig:TDotCorePartsPlot} we can further distinguish Type III thermal evolutions from non-Type III evolutions.
First, it can be seen from panel B that while Type III evolutions do tend to have higher potassium concentrations, there is still overlap with non-Type III evolutions, demonstrating that the assumption of potassium in the core is not strictly necessary for the existence of Type III evolutions.
Additionally, panel C shows that latent energy and gravitational energy due to inner core solidification contribute less to the thermal evolution of the core than decay heating and convective cooling.

Panel A demonstrates an intriguing pattern, that Type III evolutions have greater energy loss in the core from convection than non-Type III evolutions, even though Type III evolutions do not have cores that cool nearly as much.
These high convective energy flows from Type III evolutions result from maintaining core heat through time, which sustains a large temperature difference at the CMB and fuels a larger energy flow to the mantle.
However, while Type III cores lose more energy, Fig. \ref{fig:TDotCorePartsPlot}D demonstrates their scaled heat capacity is much lower than non-Type III evolutions.
As Type III evolutions have larger outer cores than non-Type III evolutions, they require significantly more energy loss to result in a change in core temperature.
So, despite a high convective energy loss term for Type III evolutions, their higher heat capacities, together with a high lower mantle viscosity, result in a core that stays hot and liquid throughout geologic time.
This phenomenon leads to two key observations: the state of a core can be a key player in its thermal evolution, and a core that remains liquid or partially liquid throughout geologic time can represent a continuous heat source to the mantle, which affects its thermal evolution.

\subsection{Type II+III Evolutions}

\begin{figure*}[t!]
    \includegraphics[width=\linewidth]{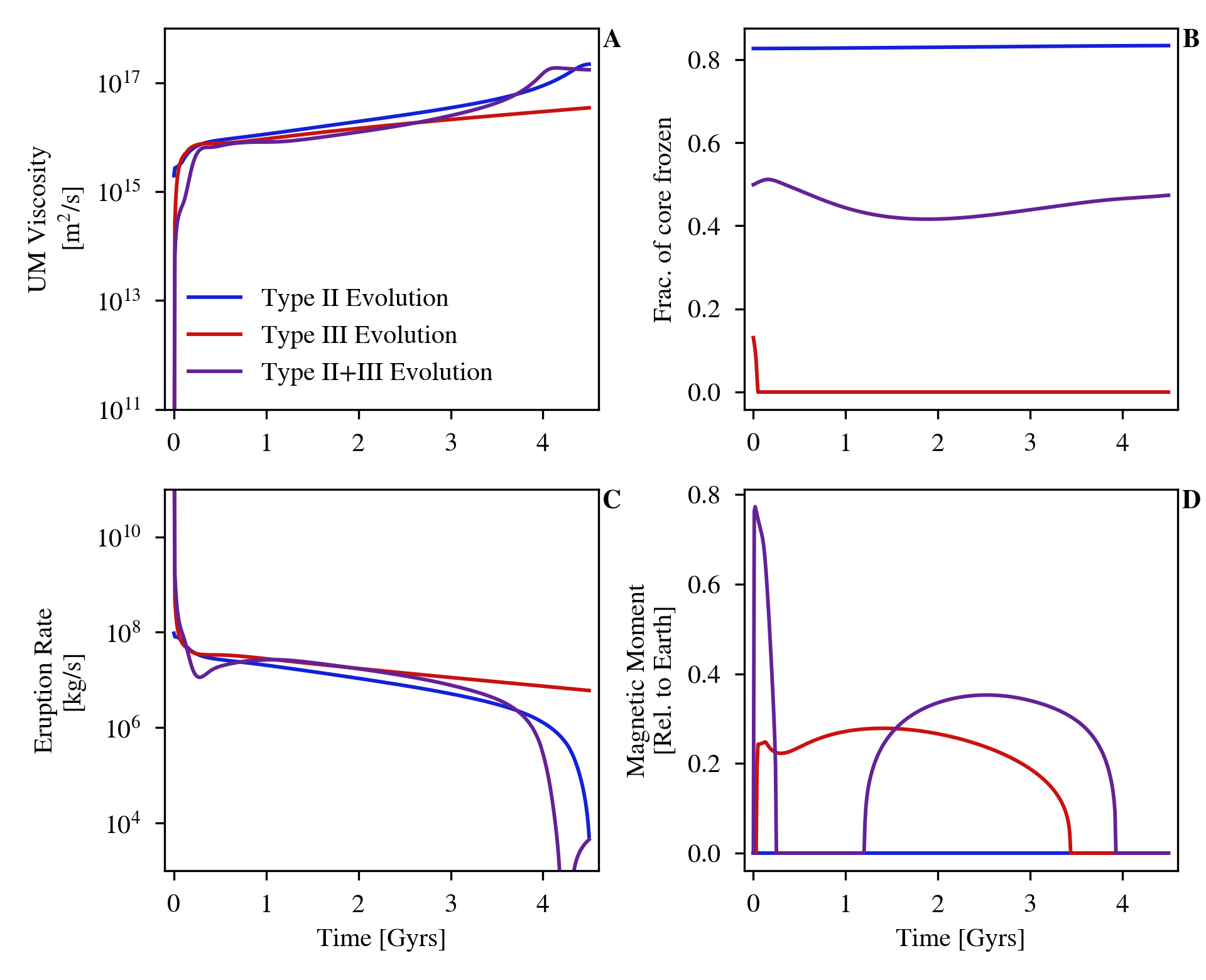}
    \caption{Time evolution of select key core and mantle variables for a sample Type II evolution (blue), a sample Type III evolution (red) and a sample evolution that is both Type II and Type III (purple). The upper left panel plots the upper mantle viscosity over time; the lower left panel plots the eruption rate over time; the upper right panel plots the fraction of the core that's frozen over time; the lower right panel plots the magnetic moment over time. The parameter values for each of these sample evolutions are given in Table \ref{TypeIIandIIIEvolParameters}. \label{fig:TypeIIandIIISampleEvol}}
\end{figure*}

While our named types do exhibit different behaviors overall, there are a number of simulations that display behavior that is characteristic of both Type II and Type III evolutions, so we label them as Type II+III evolutions.
Samples of Type II+III, Type II, and Type III evolutions are plotted in Fig.~\ref{fig:TypeIIandIIISampleEvol} for comparison.
Type II+III evolutions tend to display Type II-like behavior in the mantle, as can be seen with the characteristic decline in the sample evolution's eruption rate, and Type III-like behavior in the core, as can be seen by the sample evolution's smaller inner core.

\begin{deluxetable*}{llll}
    \tablewidth{\linewidth}
    \tablecaption{Parameters for sample Type II, Type III, and Type II+III evolutions shown in Fig. \ref{fig:TypeIIandIIISampleEvol} \label{TypeIIandIIIEvolParameters}}
    \tablehead{\colhead{Parameter} & \colhead{Type II Evolution} & \colhead{Type III Evolution} & \colhead{Type II+III Evolution}}
    \startdata
    $T_m$ [K] & 2539 & 2994 & 2951 \\
    $\Delta T_\mathrm{CMB}$ [K] & 84 & 255 & 110\\
    log$_{10}(\nu_0$ [m$^2$/s]) & 9.29 & 9.29 & 9.56 \\
    $\nu_\mathrm{jump}$ [nd] & 29.4 & 69.8 & 96.7\\
    $\Delta T_{\chi,\mathrm{ref}}$ [K] & 648 & 584 & 455\\
    $^{40}$K$_m$ [Earth] & 0.81 & 1.01 & 1.41\\
    $\eta_K$ [nd] & 0.023 & 0.479 & 0.387 \\
    k$_\mathrm{OC}$ [W/m/K] & 129 & 132 & 121\\
    $\epsilon_\mathrm{erupt}$ [nd] & 0.099 & 0.085 & 0.13\\
    $\Delta E_\mathrm{H_2O}$ [K/wt. frac.] & 3.87$\times10^6$ & 2.09$\times10^6$ & 5.70$\times10^6$\\
    $M^\mathrm{H_2O}_p$ [TO] & 5.98 & 8.26 & 8.93\\
    $\eta_\mathrm{H_2O}$ [nd] & 0.30 & 0.33 & 0.34\\
    $M^\mathrm{CO_2}_p$ [bar] & 113 & 111 & 91.8\\
    \enddata
\end{deluxetable*}

While we do not often call out Type II+III evolutions specifically, they do provide an interesting example of how the mantle can influence core dynamics.
We can see from Fig.~\ref{fig:TypeIIandIIISampleEvol} that the magnetic field in the Type II+III sample evolution shuts off at around the time that the Type II-like increase in viscosity (and corresponding decrease in melt eruption) occurs.
This demonstrates how the Type II behavior in the mantle, a shutdown of crustal recycling leading to dehydration stiffening, leads to Type III behavior in the core --- without the dehydration stiffening this Type II+III evolution would likely be able to maintain a thermal dynamo and thus not match Venus.

\subsection{Type IV (Oscillatory) Evolution}
\begin{figure*}[htp!]
    \includegraphics[width=\linewidth]{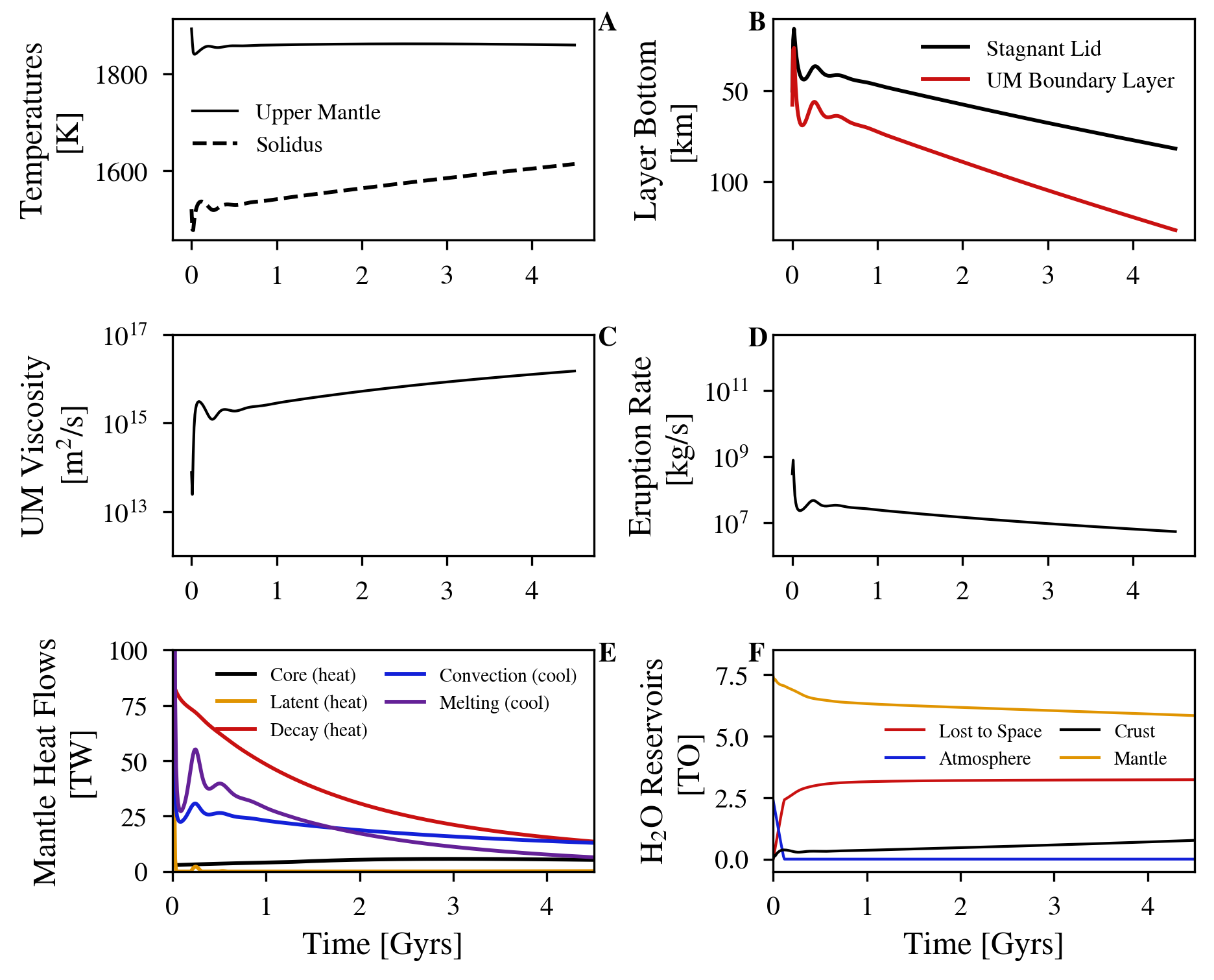}
    \caption{Time evolution of select key mantle variables for an evolution with Type IV behavior. Panel \textbf{A} plots the upper mantle and solidus temperatures over time; panel \textbf{B} shows the depth to the bottom of the stagnant lid layer and the upper mantle thermal boundary layer; panel \textbf{C} shows the upper mantle viscosity; panel \textbf{D} shows the eruption rate; panel \textbf{E} shows the different heat sources and sinks (cooling mechanisms) for the mantle, and panel \textbf{F} shows the evolution of the water reservoirs in the simulation. Note that the characteristic Type IV oscillations occur most strongly in any parameter related to the viscosity (the eruption rate, stagnant lid and thermal boundary layer thicknesses, and convective cooling). Parameter choices for this evolution can be found in Table \ref{TypeIVEvolParameters}. \label{fig:TypeIVSampleEvol}}
\end{figure*}

Type IV evolutions are characterized by transient oscillations of parameters such as mantle temperature and stagnant lid thickness during the first $\sim$500 Myrs.
Fig. \ref{fig:TypeIVSampleEvol} shows the evolution of several key properties of a sample Type IV evolution, revealing that the oscillations are quite strong in the primary mantle cooling mechanisms: convection and eruption.
As evident in Fig.~\ref{fig:evolve_plots}, the Type IV evolution also stands out with a higher present-day melt fraction than most other valid evolutions.

Because there are so few Type IV evolutions we cannot apply random forest and feature importance analyses to determine what drives this evolution, so instead we compare these cases to the other valid solutions by eye.
For example, Type IV evolutions display Type III-like behavior in their cores, starting with a partially solidified inner core but with core heating resulting in the melting of that inner core as seen in Fig. \ref{fig:evolve_plots}.
Additionally, as can be seen in Fig. \ref{fig:modern_values}, the Type IV evolutions exhibit little to no magnetic field throughout the entirety of the evolution.
Such Type III-like behavior is likely due to the high lower mantle viscosity, core temperature, and core liquidus depression, which puts Type IV evolutions firmly on the Type III side of the core freezing threshold in Fig. \ref{fig:FinalRICState}.

Type IV evolutions present an interesting example of mantle behavior but are rather underrepresented in our simulations.
Because this evolutionary type has less data to analyze, and does not appear to exhibit distinctive behavior after about 1 Gyr, we do not spend more time on it.

\begin{deluxetable}{ll}
    \tablewidth{\linewidth}
    \tablecaption{Parameters for Type IV sample evolution shown in Fig. \ref{fig:TypeIVSampleEvol}. \label{TypeIVEvolParameters}}
    \tablehead{\colhead{Parameter} & \colhead{Values}}
    \startdata
    $T_m$ [K] & 2705 \\
    $\Delta T_\mathrm{CMB}$ [K] & 502 \\
    log$_{10}(\nu_0$ [m$^2$/s]) & 9.34 \\
    $\nu_\mathrm{jump}$ [nd] & 78 \\
    $\Delta T_{\chi,\mathrm{ref}}$ [K] & 421 \\
    $T_\mathrm{ref,Lind} [K] $ & 5867 \\
    $^{40}$K$_m$ [Earth] & 1.31 \\
    $\eta_K$ [nd] & 0.71 \\
    k$_\mathrm{OC}$ [W/m/K] & 144 \\
    $\epsilon_\mathrm{erupt}$ [nd] & 0.023 \\
    $\Delta E_\mathrm{H_2O}$ [K/wt. frac.] & 1.08$\times10^6$ \\
    $M^{H_2O}_p$ [TO] & 9.8 \\
    $\eta_\mathrm{H_2O}$ [nd] & 0.24 \\
    $M^\mathrm{CO_2}_p$ [bar] & 128 \\
    \enddata
\end{deluxetable}

\section{Discussion} \label{sec:discussion}
We used a coupled one-dimensional solar-atmosphere-surface-lithosphere-mantle-core model to predict the range of plausible evolutionary scenarios for Venus that match its current atmospheric composition and lack of a magnetic field.
This approach successfully discovered 808 evolutionary tracks that match the current observables, and in doing so, revealed complementary information about past and current Venus characteristics.
In this section, we will compare our results to current hypotheses and predictions of Venus' geologic history, while also describing new predictions that can be tested with future missions to Venus.

\begin{figure*}
    \includegraphics[width = \linewidth]{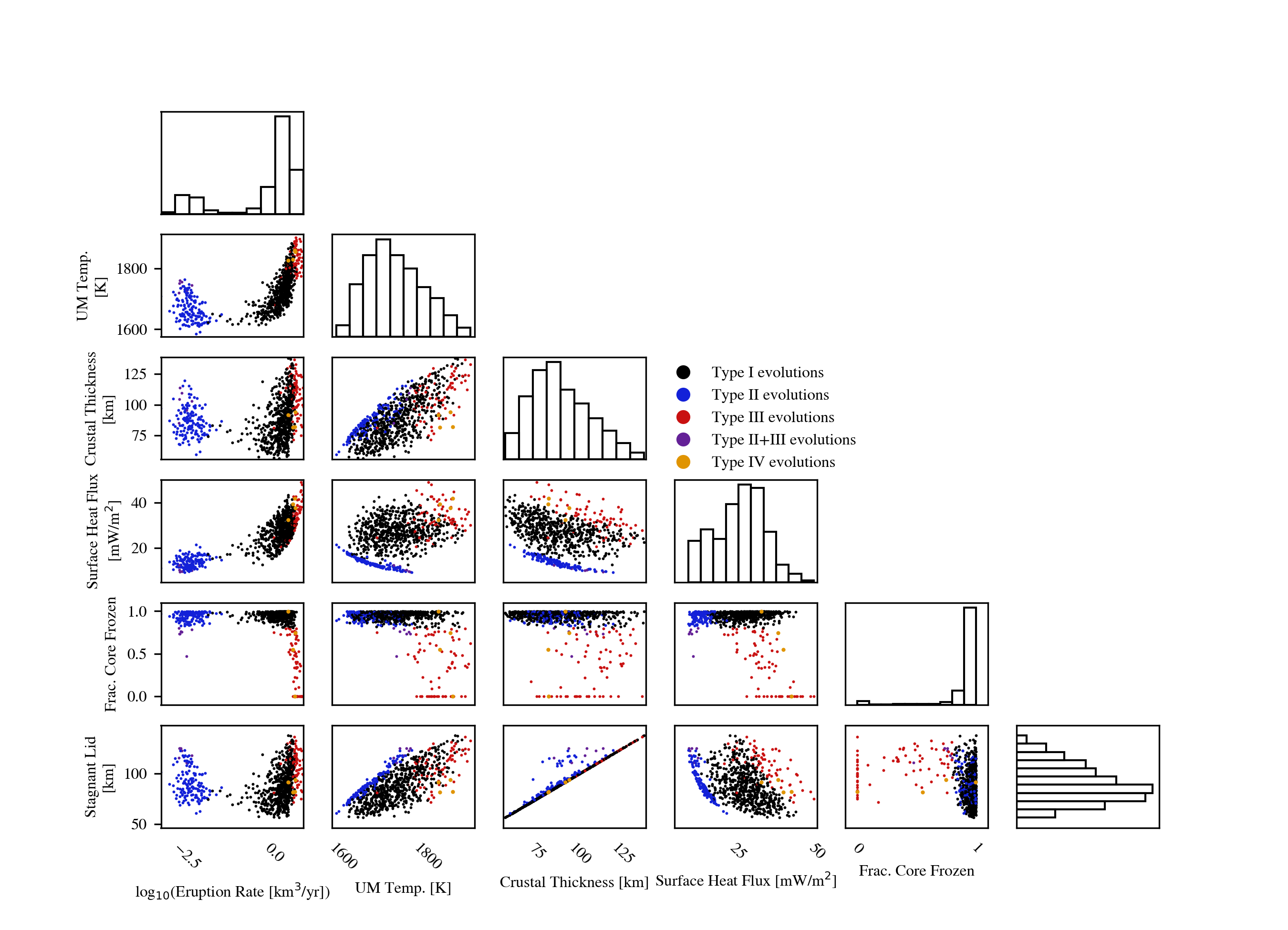}
    \caption{Final conditions (after 4.5 Gyrs) for all evolutions that reproduce Venus in a space defined by five parameters: the final temperature of the upper mantle, the final thickness of the crust, the final heat flux at the surface, the fraction of the core that ends up solid ($R_\mathrm{IC}$), and the final eruption rate. Each simulation is color-coded based on the identified evolutionary scenario as described in the text. The histograms show the distribution of all simulations marginalized over all variables except for each variable below them (or to the left). Note that Type II evolutions (the blue points) carve out a very distinctive region in any subspace containing the heat flux at the surface or the eruption rate and Type III evolutions (red points) carve out a distinctive region in any subspace containing the fraction of the core frozen.
    }
    \label{fig:outputtriangle}
\end{figure*}

To ground our results in observable predictions we display each of our simulation results in a parameter space defined by the value of certain, potential observables at 4.5 Gyrs for each simulation in Fig. \ref{fig:outputtriangle}.
This parameter space is defined by the upper mantle temperature, the crustal thickness, the surface heat flux, the fraction of the core that is frozen, the stagnant lid thickness, and the eruption rate.
All of our simulations are color-coded by the identified evolutionary scenario, any future observations of any of these values on Venus could then be used to further narrow down the possible evolutionary scenarios of Venus.

The lack of a geodynamo today permits initially larger solid cores (see Fig. \ref{fig:evolve_plots}), which is not a typical choice in most previous studies of the thermal and magnetic evolution of Venus or Earth. This bias in our results arises because magnetic field strength scales with the volume of the outer core and so the lack of a geodynamo today naturally allows this structure. If true, this result could suggest that Venus formed relatively cold. There is no direct evidence of the processes that formed Venus, so we should not dismiss the possibility that the lack of a geodynamo is a clue to its formation history. We note, for example, that some ambiguity remains regarding Earth's mantle's initial state, with observations \citep[non-arc basalts from the Archean;][]{Herzberg2010} and theory \citep[parameter sweeps of Earth's mantle's thermal history;][]{SealesLenardic2020} both finding that Earth could have formed cooler than is commonly assumed.
However, recent work incorporating large impact events into parameterized convection models has shown that a ``hot'' start for the core could still result in the lack of a geodynamo today presents other options \citet{Marchi2023}.
Regardless, should future observations find that Venus' modern core is less than 80\% solid today, then most of these ``cold'' solutions would be ruled out.

\subsection{Venus' Crust as a Constraint on the Model}
Traditionally, the stagnant lid geodynamic regime generates high crustal thickness values (greater than 60 km) for Venus \citep[e.g.][]{armann_tackley_2012}, which appear to be inconsistent with global crustal thicknesses derived from gravity measurements, which range from 20 to 45 km \citep{JamesZuber2013,Jimenez2015,Maia2022}.
Our model does not reproduce these smaller crustal thicknesses. 
From Fig.~\ref{fig:outputtriangle}, we can see that even our lowest crustal thickness evolutions (some Type I evolutions) are $>60$ km. While the actual thickness of the Venusian crust is not directly measured, the mounting evidence for a relatively thin crust warrants discussion.

Recent work has shown that Venus' surface heat flow is larger than previously thought \citep{smrekar_2022}, resulting in predictions for Venus' crustal thickness trending to lower values.
While our surface heat flows match the lower values that \citet{smrekar_2022} found for 45\% of the planet's surface (excluding tesserae) with a thick elastic lithosphere (11 -- 23 mW/m$^2$), they do not match the other 40\% of the planet' surface (again, excluding tesserae) with a heat flow of 39 -- 116 mW/m$^2$.
While our full suite of simulations do include simulations that exhibited higher surface heat flows, these higher surface heat flow simulations are do not match Venus as they maintain a dynamo at the present-day\footnote{We did find that if we employed the \citet{driscoll_thermal_2014} model that the core could  begin fully frozen and that the crust could be as thin as the recent estimates, but as an initially solid core is unlikely, we do not present those thermal histories.}.

The heat flow ranges above also translate to mechanical lithospheric thicknesses, which corresponds to our stagnant lid depth.
We do not include the upper mantle thermal boundary layer in the stagnant lid as it is defined as part of the asthenosphere in our model \citep{BreuerMoore2007}.
For the 40\% of Venus with a thin elastic lithosphere, the mechanical lithosphere is 7 to 21 km.
For the 45\% of Venus with a thick elastic lithosphere, the mechanical lithosphere is 36 to 75 km.
Again, while our results in Fig.~\ref{fig:outputtriangle} can reproduce the 45\% of Venus's surface with a thick elastic lithosphere, our model cannot produce thin lithospheres.

These discrepancies point toward some shortcomings of our model and reinforce the conclusion in \citet{smrekar_2022} that ``the concept of a single value of $T_e$ [elastic lithospheric thickness] or $F_s$ [surface heat flow] can be used to represent Venus needs to be abandoned.''
It is difficult to square the circle and reproduce every aspect of Venus's evolution in a one-dimensional model that cannot generate spatial variation: the need for no magnetic field today limits the heat flow from Venus's core, which in our global model can only result in a cooler mantle temperature and a thicker lithosphere. 
The spatial diversity of volcano-tectonic features on Venus show that the heat flow and lithospheric thickness of the planet is better seen as bimodal, where high-heat flow regions corresponding to plume-induced activity \citep{gulcher_corona_2020,smrekar_2022}, our model finds the mean in the middle of this bimodal distribution and ends up matching neither case.

Our one-dimensional model is still a useful tool to constrain the interior of Venus, but our model will naturally struggle to reproduce Venus's diverse surface features. 
Without the ability to self-consistently model convection cells, we cannot make strong predictions for the surface expression of those cells and thus it is no surprise that we do not match observed constraints on the Venusian crust, lithosphere, and surface heat flow.
More work can be done though, such as parameterizing mantle plumes \citep{grott_volcanic_2011}, which should increase surface heat flows and thin the lithosphere, but it remains unclear if any one-dimensional model can generate a thin crust.

Intrusive magmatism can also play an important role in manifesting surface features from mantle convection.
\citet{lourenco_plutonic-squishy_2020} defines a plutonic-squishy lid geodynamic regime, wherein large amounts of intrusive magmatism result in high surface heat flows and low lithospheric thicknesses. 
Such large amounts of intrusive magmatism can further drive many interesting and partially-localized surface behaviors such as surface mobility \citep{byrne_globally_2021}, high surface heat fluxes \citep{smrekar_2022}, and localized subduction \citep{lourenco_plutonic-squishy_2020,Adams2022}. Based on preliminary results, we believe that incorporating mantle plumes into parameterized convection models of both the mantle and core can potentially reproduce some of the results of plutonic-squishy lid models in a one-dimensional paradigm.

Our model cannot replace two-dimensional spatially resolved models of mantle convection, but with our wide and fast parameter sweeps we can identify useful chunks of parameter space than can be explored with these higher dimensional models.
We find that the influence of water on the viscosity of the mantle is a key controller of mantle dynamics, and thus should be considered and varied in spatially resolved mantle convection models.
Additionally, we recommend investigating variance in the lower mantle viscosity to ensure that any two-dimensional convection models do not reproduce a magnetic dynamo today.

\subsection{Catastrophic resurfacing}
\begin{figure*}
    \includegraphics[width = \linewidth]{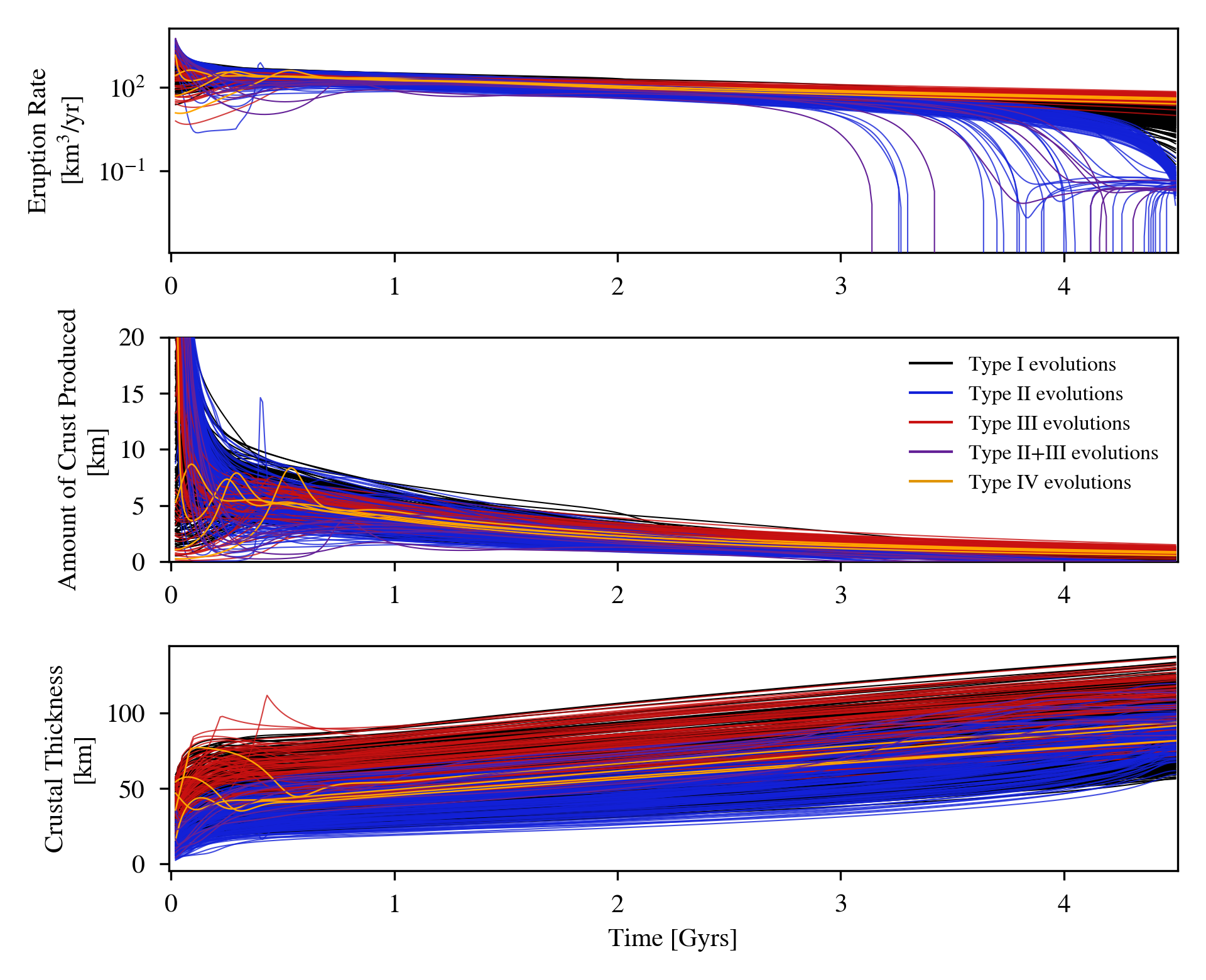}
    \caption{The evolution of the eruption rate, the amount of crust produced at each output time (10 Myrs) and the crustal thickness, color-coded by the different types of evolutionary scenarios described in the text. In plotting the time evolution of these variables we have removed the first 20 Myrs of evolution which are highly subject to transient effects. Note that the crustal thickness is not equivalent to the integrated amount of crust produced at each output time because the crustal thickness is limited by the stagnant lid thickness. Note that there are no massive eruptions (which could represent catastrophic resurfacing events) in the recent 1.5 Gyrs of any of our simulations.}
    \label{fig:eruption_rate}
\end{figure*}

None of our evolutions appear to reproduce a catastrophic resurfacing scenario for Venus.
In Fig. \ref{fig:eruption_rate} we show the eruption rate over time for all of our simulations as well as the amount of crust produced at each timestep and the crustal thickness, color-coded by evolutionary type.
Our models do not appear to yield evidence of a spike in the eruption rate, especially within the past 1 Gyr of Venusian history. Our results for the evolution of Venus' atmosphere, however, appear consistent with the catastrophic overturn model of \cite{WellerKiefer2025}, just without the jumps in atmospheric mass during major resurfacing events. So it may be that the current atmosphere offers poor leverage in distinguishing punctuated versus smooth tectonic evolution.

Our models support the hypothesis of continuous resurfacing, where relatively consistent eruption rates provide for the resurfacing of Venus' entire surface over the 700 Myr timeframe required to match crater data \citep{Phillips1992,guest_stofan_1999}. Thus our results support regional-scale volcanic and tectonic resurfacing mechanisms such as peel-back delamination \citep{Adams2022}, lithospheric dripping, subduction, and plume-lithosphere interactions at coronae \citep{gulcher_corona_2020} as well as plume-related eruptions that produce resurfacing rates that can match crater production rates \citep{smrekar_recent_2010,orourke_thermal_2015}. Our results are also consistent with
Monte Carlo models of crater production and resurfacing events that support the equilibrium resurfacing hypothesis  \citep{Bjonnes2012,orourke_venus_2014}.

Higher resolution topographic and imagery data of the Venusian surface from VERITAS and EnVision will likely provide even better constraints on crater production and resurfacing rates.
By identifying modified and unmodified craters on Venus, we can better constrain our crater counting statistics, which in turn can be used to determine whether Venus experienced catastrophic or equilibrium resurfacing \citep{HerrickReview2023}.
Additionally, noble gas abundances observed from DAVINCI will help us constrain past outgassing histories \citep{Garvin_2022,Widemann2023}, which can also help constrain our models, e.g., by refining $^{40}$Ar abundances, the decay product of $^{40}$K, and point towards catastrophic or continuous resurfacing.

\subsection{Feasibility of a past magnetic field on Venus}
A total of 88\% of our simulations of Venus exhibited a magnetic field in its past when employing buoyancy theory, with a wide range of field strength and duration across all these simulations. If we instead apply the magnetic Reynolds number criterion of $>40$ as defining dynamo-generated magnetic fields we obtain the same predictions for the possible presence of a past magnetic field. We thus restrict subsequent analyses to just the results from buoyancy theory for conciseness.
These results do not represent the probability that Venus had a magnetic field in the past, but rather is just the distribution in our own parameter space sampling.
The large fraction of cases with a remnant magnetic field occurs despite the fact that many of our successful cases begin with a mostly solidified core. As can be seen in Fig. \ref{fig:modern_values}, most of these past magnetic fields are long-lived but weaker than Earth's.

There are three different ways to sustain a dynamo in our model.
First, thermally sustained dynamos have a positive core thermal buoyancy that is larger than a non-positive core compositional buoyancy (from the core shrinking); such a large thermal buoyancy requires convection (based on the $\Delta T_\mathrm{CMB}$ to $\nu_\mathrm{LM,0}$ ratio) to be the primary heat loss mechanism in the core, as opposed to conduction (based on $k_\mathrm{OC}T_\mathrm{CMB}$), this condition is shown in Eq. (\ref{eq:cmb_heat_flux}).
Second, compositionally sustained dynamos have a positive core compositional buoyancy that is larger than a non-positive core thermal buoyancy, which requires a growing inner core as can be seen in Eq. (\ref{eq:buoyancyeqs}).
Third is a thermo-compositionally sustained dynamo in which both thermal convection and a solidifying inner core combine to create positive buoyancy in the outer core.

\begin{figure*}[htp!]
    \includegraphics[width=\linewidth]{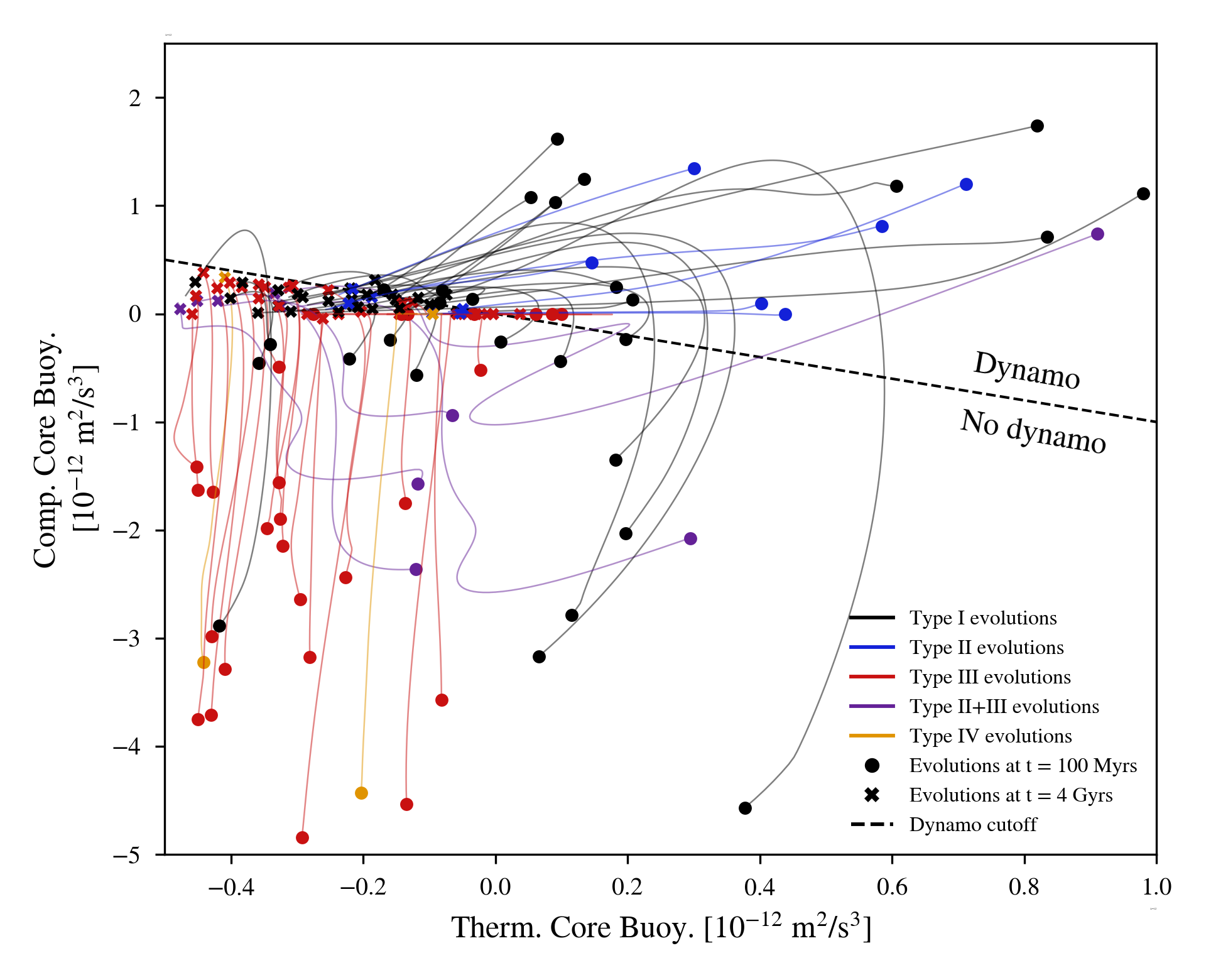}
    \caption{The evolution of the thermal and compositional core buoyancy of some representative simulations. Each evolution is color-coded based on identified typology. The thermal and compositional buoyancy state of each simulation 100 Myrs into the simulation is marked with a point. The state of each simulation at 4 Gyrs is marked with an X. The dotted black line represents the cutoff for an active dynamo (when Eq.~\eqref{dynamobuoyancycondition} is satisfied). Some evolutions exhibit long-lived magnetic fields from the start, others generate them later, and others do not ever exhibit an active dynamo.} \label{fig:BuoyancyEvolveSamples}
\end{figure*}

To better illustrate the diversity of magnetic field histories (including simulations that lack a past magnetic field), we plot the core buoyancy evolutions for a random sample of simulations, color-coded by evolution typology in Fig. \ref{fig:BuoyancyEvolveSamples}.
There is a clear difference between Type III and non-Type III evolutions, with the majority of Type III evolutions exhibiting a negative thermal core buoyancy for most of their evolution, preventing the formation of a magnetic field.
While some Type III evolutions shown do exhibit a past magnetic field (and some non-Type III evolutions do not), in general we find that the majority of Type III evolutions do not exhibit a past magnetic field and the majority of non-Type III evolutions do.
The buoyancy evolutions in Fig. \ref{fig:BuoyancyEvolveSamples} illustrate the complex interplay between dynamo generation and the thermal evolution of the core, where the thermal buoyancy is driven directly by thermal evolution and the compositional buoyancy is driven indirectly by thermal evolution (whether the inner core grows or shrinks).

\begin{figure*}[htp!]
    \includegraphics[width=\linewidth]{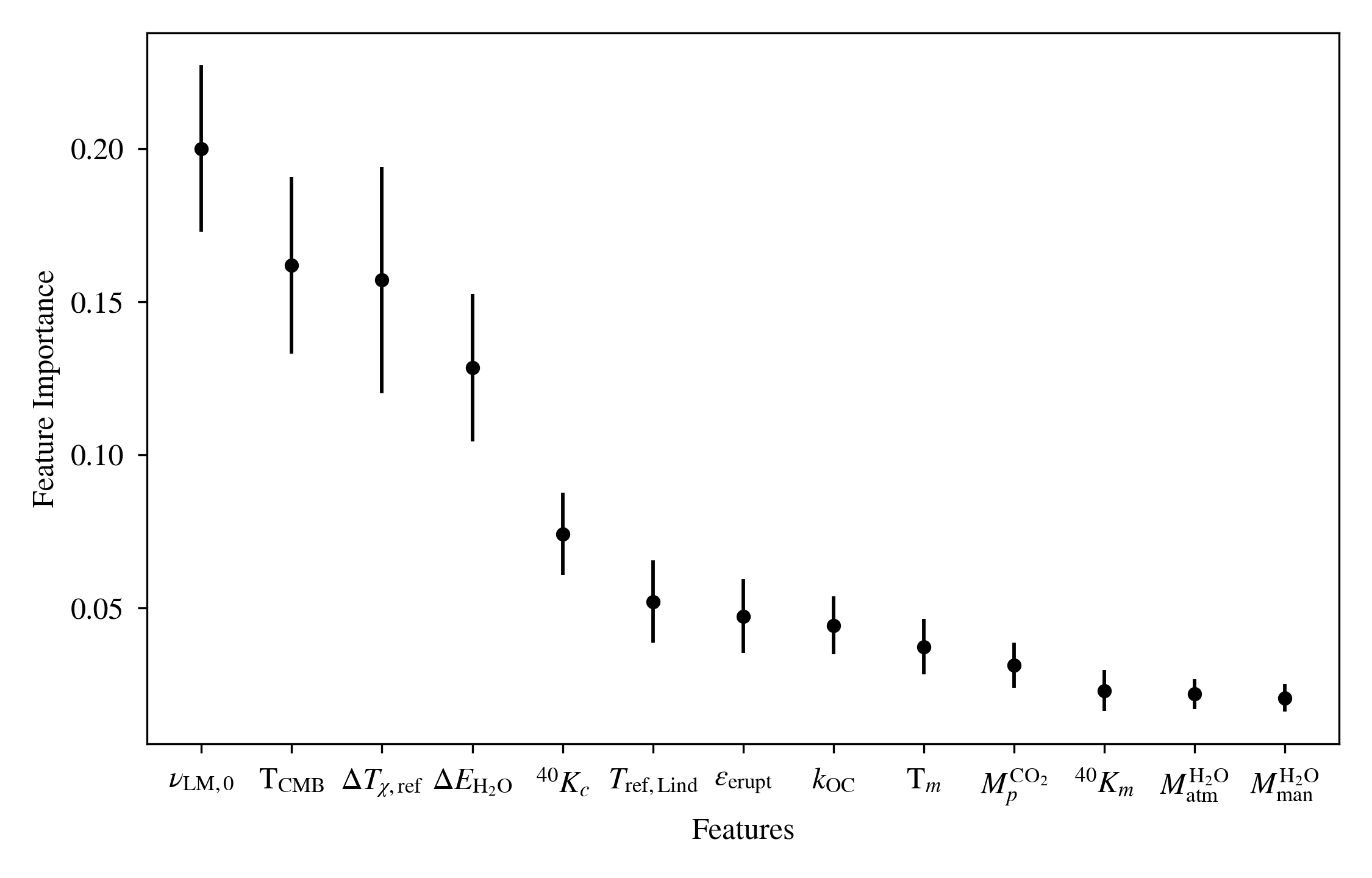}
    \caption{The relative importance of each feature for determining whether an evolution had a magnetic field at some point in its past. The feature variables on the $x$-axis are described in Table \ref{rangevary}. Error bars represent the standard deviation for the feature importance as calculated from 50 different random forest models trained on the same data. \label{fig:CoreFeatureImportance}}
\end{figure*}

To identify the controllers of thermal and compositional buoyancy evolution, we calculated feature importance values based off of a random forest classifier that categorized which evolutions had a magnetic field in the past and which didn't.
In Fig. \ref{fig:CoreFeatureImportance}, we show the feature importance values based on a random forest classifier. Not surprisingly, the core and lower mantle properties tend to be the most important, but note also the strong dependence on $\Delta E_{H_2O}$ that reveals that the role of water in setting the mantle viscosity is also critical.
This is reinforced by Fig. \ref{fig:BuoyancyEvolveSamples} where the pure Type II evolutions all exhibit a magnetic field in the past.

\begin{figure*}[htp!]
    \includegraphics[width=\linewidth,trim={0cm 0 0cm 0}]{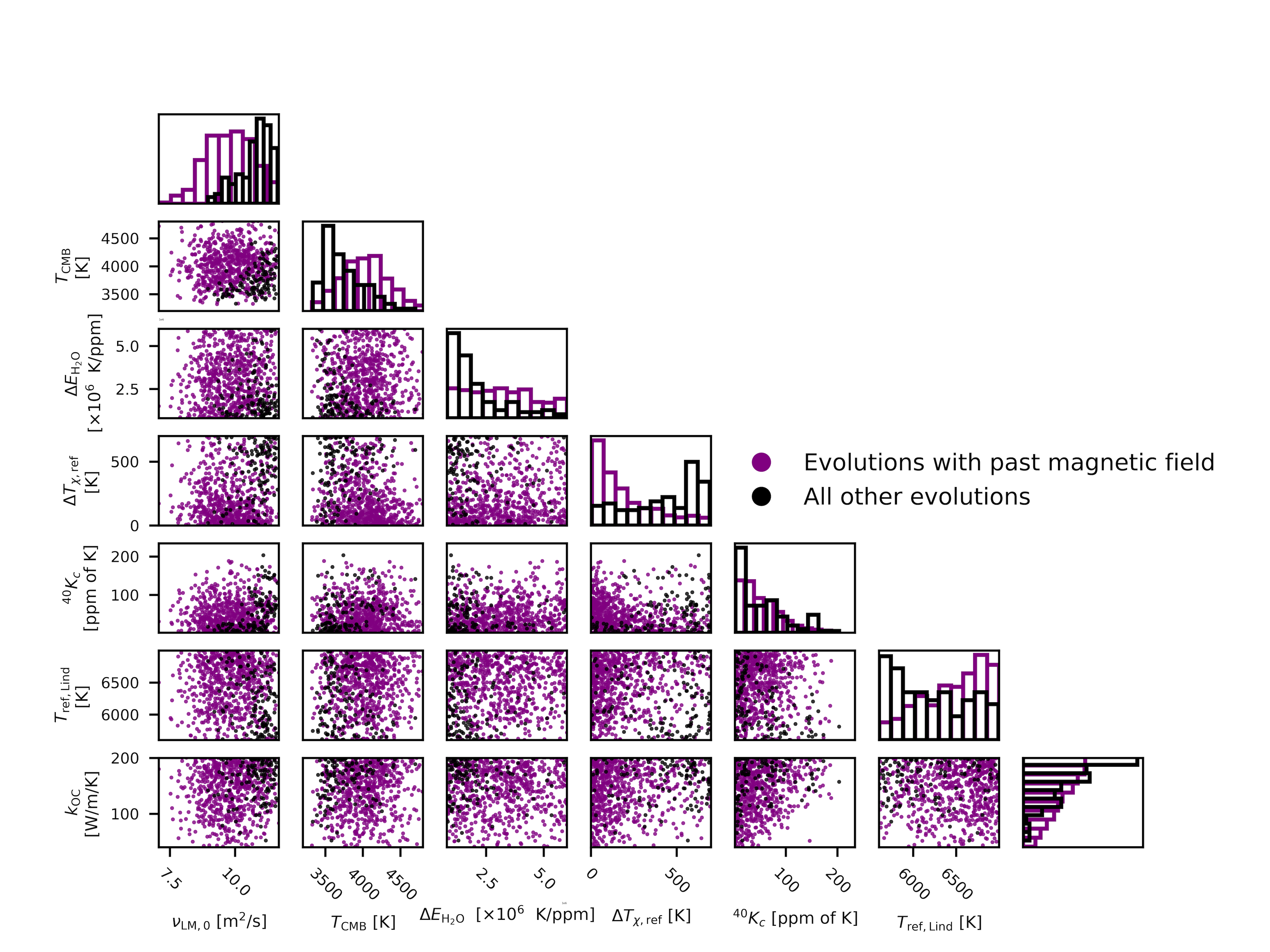}
    \caption{The scatter plots in this corner plot show where every evolution with a past magnetic field lies in a subset of the full parameter space defined by six variables: the initial temperature at the core-mantle boundary ($T_\mathrm{CMB}$), the strength of water-viscosity coupling ($\Delta E_\mathrm{H_2O}$), the reference light element temperature depression ($T_{\chi,\mathrm{ref}}$), the amount of $^{40}$K in the core ($^{40}$K$_c$), the reference core liquidus temperature ($T_\mathrm{ref,Lind}$), and the reference viscosity for the lower mantle ($\nu_\mathrm{LM,0}$). The histograms show the distribution of evolutions with a past magnetic field (purple) and those without (black) marginalized over all parameters except for each parameter below them (or to the left in the case of the reference core liquidus temperature). \label{fig:PastMagMomFeatureTriangle}}
\end{figure*}

To better understand the correlations between these different parameters, we plot their relationships in Fig. \ref{fig:PastMagMomFeatureTriangle}.
Evolutions that exhibit a magnetic field in the past compared to evolutions that don't prefer a higher initial CMB temperature ($>4000$ K), lower viscosity in the lower mantle, higher $\Delta E_{H_2O}$, and lower $\Delta T_{\chi,\mathrm{ref}}$.
We can see in Fig. \ref{fig:CoreFeatureImportance}, that the reference lower mantle viscosity, $\nu_\mathrm{LM,0}$, is incredibly important for determining whether an evolution had a past magnetic field, which is expected as viscosity determines the vigor of convection and consequently the thermal dynamo.
This dependence on viscosity is likely why the viscosity activation energy depression due to water, $\Delta E_\mathrm{H_2O}$, also appears to partially select for magnetization, with higher values (and correspondingly lower viscosities) being preferred to generate a magnetic field.
Type III evolutions, being defined by a high low mantle viscosity, exist in a parameter space that does not permit a past magnetic field as commonly as non-Type III evolutions.

The role of the mantle viscosity in the evolution of both the mantle and the core dynamo demonstrates the importance of the whole-planet approach for properly modeling a core's dynamics \citep{foley_whole_2016}.
Furthermore, since the mantle's dynamics are influenced by volatile cycling between the mantle, crust, and atmosphere, it is clear that the best way to simulate a planet's core is to couple it to the entire planet.
This whole-planet interaction is best seen in the Type II+III evolutions, where the breakdown of crustal recycling and resulting dehydration stiffening characteristic of Type II evolutions leads to a high mantle viscosity that shuts off the core dynamo.
Ultimately the viscosity of Venus' mantle remains an unknown value and will depend on a more accurate estimation of the planet's composition and water content.

Another interesting result we can see from Figs. \ref{fig:CoreFeatureImportance}--\ref{fig:PastMagMomFeatureTriangle} is that the thermal conductivity of the outer core, $k_\mathrm{OC}$ does not seem to be as important a controller of past magnetic dynamos as it is in other interior models \citep[e.g.,][]{nimmo_influence_2004,orourke_prospects_2018}.
Across a wide range of potential $k_\mathrm{OC}$ values (40-200) W/m/K \citep[e.g.,]{ohta_experimental_2016,konopkova_direct_2016} our model results in a number of evolutions that exhibit a past magnetic dynamo. 
This result may be due to our model exploring a wider range of lower mantle viscosity values than other works, which allows for convection even with high $k_\mathrm{OC}$ values due to low mantle reference viscosities, and strong coupling to water resulting in even lower mantle viscosities and more vigorous convection.

Our model has shown that the conditions required for Venus to have a largely liquid core today, a high lower mantle viscosity, are the same conditions that make it difficult to power a thermally or compositionally driven dynamo.
However, there are a number of unmodeled phenomena that must be accounted for before we can claim a non-detection of a remanent magnetic field implies Venus has a small inner core or vice-versa.
For example an insulating basalt layer as in \citet{armann_simulating_2012}, could cause a core dynamo to shut down even with parameter values that seem amenable to a dynamo in our model.
Additionally, an ancient Venusian dynamo could have, as has been hypothesized for Earth \citep{ZieglerStegman2013,Stixrude2020,Blanc2020}, been generated by a basal magma ocean \citep{Orourke2020}, the presence of which would not only change our Type III dynamo evolution, but also the core and mantle thermal evolution.
We also don't take into account the possibility that the core may have some preserved primordial compositional stratification that could prevent dynamo generation \citep{jacobson_formation_2017} regardless of any other lower mantle or core parameters.
These complications present potential degeneracies in interpretations derived from our model, degeneracies that can be alleviated by assessing both the remnant magnetization on Venus' surface, and the state of its core. Future work should consider these possibilities.

Because the current Venus mission suite does not plan to measure global remnant magnetization \citep{Widemann2023}, measurements of the core state could thus be applied to estimate the presence of a past magnetic field on Venus.
VERITAS and EnVision radio science instruments are planned to greatly decrease the uncertainties on both the moment of inertia and $k_2$ Love number measurements to the percent or sub-percent level \citep{dumoulin_tidal_2017,Cascioli2021,Widemann2023}.
These uncertainties will be enough to determine the radius of the core to 20 km \citep{Cascioli2021} as well as the state of the core, with $k_2$ uncertainties low enough to resolve whether the Venusian inner core is fully liquid, fully solid, or partially solid \citep{dumoulin_tidal_2017}.
While not a conclusive measurement, along with other anticipated data such as the crustal thickness and eruption rate, could provide greater insight into the overall thermal and dynamo evolution of Venus.

\subsection{Requirements for oxygen sinks in Venus evolutionary models}

\begin{figure*}[htp!]
    \includegraphics[width=\linewidth]{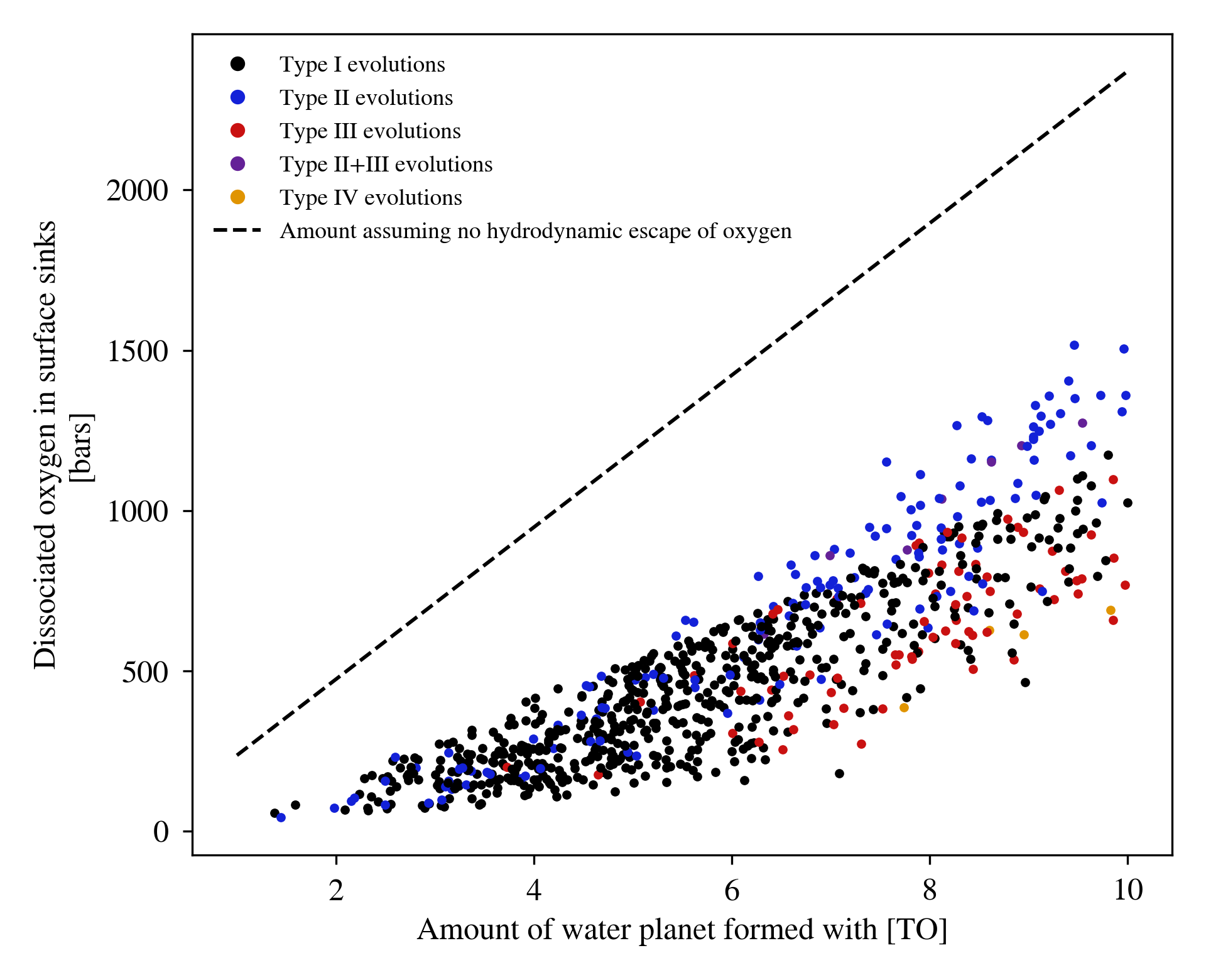}
    \caption{This scatter plot shows how much oxygen is left in the planetary system at the present day for each simulation, based on how much water that simulation initially formed with. Each simulation is color-coded by identified evolutionary type. While some oxygen is carried away by hydrogen during hydrodynamic escape, this plot shows how much is not carried away and must be absorbed by surface sinks or removed non-thermally to reproduce the oxygen-poor Venusian atmosphere. The dashed line represents the upper limit of oxygen production, showing how much oxygen would be in the atmosphere from photolysis if no oxygen was carried away during hydrodynamic escape and Venus lost all of the water it formed with. \label{fig:LeftoverOxygen}}
\end{figure*}

While our model does not explicitly account for surface oxygen sinks on Venus, we do keep track of how much oxygen photodissociated from water does not escape to space.
We plot these amounts in Fig.~\ref{fig:LeftoverOxygen}, where we show how many bars of oxygen must be absorbed by surface sinks as a function of Venus' primordial water abundance.
The amount of atmospheric oxygen not pulled to space by H in hydrodynamic escape must be lost from the atmosphere due to a combination of phenomena unmodeled in this work, including: nonthermal escape, surface oxidation, oxidation of reduced outgassed species, and oxidation of a magma ocean \citep{gillmann_consistent_2009,gillmann2020,krissansen-totton_was_2021,warren_narrow_2023}.

Magma ocean exchange is likely the most efficient way to remove hundreds of bars of oxygen \citep{gillmann_consistent_2009,gillmann2020} since surface sinks and the oxidation of reduced outgassed species can only remove on the order of tens of bars of oxygen \citep{gillmann_consistent_2009,warren_narrow_2023}.
However, even magma oceans have a limited amount of oxygen they can absorb, with estimates of magma oceans being able to absorb around 500 bars of oxygen \citep{gillmann_consistent_2009}, meaning that our higher oxygen amount simulations could potentially be ruled out as resulting in a Venus with a significant oxygen atmosphere.

Assuming that Venus' magma ocean phase lasts around 100 Myrs \citep{hamano_emergence_2013}, we find that photodissociation in our simulations produces around 2 to 450 bars of oxygen from water outgassed after the solidification of a magma ocean.
We further find that Type II simulations tend to produce more atmospheric oxygen than non-Type II simulations.
Any simulation that produces a large amount of atmospheric oxygen, especially post-magma ocean, is likely to be less compatible with Venus' evolution.
The timing and production of oxygen is especially relevant to the history of Venus as it informs discussions of late volatile delivery on Venus, which must be drier so as to not result in a significant post-magma ocean atmosphere \citep{gillmann2020} and the allowed amount of water in a habitable Venus scenario \citep{warren_narrow_2023}.

\subsection{Venus is currently volcanically active}
All of our simulations that reproduce Venus are volcanically active in the present day.
This result is consistent with both recent arguments that Venus must still be volcanically active \citep{filiberto_present-day_2020}, and potential detections of active volcanism on Venus \citep{herrick_surface_2023,sulcanese_evidence_2024}. Our results therefore support the possibility that the high-resolution surface monitoring capabilities of VERITAS \citep{smrekar_2022} and EnVision \citep{ESA2021} could detect ongoing volcanic activity.

\begin{figure*}[htp!]
    \includegraphics[width=\linewidth]{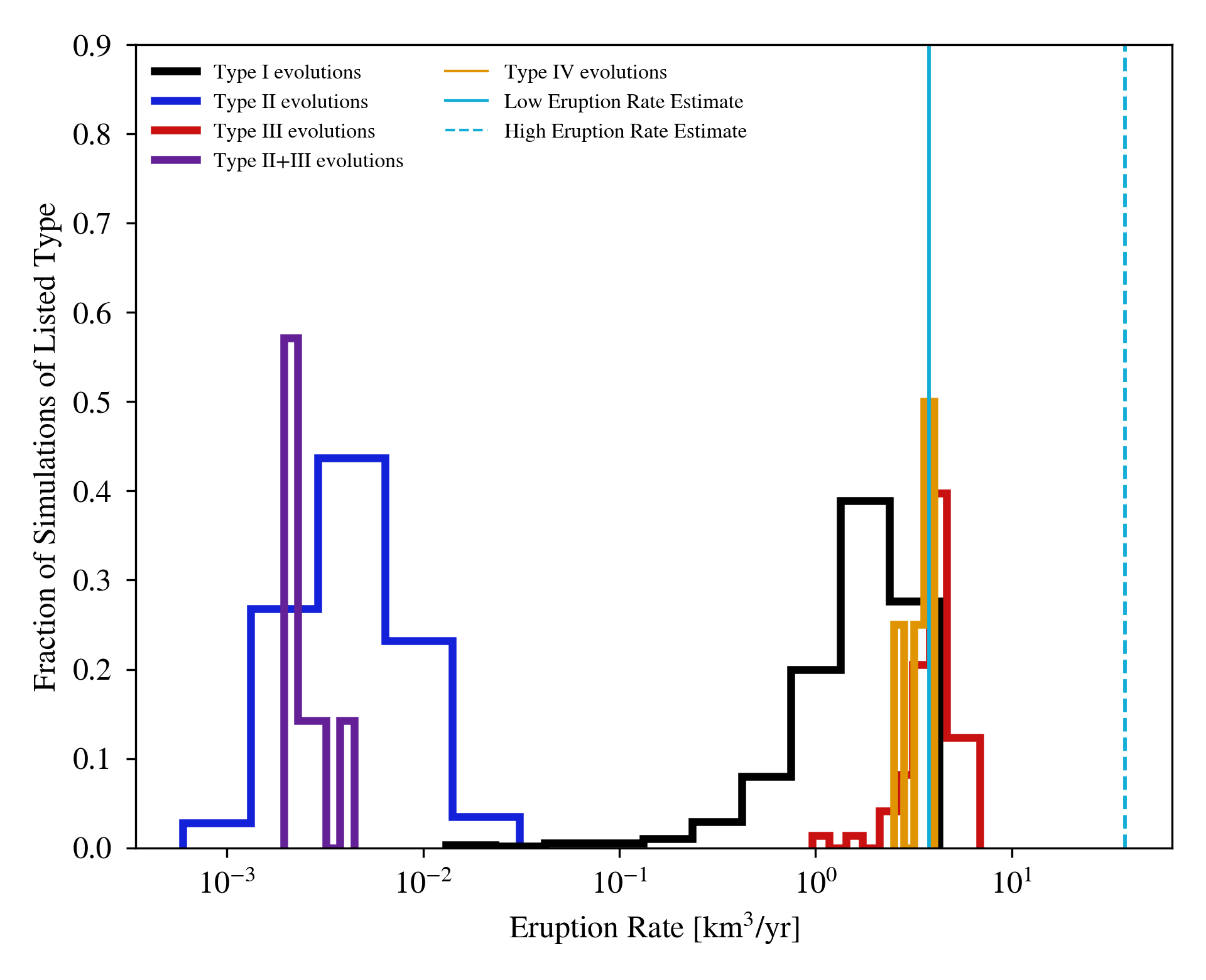}
    \caption{Histograms of the present-day eruption rate for each of our evolutionary types. Overplotted on the histograms are vertical lines corresponding to eruption rates inferred using the broad range of eruption rates calculated in \citet{sulcanese_evidence_2024} The low eruption rate (3.78 km$^3$/yr) estimate (solid cyan line) assumes a minimum thickness of 3 meters for the putative Sif Mons lava flow on Venus. The high eruption rate (37.8 km$^3$/yr) estimate assumes a maximum thickness of 20 meters for the putative Niobe Planitia lava flow on Venus.\label{fig:EruptionEventPlot}}
\end{figure*}

Reanalysis of Magellan data has found a number of surface features that appear to be the result of recent volcanic eruptions \citep{herrick_surface_2023,sulcanese_evidence_2024}.
\citet{sulcanese_evidence_2024} used two observations of likely new lava flows at Sif Mons and Niobe Planitia to derive a range of possible global eruption rates on Venus.
Using an assumption of 42 volcanic eruptions on Venus per year, they calculated a minimum eruption rate range of 3.78-5.67 km$^3$/yr and a maximum eruption rate range of 25.2-37.8 km$^3$/yr. These eruptive mass flux estimates can only be cautiously compared to our simulations because the surface deformations occurred over just a few years, while our model effectively outputs kyr averages. Nonetheless, some interesting implications emerge from such a comparison.

Some Types end with eruptive mass fluxes that are consistent with observations, as shown in Fig.~\ref{fig:EruptionEventPlot}.  The \citep{sulcanese_evidence_2024} minimum eruption rate range matches the upper end of our Type I, Type III, and Type IV simulations, while our Type II and Type II+III simulations are not compatible. Should this result be borne out, it would imply that Venus' mantle is not in the final stages of desiccation. Although we did not include eruption rate in our list of requirements for success, it is reassuring that the most common Type in our sample is consistent with this emerging observable.

With only a few percent of Venus' surface examined, there is still a significant opportunity to discover dozens of active volcanoes and refine this comparison. Should many detections be made, then comparisons between observations and 1D models would be more meaningful. A high frequency of events would suggest that long-term averages are a reasonable representation of the behavior on shorter timescales. Large, rare events are of course still possible, but a large number of active regions would suggest that the internal pressure is being relieved in a more continuous than discrete manner. We conclude that a complete accounting of current volanic activity could offer new and important constraints on 1D models of Venus.

While all of our different types of evolutions are volcanically active today, Type II evolutions exhibit a recent sharp decline in volcanism that will continue towards melt cessation in the near future, as can be seen in Fig. \ref{fig:TimeComparison}.
This result is not due to a cooler mantle but instead due to dehydration stiffening that results in a significantly more viscous mantle, which increases the thickness of the stagnant lid and thermal boundary layer, which in turn raises the solidus temperature.
We find that this process is likely irreversible: the timescale required to radiogenically heat the mantle enough to sufficiently decrease the viscosity is larger than the radioactive decay timescale. In the case that Venus is ultimately not fit by a Type II evolution, this scenario is still a useful template for future rocky planet thermal evolution studies.

\subsection{Implications for Exoplanet Science}
A better understanding of Venus is not just valuable for investigating the history of our nearest planetary neighbor, it also provides insight into the evolution of rocky planets in general \citep{Way2023}.
While only one evolutionary scenario can ever be true for Venus, terrestrial exoplanets could potentially exhibit any of the evolutionary types identified in Section 4.
Modeling terrestrial exoplanets as Venus analogs is motivated by the sheer number of terrestrial exoplanets detected that either currently orbit \citep{Ostberg2023}, or once orbited \citep{godolt_habitability_2019} interior to the habitable zone.
These planets either are currently in the runaway greenhouse stage of atmospheric loss or were once subject to it \citep{Kopparapu2013,luger_extreme_2015}, just like Venus.
In this work we have created a template of parameter values that we can use to model rocky exoplanets as Venus-like.

The stagnant lid geodynamic regime presents a useful endmember case to model the evolution of exoplanets: the strong dependence of viscosity on temperature and the dearth of mobile lid planets in our own solar system lend credence to the argument that stagnant lids will be common \citep{Stern2018}.
Thus, Venus provides a useful testbed to calibrate investigations of the influences of a stagnant lid on exoplanetary processes, especially the cycling of volatiles on stagnant lid exoplanets \citep{noack_coupling_2012,tosi_habitability_2017,foley_carbon_2018,godolt_habitability_2019,Honing_2021}.
Understanding the cycling of volatiles on an exoplanet is a key aspect of determining that planet's potential habitability \citep{foley_whole_2016,Noack2017,VanHoolst2019,kane_venus_2019,Dehant2019}.

We found across all our evolutions that stagnant lid-type crustal recycling \citep{foley_carbon_2018} is efficient enough to prevent complete desiccation.
This result is especially important for exoplanets around low-mass M stars, that currently orbit in the habitable zone but earlier orbited interior to the habitable zone \citep{Tuchow2023} due to the extended pre-Main sequence phase present on these low-mass stars \citep{luger_extreme_2015}.
Based on our results for Venus, we expect that stagnant lid exoplanets should be able to sequester water in their interiors before the planet enters the habitable zone, helping maintain a water content amenable to habitability once they enter the habitable zone \citep{moore2023}.
This possibility is especially relevant for many of the habitable zone TRAPPIST-1 planets (especially d and e), which spent hundreds of million years interior to their star's habitable zone \citep{Lincowski2018}.

Among our identified evolutionary types, the Type II evolution appears to have the strongest implications for a planet's volatile cycling and interior history, especially if the planet is older and melting has ceased.
While our simulation length of 4.5 Gyrs is relevant to reproducing Venus, the age of the planet is a relatively unconstrained parameter when it comes to exoplanet observations, especially for planets around potentially long-lived M dwarfs.
The TRAPPIST-1 system, for example, may be older than the solar system, with estimations of its age at around 7 Gyrs \citep{Burgasser2017,Fleming2020} and uncertainties on the order of 1-2 Gyrs \citep{Birky2021}. 
These wide age uncertainties span the half-lives of key heat-producing radioisotopes in our model, such as $^{40}$K and $^{235}$U.
If the Type II evolution represents a general trend that planets follow when they cool enough to stop producing melt, it may be likely that such older exoplanetary systems, whose radioisotope supply has decayed significantly, could follow these kinds of evolutionary paths and cease melting.

\subsection{Outlook}
While we have successfully created a model that simultaneously reproduces core, surface, and atmospheric properties, our model is not completely self-consistent as it leaves some planetary phenomena unmodeled.
In this section, we highlight both future potential additions to our model and the potential influence of unmodeled factors on our results. Future work could address these issues to create a better terrestrial planet model and provide a clearer understanding of Venus in particular.

Our model currently lacks a radiative transfer model to dynamically calculate the surface temperature as Venus outgasses more water and carbon dioxide into its atmosphere.
For this reason, we impose a constant surface temperature of 700 K in all of our simulations.
Much work on Venus has pointed out the importance that the surface temperature plays as a boundary condition for mantle convection \citep{bullock_grinspoon_2001,armann_simulating_2012,noack_coupling_2012,driscoll_divergent_2013,weller_physics_2020,Honing_2021}.
The surface temperature, evolving with the outgassing of greenhouse gases such as water and carbon dioxide, is even hypothesized to play a role in a transitional regime of Venus' geodynamics \citep{weller_physics_2020}.
While it's likely that self-consistently simulating the surface temperature with a radiative transfer model could change our results, given the little influence of the first few million years of our simulations (i.e., the transient behavior) on the rest of the evolution, relaxing this assumption would probably not dramatically affect our results.

Given our lack of a self-consistent surface temperature, we do not explicitly modeling the solidification of the magma ocean and thus also do not model the geochemistry of the magma ocean.
The length of the magma ocean stage can dramatically affect the partitioning of volatiles at the beginning of a planet's evolution, particularly CO$_2$ that is quickly outgassed and would generate a large atmosphere at the time of mantle solidification \citep{elkins-tanton_linked_2008,hamano_emergence_2013,krissansen-totton_was_2021,Barth2021,Carone2025}.
In an extended magma ocean case, it is likely that Venus may lose significantly more water, resulting in less water being maintained in the mantle \citep{hamano_emergence_2013}.
Such a drier mantle would likely result in generally higher viscosities, with correspondingly higher mantle temperatures.
Higher viscosities could also prevent core cooling as well as a thermal dynamo, and thus we would potentially see more simulations that look like Type III evolutions.
Relatedly, we do not model a basal magma, which could drastically change the heat flow from the core and change our magnetic field results both due to thermal effects and because the basal magma ocean itself could generate a dynamo \citep{Orourke2020}.

Lastly, future research should incorporate statistical methodologies in order to assign probabilities to each evolutionary track.
By considering a probability distribution of evolutionary scenarios, we could more easily incorporate new data into our models and results via Bayesian updating.
Additionally, the probability of any particular evolution can incorporate softer constraints, for example while all of our simulations match Venus based on our \textit{a priori} constraints set in Table \ref{constraint_table}, we have made the argument here that the Type II evolutions are less likely as they result in higher crustal thicknesses and lower eruption rates than are expected for Venus.
Having a corpus of probabilities will also better enable future work as we can create likely Venus templates that could be applied to model exoplanets, see e.g., \cite{SealesLenardic2020}. Given the large number of free parameters in the model, Markov chain Monte Carlo methods may struggle to converge and/or be tractable, but at least the results presented herein have shown that 1-D whole planet models can reproduce core, surface, and atmospheric properties. Moreover, these statistical methods may be able to determine if the lack of a geodynamo today actually favors a mostly solid initial core, or if the large fraction of cases we found in this state was just chance.

\section{Conclusions} \label{sec:conclusion}
To investigate the diversity of evolutionary scenarios around Venus, we have developed a whole-planet model of Venus that simulates the thermal, magnetic, surface, and volatile evolution of Venus with a stagnant lid.
With our model we performed simulations that varied a number of different parameters that dictate the dynamics of the atmosphere, crust, mantle, and core.
This approach successfully reproduced modern Venus' magnetic moment, the atmospheric abundances of CO$_2$ and H$_2$O, and is consistent with recent estimates of the eruptive mass flux.

From the simulations that reproduce current observations of Venus, we identified four distinct evolutionary scenarios that underscore qualitatively different planetary dynamics, see Sec. \ref{sec:results}.
The Type I evolutions experience a smooth decline in mantle temperature and eruption rate dictated by the temperature-viscosity feedback characteristic of previous parameterized convection models.
The Type II evolutions are characterized by a shutdown in crustal recycling of water to the mantle, resulting in significant dehydration stiffening due to the water-viscosity feedback, which results in very little present day melt production.
The Type III evolutions are characterized by a smaller inner core ($<0.8R_c$) at present day, with hot a outer liquid core that cannot cool efficiently or generate a dynamo due to significantly high mantle viscosities.
The Type IV evolutions are characterized by apparently damped oscillatory behavior in the first 500 Myrs.
We also identified a hybrid Type II+III scenario in which Venus' mantle experienced approximately Type II evolution, while the core exhibited characteristically Type III behavior.

In all of our evolutionary scenarios Venus is volcanically active today.
Our models are thus in agreement with the growing evidence for active volcanism on Venus \citep{filiberto_present-day_2020,gulcher_corona_2020,herrick_surface_2023,sulcanese_evidence_2024}.
While all of our evolutionary scenarios exhibit active volcanism, the modern-day eruption rate of Type II evolutions is significantly lower than all other evolutionary scenarios, with melt production in Type II evolutions likely shutting off in the near future.
Estimating the current rate of Venus volcanism from the observation in \citet{sulcanese_evidence_2024} appears to rule out Type II evolutions as resulting in eruption rates that are too low, see Fig.~\ref{fig:EruptionEventPlot}, but we stress that this result is preliminary and would greatly benefit from more spatial and temporal data. 
Our results also appear to rule out the maximal eruption rate model of \citet{sulcanese_evidence_2024}.

We found that self-consistently modeling the flow of water throughout Venus' mantle, crust, and atmosphere was critical to accurately model the water-viscosity feedback, which is a key controller of mantle dynamics.
In all of our evolutionary scenarios we found that an eclogite-driven crustal recycling mechanism \citep{foley_carbon_2018} enabled at least a terrestrial ocean of water to remain in Venus' mantle today.
While Venus' D/H ratio provides evidence that it lost a significant amount of water in its past, our model results predict that it could still be harboring water in its mantle and so may not be a truly desiccated world.

We found a variety of different magnetic field evolutionary scenarios in our model results, with about 85\% of our models producing magnetic fields in the past, a hypothesis that could be tested by analyzing Venusian surface rocks \citep{orourke_2019}.
We found that the existence of past magnetic fields spanned the full ranges we sampled of core conductivity values and the amount of potassium in the core, contrasting with previous work that emphasized the importance of the conductivity in allowing a dynamo.

Our results highlight the value of the whole planet approach to simulating terrestrial planets. Not only does our model demonstrate how the atmosphere affects the core, the different viable histories predict different values for the Venusian inner core radius, eruption rate, and crustal thickness, all of which will be better constrained by future orbiters, landers, and potentially aerial platforms. We also find that core heating of the mantle does not necessarily follow a simple exponential decay, as assumed for example in \citep{krissansen-totton_was_2021}, and thus not self-consistently modeling core dynamics can result in an incomplete picture of how a planet's mantle and atmosphere have evolved.
Moreover, the validation of such a model against a nearby and well-characterized planet means that this model can now be applied to other worlds, both in the Solar System and beyond.
Ultimately models such as the one presented here can be used to interpret spectral data from any terrestrial planet and aid in constraining interior properties, which are extremely challenging to constrain across interstellar distances.
As terrestrial planets are also expected to be the most likely abodes of extra-terrestrial life, whole planet models of the future could also include metabolic processes and eventually be employed to characterize potential biosignatures on exoplanets.

\section{Acknowledgments}
We thank Josh Krissansen-Totton and Nicholas Wogan for helpful comments and thoughts throughout the research and writing process. RG would also like to acknowledge Zoe Snape for her support. RG was financially supported by the National Science Foundation Graduate Research Fellowship Program No. DGE-1762114 as well as the NASA grant 80NSSC20K0229 and the Virtual Planetary Laboratory NASA grant 80NSSC18K0829. RB was similarly supported by both NASA grants as well as NASA award 80NSSC23K02, while VSM and MTG were supported by the Virtual Planetary Laboratory grant. The simulations in this paper were run on the Hyak supercomputer system at the University of Washington.

\bibliographystyle{aasjournal.bst}
\bibliography{References}
\end{document}